%% file: INGRIDCCIncPaper.tex
\documentclass[
aps,
prd,
twocolumn,
amsmath,
amssymb,
superscriptaddress,
altaffilletter,
groupedaddress
reprint
]{revtex4-1}  
\usepackage{graphicx}
\usepackage{url}
\usepackage{subfigure}
\usepackage{lineno}
\usepackage{epstopdf}
\usepackage[colorlinks=true,linkcolor=black]{hyperref}

\graphicspath{{figs/}}

\begin{document}

\title{
Measurement of the 
muon neutrino
inclusive charged-current cross section in the energy range of 1-3~GeV
with the T2K INGRID detector}

\date{\today}
\input{authors.tex}

\begin{abstract}
We report a measurement of the 
$\nu_\mu$-nucleus inclusive
charged current 
cross section (=$\sigma^{cc}$)
on iron using data from 
exposed to the J-PARC neutrino beam.
The detector consists of 14~modules in total,
which are spread over a range of off-axis angles from
0$^\circ$ to 1.1$^\circ$.
The variation in the neutrino energy spectrum
as a function of
the off-axis angle,
combined with event topology information, is used
to calculate this cross section as a function of neutrino energy.
The cross section 
is measured to be
$\sigma^{cc}(1.1\text{~GeV}) = 1.10 \pm 0.15$ $(10^{-38}\text{cm}^2/\text{nucleon})$,
$\sigma^{cc}(2.0\text{~GeV}) = 2.07 \pm 0.27$ $(10^{-38}\text{cm}^2/\text{nucleon})$, and 
$\sigma^{cc}(3.3\text{~GeV}) = 2.29 \pm 0.45$ $(10^{-38}\text{cm}^2/\text{nucleon})$,
at energies of 1.1, 2.0, and 3.3~GeV, respectively.
These results are consistent with the
cross section calculated by the
neutrino interaction generators currently used by T2K.  More importantly, the method 
described here opens up a new way to determine the energy dependence of
neutrino-nucleus cross sections.
\end{abstract}

\maketitle

\input{intro}

\input{ingrid}

\input{mc}

\input{analysis}

 \input{systematics}

\input{result}

\input{conclusion}

\begin{acknowledgments}
We thank the J-PARC staff for superb accelerator performance and the CERN NA61 
collaboration for providing valuable particle production data.
We acknowledge the support of MEXT, Japan; 
NSERC (grant SAPPJ-2014-00031), NRC and CFI, Canada;
CEA and CNRS/IN2P3, France;
DFG, Germany; 
INFN, Italy;
National Science Centre (NCN), Poland;
RSF, RFBR and MES, Russia; 
MINECO and ERDF funds, Spain;
SNSF and SERI, Switzerland;
STFC, UK; and 
DOE, USA.
We also thank CERN for the UA1/NOMAD magnet, 
DESY for the HERA-B magnet mover system, 
NII for SINET4, 
the WestGrid and SciNet consortia in Compute Canada, 
and GridPP, UK.
In addition participation of individual researchers
and institutions has been further supported by funds from: ERC (FP7), H2020 RISE-GA644294-JENNIFER, EU; 
JSPS, Japan; 
Royal Society, UK; 
DOE Early Career program, USA.
\end{acknowledgments}

\bibliographystyle{unsrt}
\bibliography{INGRIDCCIncPaper}

\end{document}

%% file: authors.tex

\newcommand{\INSTEE}{\affiliation{University of Bern, Albert Einstein Center for Fundamental Physics, Laboratory for High Energy Physics (LHEP), Bern, Switzerland}}
\newcommand{\INSTFE}{\affiliation{Boston University, Department of Physics, Boston, Massachusetts, U.S.A.}}
\newcommand{\INSTD}{\affiliation{University of British Columbia, Department of Physics and Astronomy, Vancouver, British Columbia, Canada}}
\newcommand{\INSTGA}{\affiliation{University of California, Irvine, Department of Physics and Astronomy, Irvine, California, U.S.A.}}
\newcommand{\INSTI}{\affiliation{IRFU, CEA Saclay, Gif-sur-Yvette, France}}
\newcommand{\INSTGB}{\affiliation{University of Colorado at Boulder, Department of Physics, Boulder, Colorado, U.S.A.}}
\newcommand{\INSTFG}{\affiliation{Colorado State University, Department of Physics, Fort Collins, Colorado, U.S.A.}}
\newcommand{\INSTFH}{\affiliation{Duke University, Department of Physics, Durham, North Carolina, U.S.A.}}
\newcommand{\INSTBA}{\affiliation{Ecole Polytechnique, IN2P3-CNRS, Laboratoire Leprince-Ringuet, Palaiseau, France }}
\newcommand{\INSTEF}{\affiliation{ETH Zurich, Institute for Particle Physics, Zurich, Switzerland}}
\newcommand{\INSTEG}{\affiliation{University of Geneva, Section de Physique, DPNC, Geneva, Switzerland}}
\newcommand{\INSTDG}{\affiliation{H. Niewodniczanski Institute of Nuclear Physics PAN, Cracow, Poland}}
\newcommand{\INSTCB}{\affiliation{High Energy Accelerator Research Organization (KEK), Tsukuba, Ibaraki, Japan}}
\newcommand{\INSTED}{\affiliation{Institut de Fisica d'Altes Energies (IFAE), The Barcelona Institute of Science and Technology, Campus UAB, Bellaterra (Barcelona) Spain}}
\newcommand{\INSTEC}{\affiliation{IFIC (CSIC \& University of Valencia), Valencia, Spain}}
\newcommand{\INSTEI}{\affiliation{Imperial College London, Department of Physics, London, United Kingdom}}
\newcommand{\INSTGF}{\affiliation{INFN Sezione di Bari and Universit\`a e Politecnico di Bari, Dipartimento Interuniversitario di Fisica, Bari, Italy}}
\newcommand{\INSTBE}{\affiliation{INFN Sezione di Napoli and Universit\`a di Napoli, Dipartimento di Fisica, Napoli, Italy}}
\newcommand{\INSTBF}{\affiliation{INFN Sezione di Padova and Universit\`a di Padova, Dipartimento di Fisica, Padova, Italy}}
\newcommand{\INSTBD}{\affiliation{INFN Sezione di Roma and Universit\`a di Roma ``La Sapienza'', Roma, Italy}}
\newcommand{\INSTEB}{\affiliation{Institute for Nuclear Research of the Russian Academy of Sciences, Moscow, Russia}}
\newcommand{\INSTHA}{\affiliation{Kavli Institute for the Physics and Mathematics of the Universe (WPI), The University of Tokyo Institutes for Advanced Study, University of Tokyo, Kashiwa, Chiba, Japan}}
\newcommand{\INSTCC}{\affiliation{Kobe University, Kobe, Japan}}
\newcommand{\INSTCD}{\affiliation{Kyoto University, Department of Physics, Kyoto, Japan}}
\newcommand{\INSTEJ}{\affiliation{Lancaster University, Physics Department, Lancaster, United Kingdom}}
\newcommand{\INSTFC}{\affiliation{University of Liverpool, Department of Physics, Liverpool, United Kingdom}}
\newcommand{\INSTFI}{\affiliation{Louisiana State University, Department of Physics and Astronomy, Baton Rouge, Louisiana, U.S.A.}}
\newcommand{\INSTJ}{\affiliation{Universit\'e de Lyon, Universit\'e Claude Bernard Lyon 1, IPN Lyon (IN2P3), Villeurbanne, France}}
\newcommand{\INSTHB}{\affiliation{Michigan State University, Department of Physics and Astronomy,  East Lansing, Michigan, U.S.A.}}
\newcommand{\INSTCE}{\affiliation{Miyagi University of Education, Department of Physics, Sendai, Japan}}
\newcommand{\INSTDF}{\affiliation{National Centre for Nuclear Research, Warsaw, Poland}}
\newcommand{\INSTFJ}{\affiliation{State University of New York at Stony Brook, Department of Physics and Astronomy, Stony Brook, New York, U.S.A.}}
\newcommand{\INSTGJ}{\affiliation{Okayama University, Department of Physics, Okayama, Japan}}
\newcommand{\INSTCF}{\affiliation{Osaka City University, Department of Physics, Osaka, Japan}}
\newcommand{\INSTGG}{\affiliation{Oxford University, Department of Physics, Oxford, United Kingdom}}
\newcommand{\INSTBB}{\affiliation{UPMC, Universit\'e Paris Diderot, CNRS/IN2P3, Laboratoire de Physique Nucl\'eaire et de Hautes Energies (LPNHE), Paris, France}}
\newcommand{\INSTGC}{\affiliation{University of Pittsburgh, Department of Physics and Astronomy, Pittsburgh, Pennsylvania, U.S.A.}}
\newcommand{\INSTFA}{\affiliation{Queen Mary University of London, School of Physics and Astronomy, London, United Kingdom}}
\newcommand{\INSTE}{\affiliation{University of Regina, Department of Physics, Regina, Saskatchewan, Canada}}
\newcommand{\INSTGD}{\affiliation{University of Rochester, Department of Physics and Astronomy, Rochester, New York, U.S.A.}}
\newcommand{\INSTBC}{\affiliation{RWTH Aachen University, III. Physikalisches Institut, Aachen, Germany}}
\newcommand{\INSTFB}{\affiliation{University of Sheffield, Department of Physics and Astronomy, Sheffield, United Kingdom}}
\newcommand{\INSTDI}{\affiliation{University of Silesia, Institute of Physics, Katowice, Poland}}
\newcommand{\INSTEH}{\affiliation{STFC, Rutherford Appleton Laboratory, Harwell Oxford,  and  Daresbury Laboratory, Warrington, United Kingdom}}
\newcommand{\INSTCH}{\affiliation{University of Tokyo, Department of Physics, Tokyo, Japan}}
\newcommand{\INSTBJ}{\affiliation{University of Tokyo, Institute for Cosmic Ray Research, Kamioka Observatory, Kamioka, Japan}}
\newcommand{\INSTCG}{\affiliation{University of Tokyo, Institute for Cosmic Ray Research, Research Center for Cosmic Neutrinos, Kashiwa, Japan}}
\newcommand{\INSTGI}{\affiliation{Tokyo Metropolitan University, Department of Physics, Tokyo, Japan}}
\newcommand{\INSTF}{\affiliation{University of Toronto, Department of Physics, Toronto, Ontario, Canada}}
\newcommand{\INSTB}{\affiliation{TRIUMF, Vancouver, British Columbia, Canada}}
\newcommand{\INSTG}{\affiliation{University of Victoria, Department of Physics and Astronomy, Victoria, British Columbia, Canada}}
\newcommand{\INSTDJ}{\affiliation{University of Warsaw, Faculty of Physics, Warsaw, Poland}}
\newcommand{\INSTDH}{\affiliation{Warsaw University of Technology, Institute of Radioelectronics, Warsaw, Poland}}
\newcommand{\INSTFD}{\affiliation{University of Warwick, Department of Physics, Coventry, United Kingdom}}
\newcommand{\INSTGE}{\affiliation{University of Washington, Department of Physics, Seattle, Washington, U.S.A.}}
\newcommand{\INSTGH}{\affiliation{University of Winnipeg, Department of Physics, Winnipeg, Manitoba, Canada}}
\newcommand{\INSTEA}{\affiliation{Wroclaw University, Faculty of Physics and Astronomy, Wroclaw, Poland}}
\newcommand{\INSTH}{\affiliation{York University, Department of Physics and Astronomy, Toronto, Ontario, Canada}}

\INSTEE
\INSTFE
\INSTD
\INSTGA
\INSTI
\INSTGB
\INSTFG
\INSTFH
\INSTBA
\INSTEF
\INSTEG
\INSTDG
\INSTCB
\INSTED
\INSTEC
\INSTEI
\INSTGF
\INSTBE
\INSTBF
\INSTBD
\INSTEB
\INSTHA
\INSTCC
\INSTCD
\INSTEJ
\INSTFC
\INSTFI
\INSTJ
\INSTHB
\INSTCE
\INSTDF
\INSTFJ
\INSTGJ
\INSTCF
\INSTGG
\INSTBB
\INSTGC
\INSTFA
\INSTE
\INSTGD
\INSTBC
\INSTFB
\INSTDI
\INSTEH
\INSTCH
\INSTBJ
\INSTCG
\INSTGI
\INSTF
\INSTB
\INSTG
\INSTDJ
\INSTDH
\INSTFD
\INSTGE
\INSTGH
\INSTEA
\INSTH

\author{K.\,Abe}\INSTBJ
\author{C.\,Andreopoulos}\INSTEH\INSTFC
\author{M.\,Antonova}\INSTEB
\author{S.\,Aoki}\INSTCC
\author{A.\,Ariga}\INSTEE
\author{S.\,Assylbekov}\INSTFG
\author{D.\,Autiero}\INSTJ
\author{M.\,Barbi}\INSTE
\author{G.J.\,Barker}\INSTFD
\author{G.\,Barr}\INSTGG
% ...

\collaboration{The T2K Collaboration}\noaffiliation

%% file: intro.tex
\section{Introduction}

Many recent 
long baseline neutrino oscillation experiments 
use muon neutrino beams, with neutrino energies ranging
from sub-GeV to a few GeV.  
The observed
neutrino-nucleus charged-current (CC) interactions
are then used to infer neutrino oscillation parameters.
In this energy region, 
CC quasi-elastic 
and CC single-pion production reactions
dominate the total cross section, and so 
understanding 
these channels is 
essential for precision
measurements
of the oscillation parameters.
Measurements of exclusive cross sections, however,
are complicated by re-interactions of 
the final state hadrons
as they exit the nucleus, known as final state interactions (FSI).
FSI can absorb or produce particles, resulting in a different set of particles
entering the detector than would be expected from the initial interaction.  For example,
the pion from a CC single-pion interaction might be absorbed in the nucleus, so the
observable final state 
is similar to
that from a CC quasi-elastic event.
The CC inclusive channel is much less sensitive 
to these effects, 
since it only requires the detection
of a charged lepton (muon) from the interaction.
A precise measurement of this channel, combined with exclusive measurements, 
will help improve our understanding of
neutrino interactions
in this energy region.

So far,
the MINOS and T2K experiments
have measured  the CC inclusive $\nu_\mu$ 
cross section on iron~\cite{Adamson:2009ju,Abe:2014nox}
using neutrino beams which cover the few-GeV region.
The former experiment measured the CC inclusive cross section for neutrinos with energies
above 3.5~GeV using the MINOS near detector.
The latter used the T2K
 near detector, INGRID,
to measure a flux-averaged CC inclusive cross section
at a mean energy of 1.51~GeV, 
where the r.m.s spread of the neutrino energy was 0.76 GeV (0.84 GeV) below (above) this
mean energy. 
The MINER$\nu$A experiment also measured the CC inclusive cross section on iron,
but only the cross-section ratio of iron to CH has been
published~\cite{Tice:2014pgu}.
The CC inclusive cross section on iron 
has not yet been measured in the 2-3~GeV energy range,
and a measurement covering 1-3~GeV
would provide a consistency check between the T2K and MINOS results.

The T2K experiment is 
a long baseline neutrino oscillation experiment in Japan
\cite{Abe:2011ks}.
T2K utilizes an almost pure $\nu_{\mu}$ 
beam produced as the decay product of $\pi^+$'s and $K^+$'s.
The beam is first measured by the near detector, ND280, 
located 280~m downstream from the pion production target.
After traveling 295~km, the
neutrinos are then
observed at the far detector, Super-Kamiokande.
Oscillation parameters are determined by 
comparing 
the neutrino interactions observed
at the near and far detectors.

\begin{figure}[htbp] 
   \centering
   \includegraphics[width=0.4\textwidth]{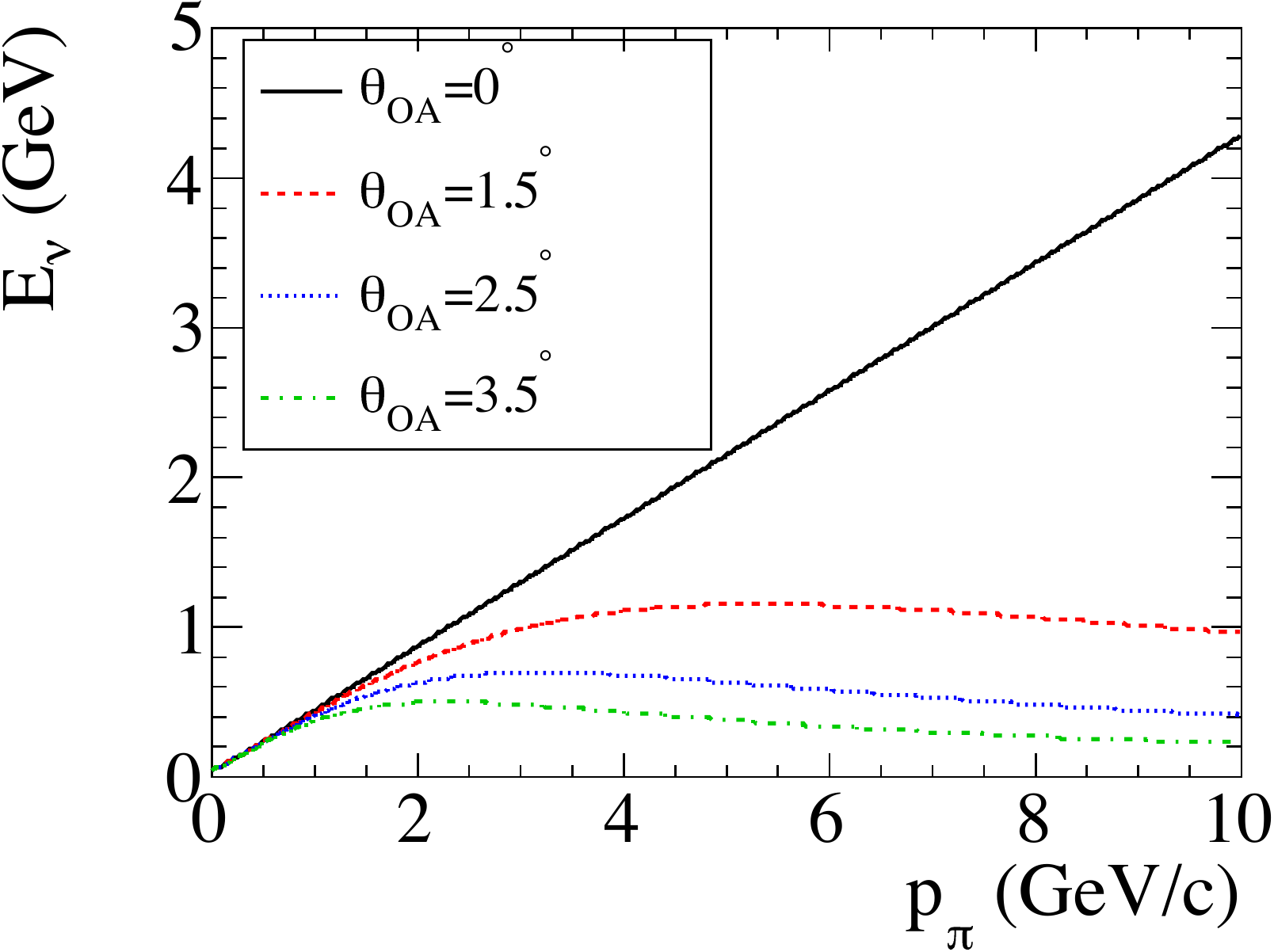} 
   \caption{Neutrino energy as a function of the pion momentum for different $\theta_{\text{OA}}$s.}
  \label{fig:enu_ppi}
\end{figure}

T2K was built around 
the ``off-axis beam method'',
where detectors are intentionally
placed off the central axis of the neutrino beam (hereafter beam-axis).
The angle with respect to the beam-axis is called the off-axis angle: $\theta_{\text{OA}}$.
The direction
of the neutrino parent particles
is 
distributed around the
beam-axis, so
$\theta_{\text{OA}}$ is 
approximately equal to
the angle between the parent particle and neutrino directions.
In this case, the energy
of a neutrino
produced from the two-body decay 
($\pi \rightarrow \nu_\mu + \mu$)
can be expressed as follows:
\begin{align}
E_\nu =
\frac{m^2_\pi - m^2_\mu}{2(E_\pi - p_\pi \cos \theta_{\text{OA}})}\;, 
\label{eq:enu_ppi}
\end{align}
where $m_\pi$ and $m_\mu$ are the masses of the pion and muon whilst
$E_\pi$ and $p_\pi$ are the 
energy and momentum of the pion.
The relation between $E_\nu$ and $p_\pi$ 
for different $\theta_{\text{OA}}$s
is plotted in Fig.~\ref{fig:enu_ppi},
showing the maximum neutrino energy 
reducing as $\theta_{\text{OA}}$ increases. 
This indicates that
the energy spectrum of the neutrino beam peaks at a lower energy and has a narrower width
as $\theta_{\text{OA}}$ increases.
In the T2K experiment, 
$\theta_{\text{OA}}=2.5^\circ$ was chosen so that 
the neutrino flux peaks around 0.6~GeV, an energy
which 
maximizes the oscillation
probability of the muon neutrino
at the far detector.

The T2K INGRID detector is installed on the beam-axis at the near site.
It consists of 14 identical modules,
which are spread over a range of $\theta_{\text{OA}}$ 
from 0$^{\circ}$ to 1.1$^{\circ}$.
Thus, the peak of the neutrino energy spectrum
differs among the modules
as in Eq.~(\ref{eq:enu_ppi}).

In this paper
we present a measurement of the $\nu_\mu$ inclusive CC cross section on iron
in the energy range of 1-3~GeV 
with INGRID.
%
This analysis uses
data collected from 2010 to 2013,
corresponding to $6.27\times10^{20}$ protons on target (p.o.t.).
The neutrino interactions at different INGRID modules,
 which are distributed at different positions and thus observed different beam spectra,
 is used  to extract the energy dependence of the cross section.
The topology of each event,
which is based on 
the kinematics of the outgoing muon,
is also used to further improve the sensitivity 
of this measurement to the neutrino energy, since the two are directly related.
The different energy spectra 
 and event topologies are combined to construct a probability density function (PDF), which is used to measure the cross section
using the least $\chi^2$ method.

The paper is organized as follows.
In Sec.~\ref{sec:t2k},
we describe
the T2K 
near
detector, INGRID.
Section~\ref{sec:mc} introduces
the Monte Carlo (MC) simulation 
used to predict neutrino event rates at the INGRID detector and
describes the systematic uncertainties associated with this model.
Next, the analysis method used to extract the energy 
dependence of the cross section is explained
in Sec.~\ref{sec:analysis} 
with a discussion of the remaining systematics 
in Sec.~\ref{sec:systematics}.
Finally, the result of the analysis is presented in Sec.~\ref{sec:result}.

%% file: ingrid.tex
\section{The T2K near detector: INGRID~\label{sec:t2k}}

\begin{figure}[h] 
   \centering
   \includegraphics[width=0.45\textwidth]{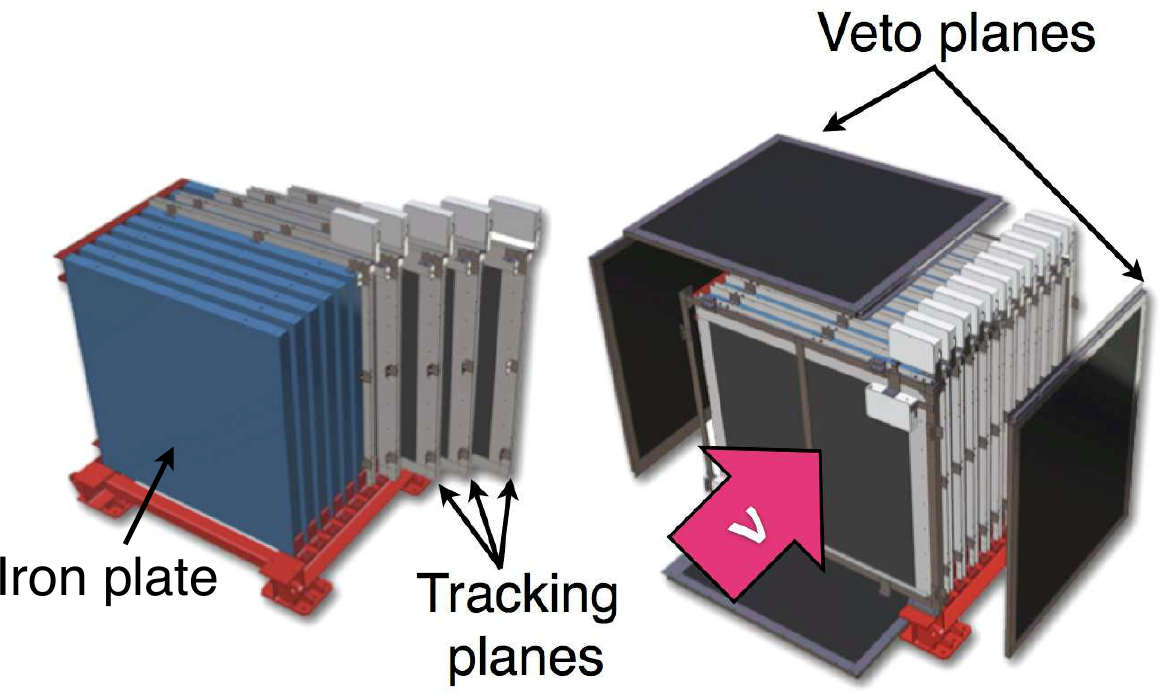} 
   \caption{Schematic drawing of an INGRID module. 
   Each module has 9 iron target plates and
   11 tracking planes, with 4 veto planes covering the side surfaces.
   }
   \label{fig:ingrid_module}
\end{figure}

Situated 280~m downstream from the pion production target,
the INGRID detector monitors the 
neutrino beam direction and intensity.
It consists of 
14 identical modules, each of 
which is composed of 
9 iron target plates
and 11 scintillator tracking planes.
The iron plates and the tracking planes
are stacked in alternating
layers forming a 
sandwich structure, as shown in Fig.~\ref{fig:ingrid_module}.

Each of the iron plates has dimensions of $124 \times 124$~cm$^2$ 
and a thickness of 6.5~cm, providing a total iron mass of 7.1~ton per module.
The module is surrounded by 
scintillator veto planes,
which detect charged particles coming from outside of the module.
Each tracking plane has two layers 
of scintillator bars aligned orthogonally to one another, enabling
particles to be tracked in all 3 dimensions as they pass through the plane.
The veto planes are also formed from scintillator bars.
The bars are coated in $\textrm{TiO}_{2}$ reflectors to help contain scintillation light,
which is then captured by 
wavelength shifting (WLS) fibers which run through the center of the bars.
This light is then read out by a Multi-Pixel Photon Counter (MPPC)~\cite{Yokoyama:2008hn,Yokoyama:2010qa}
and the resultant  
signal is digitized by the Trip-t front-end board (TFB)~\cite{Vacherete:2007}, to 
give the integrated charge and timing information.
The integration cycle of the electronics is synchronized with the neutrino 
beam pulse structure, ensuring all data is captured.

\begin{figure}[!b] 
   \centering
   \includegraphics[width=0.3\textwidth]{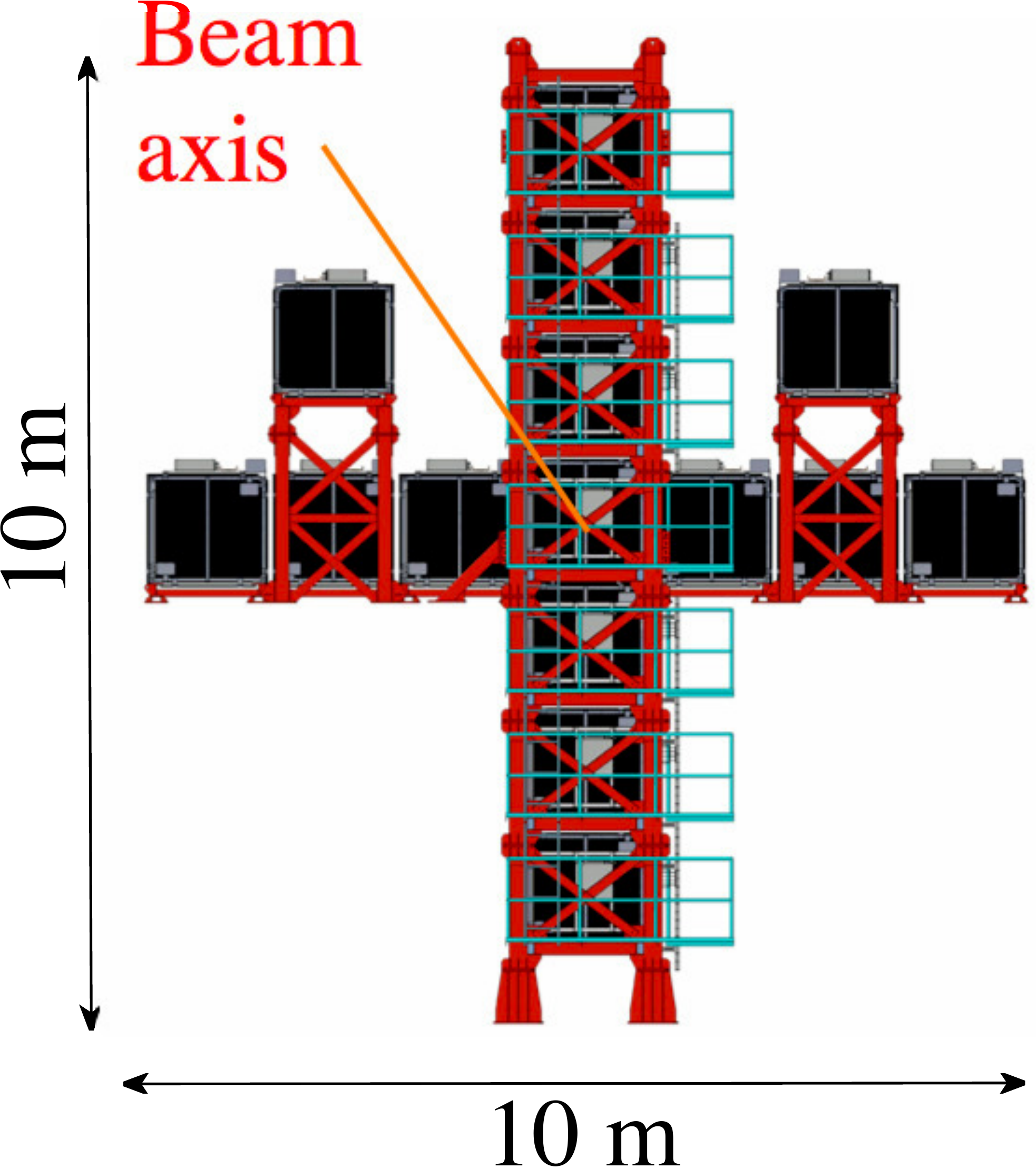} 
   \includegraphics[width=0.35\textwidth]{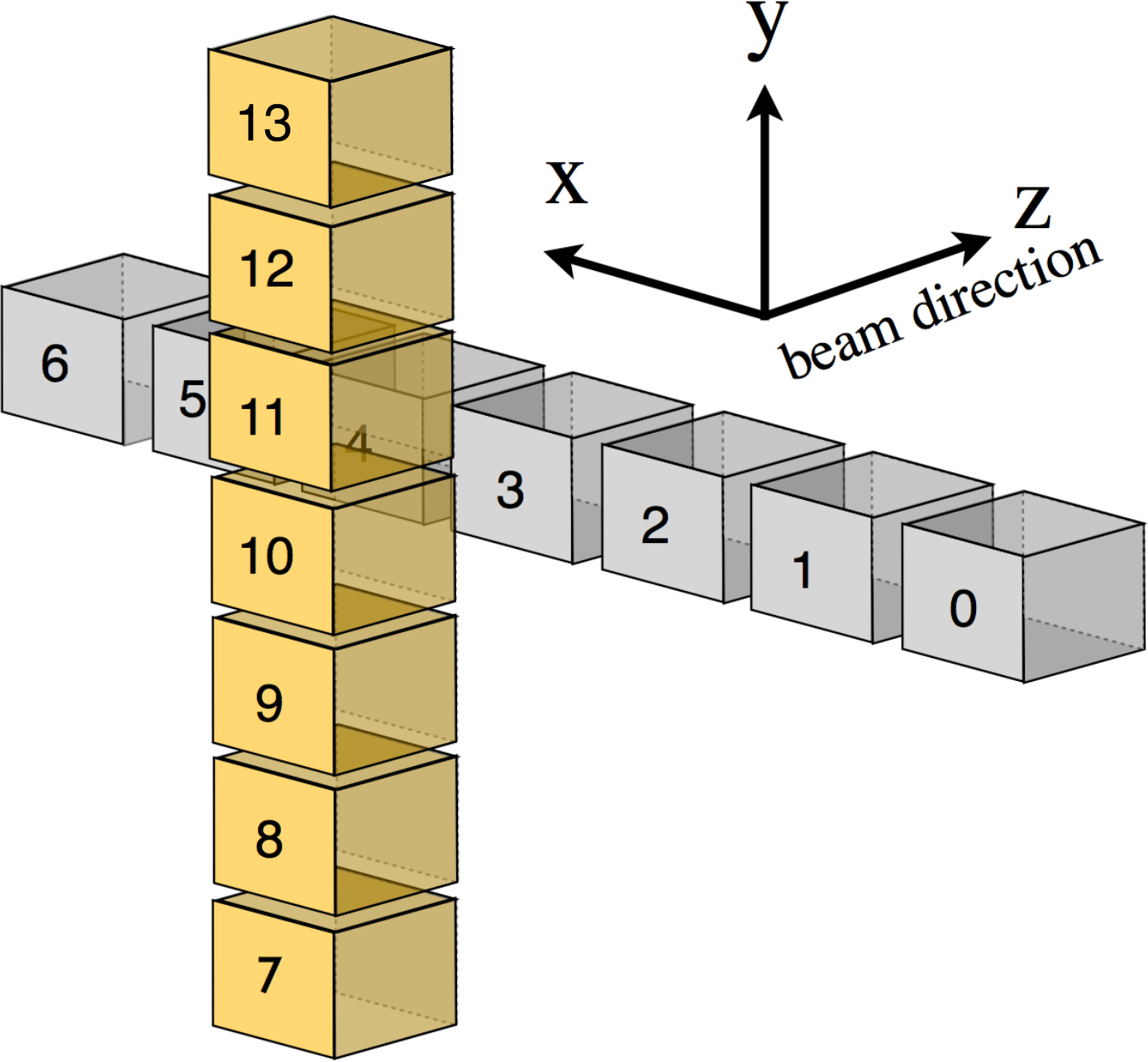} 
   \caption{Top: The INGRID detector. 
   The modules are arranged in a 10~m x 10~m 
cross. The two off-axis modules not located in the arms of the cross
are not used in this analysis.
Bottom: Module ID given to each module.}
   \label{fig:ingrid}
\end{figure}

The modules are installed in a cross shape
centered on the beam-axis.
An overview of the INGRID detector
 is shown on the top in Fig.~\ref{fig:ingrid}. 
 An ID is assigned to each module 
as shown on the bottom in Fig~\ref{fig:ingrid}.
Further details of the detector and the basic performance of INGRID can be found 
in Ref.~\cite{Abe:2011xv}.



%% file: mc.tex
\section{Simulating neutrino interactions in the INGRID detector\label{sec:mc}}

In this analysis the cross section is measured
by comparing data to a prediction,
which is calculated 
using three sequential MC simulations:
\begin{enumerate}
\item Prediction of the neutrino flux at the INGRID modules.
\item Simulation of the neutrino-nucleus interactions in the iron target ($^{56}$Fe).
\item Propagation of final state particles through
the detector and modeling of its response.
\end{enumerate}
After these processes,
an event selection, which is detailed in Sec.~\ref{sec:nu_event_recon},
is applied to the output in the same way as the data.

\input{ingrid_nuflux}

\input{neut}

\input{geant4}

%% file: ingrid_nuflux.tex
\subsection{Neutrino flux\label{subsec:ingrid_nuflux}}
\subsubsection{Flux prediction}
A detailed description of the neutrino flux prediction can be found in Ref.~\cite{Abe:2012av}.
In the simulation, 
protons are impinged upon the carbon target
to produce hadrons, which decay into neutrinos.
FLUKA2008~\cite{Ferrari:2005zk}
and GEANT/GCALOR~\cite{Zeitnitz:1992vw}
are used
to model hadron production in the target and surrounding material.
Propagation of the resulting particles through the electromagnetic horns,
which focus the charged hadrons along the beam-axis, is simulated using dedicated, 
GEANT3~\cite{Brun:1994aa}-based code, which also models the subsequent decay
of the particles.
For each hadron decay mode which produces neutrinos,
the probability of the neutrinos to be emitted
in the direction of the INGRID detector is calculated.
The flux prediction is obtained by weighting the generated neutrinos
with these probabilities.
The flux is then tuned
using hadron interaction data,
primarily
 from the NA61/SHINE experiment~\cite{Abgrall:2014xwa}.
Other hadron production data (Eichten~{\it et al.}~\cite{Eichten:1972nw} 
and Allaby~{\it et al.}~\cite{Allaby:1970jt})
are also used to
tune the simulation in regions of the 
hadron production phase space that are not covered by the current NA61/SHINE measurement.
In this analysis, the NA61/SHINE data taken in 2007 is used to
correct the neutrino flux~\cite{Abgrall:2011ae,Abgrall:2011ts}.

The neutrino flavor content across 
different energy regions at module~3 
(one of the center modules)
is summarized in Table~\ref{tab:nuflux_fraction}.
\begin{table}[htb]
\centering
\caption{Fraction of the integrated flux by neutrino flavor
in each energy range at module~3.}
\begin{tabular}{lrrrrr}
\hline\hline
&\multicolumn{5}{c}{Neutrino energy range (GeV)}\\
Flavor & 0--1 & 1--2 & 2--3 & 3--4 & $>$4 \\
\hline
$\nu_\mu$ & 94.2\%  & 96.8\% & 95.4\% & 89.7\% & 86.5\% \\
$\bar{\nu}_\mu$ & 4.8\% & 2.7\% & 3.8\% & 7.9\% & 9.3\% \\
$\nu_e$ & 0.9\% & 0.5\% & 0.7\% & 2.0\% & 3.5\% \\
$\bar{\nu}_e$ & 0.1\% & 0.0\% & 0.1\% & 0.3\% & 0.6\% \\
\hline\hline
\end{tabular}
\label{tab:nuflux_fraction}
\end{table}
This shows that 
muon neutrinos account
for $>\sim$95\% of the total flux for
 $E_\nu < 3$~GeV.
The muon neutrino flux fraction 
then falls to less than 90\% 
for $E_\nu>$3~GeV, where
$\bar{\nu}_\mu + \nu_e$ account for $>\sim$10\% of the total flux.
The flavor content of the neutrino flux at the other modules
is similar to that at module~3.

\begin{figure}[!h] 
   \centering
      \includegraphics[width=0.4\textwidth]{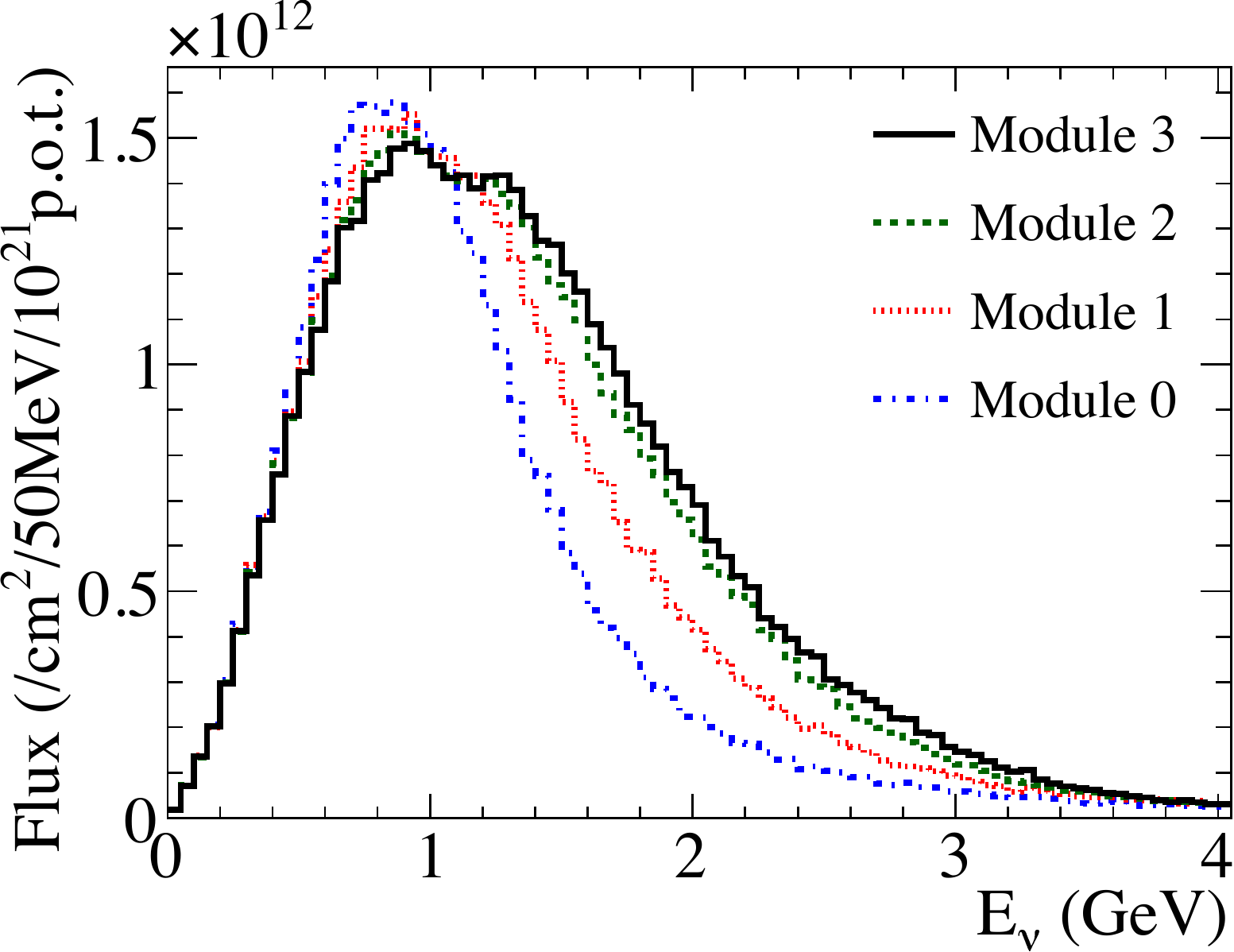} 
   \caption{
   Muon neutrino fluxes 
   at modules~0, 1, 2 and 3.
   }
   \label{fig:ingrid_flux}
\end{figure}

Figure~\ref{fig:ingrid_flux}
shows the obtained muon neutrino flux spectra at the INGRID modules.
The neutrino energy spectrum changes with module position,
with the spectrum at module~0 softer than that at module~3.
This is because module~0 is located at 
$\theta_{OA}=$1.2$^\circ$
from the neutrino beam-axis, and 
so the neutrino flux passing through it 
is shifted to lower energies due to the off-axis beam effect.
This feature is, indeed, essential for this cross-section measurement.

\vspace{-0.5cm}
\subsubsection{Flux uncertainties}
%

\begin{figure}[!h] 
   \centering
 \includegraphics[width=0.45\textwidth]{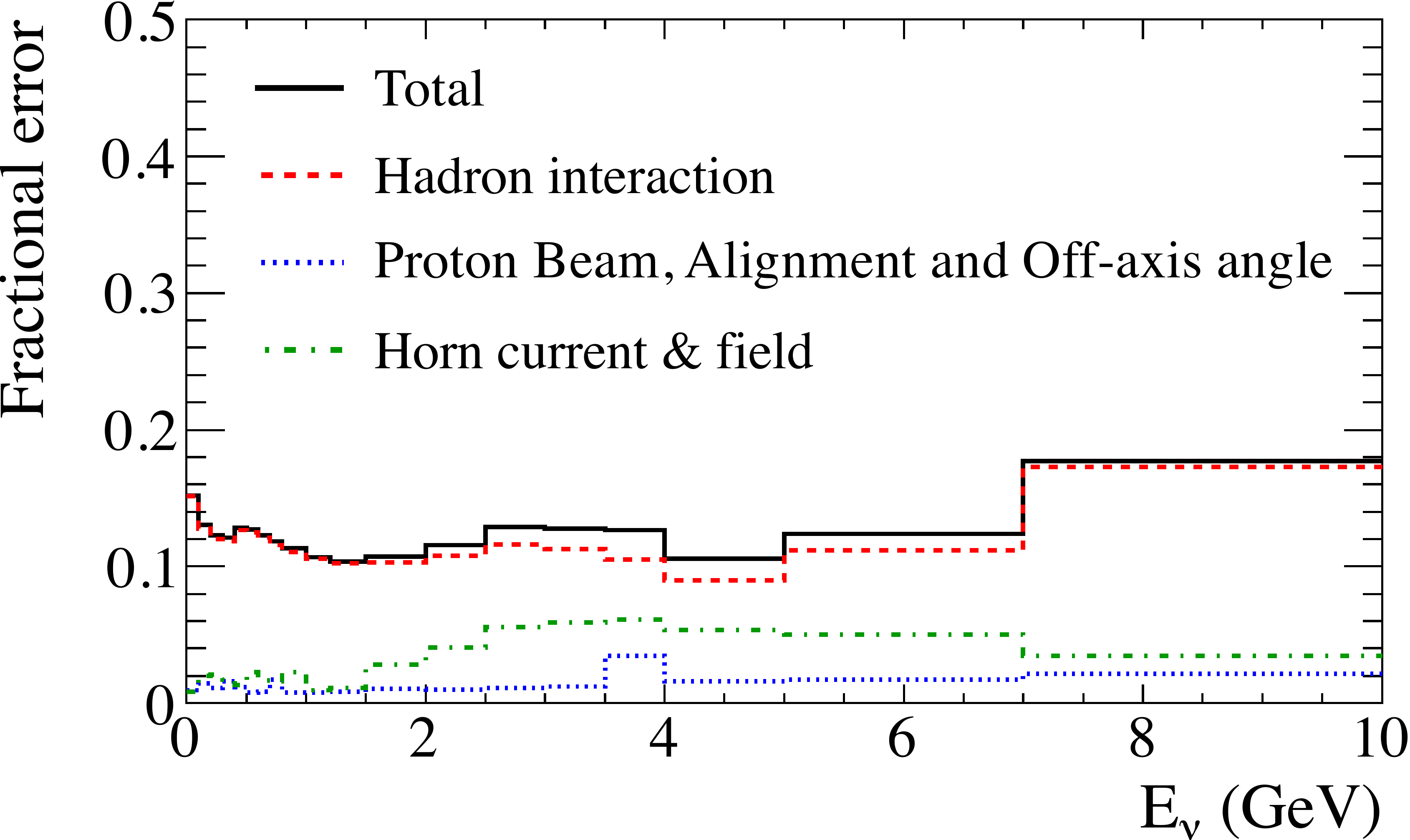} 
   \caption{
   Fractional flux error at module~3,
   including all sources of uncertainty. }
   \label{fig:fluxerr}
\end{figure}

The systematic error on the neutrino flux prediction
comes
from uncertainties in hadron production and from errors in the measurement of the proton beam, the horn current, and the target alignment.
The uncertainties in the hadron production 
are mainly driven by uncertainties in the NA61/SHINE measurements.
and those in measurements by Eichten {\it et al.}
and Allaby {\it et al.}.
The second category of flux errors is associated with 
inherent uncertainties
and operational variations in the beamline conditions. 
They include uncertainties in the proton beam position, 
the beam direction,
the absolute horn current, the horn angular alignment, the horn field asymmetry, the target alignment and the proton beam intensity. 
The method used to estimate these flux uncertainties is described in Ref.~\cite{Abe:2012av}.
Figure~\ref{fig:fluxerr} shows the flux error at module~3,
which includes all sources of uncertainty, and demonstrates that
the systematic error on the neutrino flux
is dominated by the uncertainties in the hadron production model.
The propagation of the flux error in this analysis
is described in Sec.~\ref{subsec:flux_unc}.


%% file: neut.tex
\subsection{Neutrino-nucleus interaction simulation\label{subsec:neut}}

Neutrino-nucleus (Fe) interactions in the INGRID module
are simulated by a neutrino event generator,
which is a composite of different neutrino interaction models.
NEUT (ver.5.1.4.2)~\cite{Hayato:2009zz} is 
used as the primary event generator in this analysis.
GENIE (ver.2.8.0)~\cite{Andreopoulos:2009rq}, 
a different neutrino interaction simulation package,  
is also used for comparison.
This section first describes
the various interaction models simulated in the NEUT event generator, then explains
the systematic uncertainties associated with each model.
Details of the event generators used in T2K 
can be found in Ref.~\cite{Abe:2013jth}.

\subsubsection{The NEUT neutrino event generator}

Given a neutrino energy and a detector geometry,
NEUT determines the interaction mode 
of an event
and calculates the kinematics of the final state particles.
It also simulates FSI for hadrons as they traverse the target nucleus.
The following interaction modes are provided for both charged-current (CC) 
and neutral-current (NC)
interactions by
NEUT:
\begin{itemize}
\item Quasi-elastic scattering (CCQE or NCQE)
\item Resonant pion production (CC1$\pi$ or NC1$\pi$)
\item Deep inelastic scattering (CCDIS or NCDIS)
\item Coherent pion production
\end{itemize}

Here $N$ and $N'$ denote nucleons, $l$ is the lepton, 
and $A$ is the target nucleus.

\subsubsection{Neutrino interaction model uncertainties}

\begin{table*}[t]
\centering
\caption{Neutrino interaction systematic parameters, nominal values, uncertainties (1$\sigma$),   
and interaction types (CC, NC or CC+NC).
The first, second, and third groups represent the model parameters, the ad hoc parameters, and the pion FSI parameters respectively.
A 1 or 0 in the nominal value column means that the effect of 
the systematic parameter is 
implemented or not implemented by default~\cite{Abe:2013xua, Abe:2013fuq}.
}
\label{tbl:nuint_syst}
\scalebox{0.8}{
\begin{tabular}{c c c c}
\hline\hline
Parameter & Nominal value & Uncertainty (1$\sigma$) & Interaction type\\
\hline
$M^{QE}_{A}$   & 1.21 GeV & 0.20 GeV& CC \\
$M^{RES}_{A}$ & 1.21 GeV & 0.20 GeV & CC+NC \\
Fermi momentum (Fe) & 250 MeV/c & 30 MeV/c & CC \\
Binding energy (Fe) & 33 MeV & 9 MeV/c & CC \\
Spectral function  & 0 (off) & 1 (on) & CC \\
$\pi$-less $\Delta$ decay & 0.2 & 0.2 & CC + NC \\
W shape & 87.7 MeV & 45.3 MeV & CC+NC\\
\hline
CCQE normalization ($E_{\nu}\leq1.5$ GeV) & 1 & 0.11 & CC \\
CCQE normalization ($1.5\leq E_{\nu}\leq3.5$ GeV) & 1 & 0.30 & CC\\
CCQE normalization ($E_{\nu}\geq3.5$ GeV) & 1 & 0.30 & CC \\
CC1$\pi$ normalization ($E_{\nu}\leq2.5$ GeV) & 1 & 0.21 & CC\\
CC1$\pi$ normalization ($E_{\nu}\geq2.5$ GeV) & 1 & 0.21 & CC\\
CC coherent normalization & 1 & 1.0 & CC\\
CC other shape & 0 & 
0.1 at $E_\nu$=4.0~GeV & CC\\
NC 1$\pi^0$ normalization & 1 & 0.31 & NC \\
NC coherent $pi$ normalization & 1 & 0.30 & NC\\
NC 1$\pi^{\pm}$ normalization & 1 & 0.30 & NC\\
NC other normalization & 1 & 0.30 & NC\\
1$\pi$ $E_{\nu}$ shape & 0 (off) & 0.50 & CC + NC\\
\hline
Pion absorption & 1 & 0.5 & CC+NC\\
Pion charge exchange ($P_\pi<500$~MeV/c) & 1 & 0.5 & CC+NC\\
Pion charge exchange ($P_\pi>400$~MeV/c) & 1 & 0.3 & CC+NC\\
Pion QE scattering ($P_\pi<500$~MeV/c) & 1 & 0.5 & CC+NC\\
Pion QE scattering ($P_\pi>400$~MeV/c) & 1 & 0.3 & CC+NC\\
Pion inelastic scattering & 1 & 0.5 & CC+NC\\
\hline\hline
\end{tabular}
}
\end{table*}

Table~\ref{tbl:nuint_syst} summarizes the parameters 
used for modeling neutrino interactions in NEUT.
The systematic parameters listed in the table
were 
evaluated
 in the previous analyses of neutrino oscillation from T2K~\cite{Abe:2013xua, Abe:2013fuq}
and fall into the following four categories:
\begin{description}
\item[$M^{QE}_A$, $M^{RES}_A$, and the nuclear model]\mbox{}\\
These parameters are used for modeling CCQE and CC1$\pi$ interactions.
A 20\% error is assigned to the axial vector masses, which comes from a comparison of external measurements of these parameters.
The uncertainty on the Fermi momentum and binding energy are 
estimated from electron scattering data~\cite{Smith:1972xh}.
The uncertainty in the nuclear model
is evaluated 
by exchanging
the RFG nuclear model 
with the spectral function model
described in Ref.~\cite{Benhar:1994af}.

\item[$\pi$-less $\Delta$ decay and W~shape]\mbox{}\\
In the resonant pion production process, baryon resonances, mainly $\Delta$s,
can interact with other nucleons 
in the target nucleus,
and disappear without pion emission. 
The $\pi$-less $\Delta$ decay parameter is introduced 
to take into account the uncertainties in this process.
The W shape parameter is introduced to modify the shape of the
momentum distribution of pions from NC single pion production interactions
so that it matches MinibooNE data~\cite{AguilarArevalo:2009ww}.
\item[Normalization parameters]\mbox{}\\
Normalization parameters are used to change
the overall normalization of the cross section.
The normalizations for CCQE and CC1$\pi$
are defined separately for different energy regions.
The uncertainties on the normalizations are mostly determined from 
the MiniBooNE data.
The CC other shape parameter is introduced
as an energy dependent uncertainty 
on CC DIS and CC resonant interactions,
where the resonance decays into a nucleon and photon, kaon or eta.
According to the MINOS measurement~\cite{Adamson:2009ju},
the relative uncertainty on the CC inclusive cross section on iron,
which is dominated by CCDIS,
is 
approximately 10\% at 4~GeV.
Using this as a reference point, 
the error on the CCDIS and CC resonant cross section is 
scaled using the following formula:
\begin{equation}
\frac{\delta \sigma_{CCother}}{\sigma_{CCother}}
 = \frac{0.4\text{~GeV}}{E_\nu \text{~(GeV)}}\;
\end{equation}
Although the error goes to infinity as $E_\nu$ approaches 0~GeV,
this error does not have a significant contribution
to the total cross section error for the lower neutrino energy region
because
the interactions have a threshold energy of approximately 0.6~GeV
and a small cross section in the 1~GeV energy region.
Finally,
the $1\pi$~$E_{\nu}$ shape parameter
is a weighting factor as a function of neutrino energy.
This is introduced to cover
the discrepancy between the MiniBooNE measurement of 
the CC1$\pi$
cross section versus $E_\nu$ and the NEUT prediction 
using the best fit parameters obtained from the fit to the MinibooNE data~\cite{AguilarArevalo:2010bm}.
This discrepancy is as large as 50\% at 600~MeV, so
an error of 50\% is assigned to this weighting factor.
In the nominal NEUT, 
this weighting is not 
applied (``off'' as in Table~\ref{tbl:nuint_syst}).
\item[Pion FSI]\mbox{}\\
There are three pion FSI processes
of interest in the T2K energy range:
absorption, charge exchange and QE scattering.
In addition to these interactions, 
the particle production process, 
defined as ``inelastic scattering'', was considered, since it is the dominant process at higher pion energies.
Uncertainties on the FSI parameters are estimated using 
external data sets~\cite{dePerio:2011zz}.
\end{description}

Propagation of these uncertainties 
is described in Sec.~\ref{subsec:neut_unc}.

%% file: geant4.tex
\subsection{Detector simulation\label{subsec:geant4}}

The particle type and kinematic information provided by NEUT is passed to the detector simulation
built within a GEANT4 framework~\cite{Agostinelli:2002hh}.
All the detector components
are modeled in the simulation.
The energy deposited by each particle in the scintillator planes
is converted into a number of photo-electrons, taking into account
the non-linear response of the scintillator,
light collection efficiency and attenuation of the WLS fiber,
and the non-linearity of the MPPC response.
The non-linear response of the ADCs
on the front-end electronics
is also modeled 
based on the results of charge injection tests.

Particles generated in the wall upstream of the INGRID detectors
are also propagated into the detector simulation,
and are treated as a background (BG) source.

%

The hadronic interaction of particles in the detector is simulated by GEANT4 using the FTFP\_BERT physics list.
In this physics list, the GEANT4 Bertini cascade model is used 
to simulate nuclear reactions by hadrons 
with kinetic energies below 5.5~GeV.
For particles with kinetic energies above 5~GeV,
the list uses the Fritiof model~\cite{Andersson:1986gw,NilssonAlmqvist:1986rx}.

%% file: analysis.tex
\section{Analysis method~\label{sec:analysis}}

\subsection{Overview}

First, neutrino interactions are selected in each INGRID module.
The different neutrino energy spectra 
sampled by the different modules
provide a way of extracting the energy dependent cross section.
The selected events are further categorized according to the
topology of their final-state muon in order
to improve the sensitivity of the samples to the energy of the incoming neutrino.
A PDF is then constructed 
relating the different INGRID modules and event topologies
to the neutrino energy.
Finally, to extract the CC inclusive cross section,
a $\chi^2$ fit is performed 
between the selected events and the PDFs.

\input{selection}


\input{categorization}


\input{grouping}


\input{ana_detector_unc}


\input{xsec_extraction}


\input{binning}


%% file: selection.tex
\subsection{Neutrino event selection\label{sec:nu_event_recon}}
\begin{figure}[ht] 
   \centering
   \includegraphics[width=0.45\textwidth]{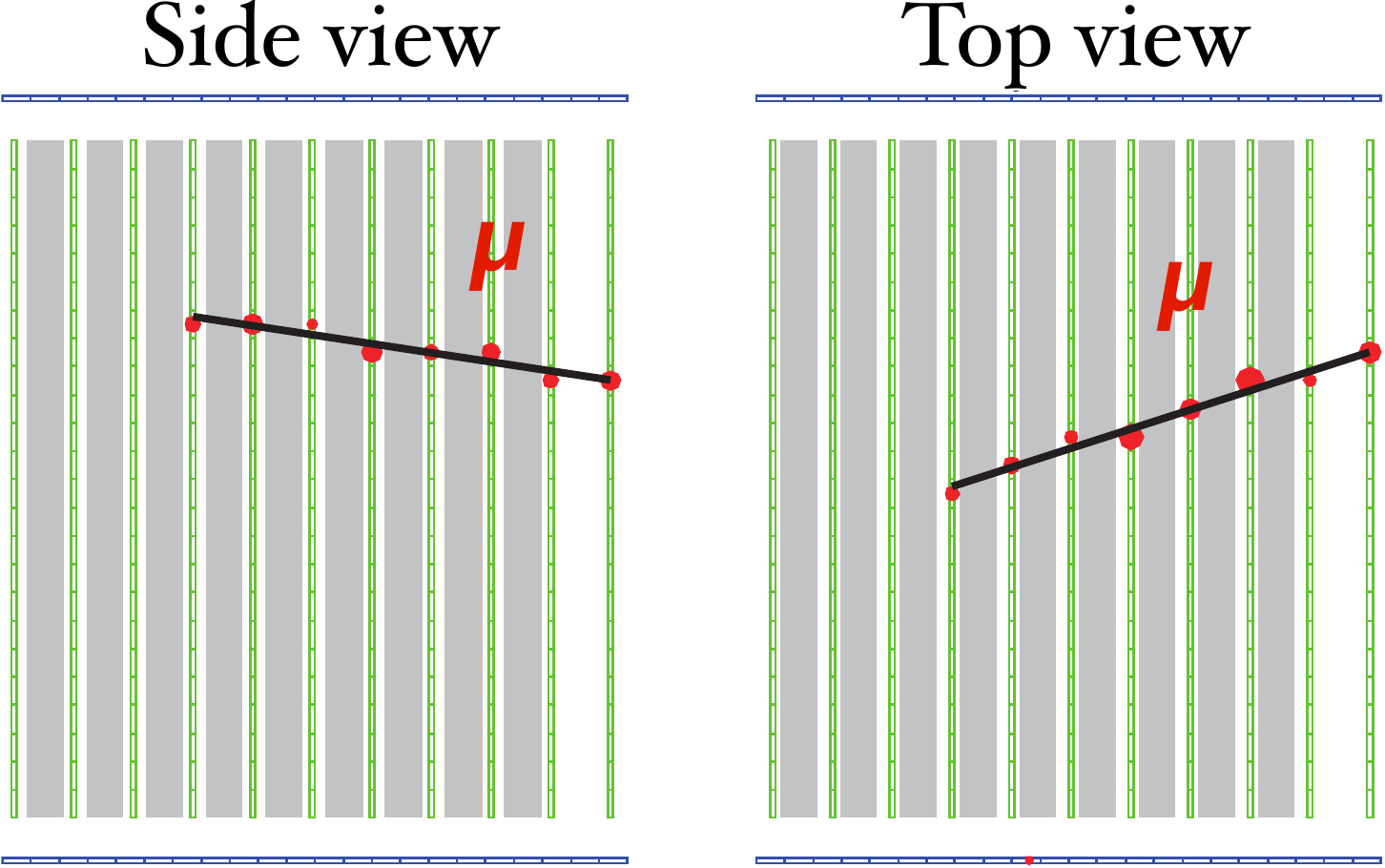} 
   \caption{Event display of a muon-neutrino event candidate.
   Circles and solid lines represent 
   the hits at the scintillator planes
   and reconstructed tracks, respectively.
   The neutrino beam enters from the left side.
   }
   \label{fig:evt_disp}
\end{figure}

A detailed description of the event selection for neutrino interactions at INGRID
can be found in Refs.~\cite{Abe:2014nox,Abe:2015awa}.
A brief explanation of 
each step of the selection is given here.
A typical selected muon-neutrino interaction candidate is shown in Fig.~\ref{fig:evt_disp}.


\begin{description}
\item[1. Pre-selection]\mbox{}\\
The integrated charge and timing of hits in each channel are
recorded with a 2.5 photo-electron threshold.
If there are more than 3 hit channels within a 100~nsec time window,
these hits are combined to form a hit cluster.
The scintillator planes are then
searched to find those with at least one 
hit in both their X and Y oriented layers.
Such a plane is called an ``active plane'',
and events with 2 or more active planes are selected.
\item[2. 2D track reconstruction]\mbox{}\\
A ``cellular automaton'' tracking algorithm~\cite{Wolfram:1982me}
is applied to hits in the X and Y planes
to obtain tracks in the XZ view and YZ view respectively.
\item[3. 3D track matching]\mbox{}\\
The difference in the most upstream layer hit between any two
tracks in the XZ view and YZ view
is used to determine if they originate from the same vertex. 
If this difference is greater than 2 planes then
the tracks are not matched.
\item[4. Vertexing]\mbox{}\\
The vertex is defined as the most upstream hit
of the track.
If there are two or more 
3D tracks,
a check is performed to see if they originate from the same vertex or not.
\item[5. Timing cut]\mbox{}\\
The T2K neutrino beam has eight bunches in each beam spill,
and each bunch has a width of 58~nsec.
The selected events are required to lie within 100~nsec of the expected time
of each bunch.
\item[6. Veto cut]\mbox{}\\
The reconstructed track is extrapolated to 
upstream positions in the side veto planes
and upstream veto planes, and the
event is cut 
if any hits are found near the extrapolated entry point.
\item[7. Fiducial volume cut]\mbox{}\\
The fiducial volume (FV) is defined as a cube with a ($\pm$50)$\times$($\pm$50)~cm$^2$ transverse area, corresponding to the scintillator bars from the 3rd to 22nd channel in the X and Y direction, and from tracking plane\#1 to 8
in the Z direction
(see Fig.~\ref{fig:fv_diagram}).
Events with a vertex inside the FV 
are selected.
\vspace{6mm}
\begin{figure}[!h] 
   \centering
   \includegraphics[width=0.38\textwidth]{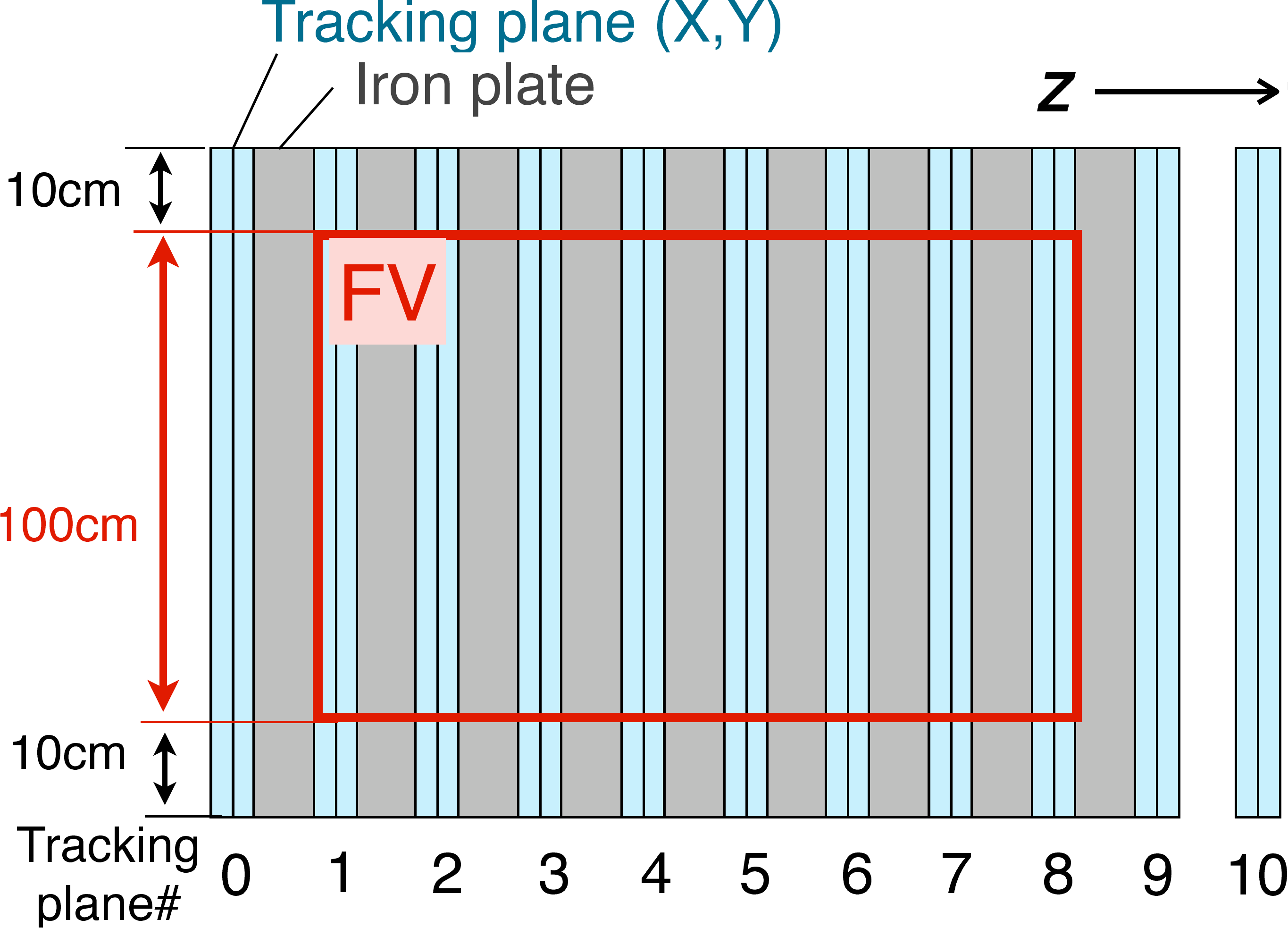}
   \includegraphics[width=0.23\textwidth]{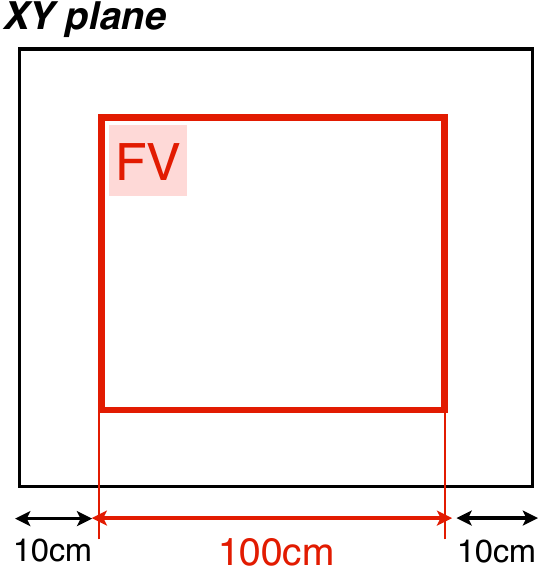}
   \caption{Fiducial volume of the INGRID module
   viewed from of the side (top) and front (bottom).}
   \label{fig:fv_diagram}
\end{figure}
\end{description}


\begin{table*}[!t]
\centering
\caption{Summary of the neutrino event selection. 
The p.o.t. used in the MC simulation
is normalized to 
correspond to the data set 
used in this analysis, i.e. 
$6.27\times10^{20}$.
}
\begin{tabular}{l | c | c c c c}
\hline\hline
& Data & \multicolumn{4}{c}{MC} \\
& & $\nu_\mu$ & $\bar{\nu}_\mu + \nu_e$ & beam-related BG & Total \\
\hline
Vertexing  & $3.993\times10^7$ & $1.655\times10^7$ &$0.039\times10^7$ &$2.294\times10^7$ & $3.987\times10^7$\\
Timing cut & $3.992\times10^7$& $1.655\times10^7$ & $0.039\times10^7$& $2.294\times10^7$& $3.987\times10^7$\\
Veto cut   & $1.725\times10^7$& $1.458\times10^7$ & $0.036\times10^7$& $0.239\times10^7$& $1.733\times10^7$\\
FV cut      & $1.103\times10^7$& $1.098\times10^7$ & $0.027\times10^7$& $0.006\times10^7$& $1.131\times10^7$\\
\hline\hline
\end{tabular}
\label{tab:event_reduc}
\end{table*}

After the event selection above,
corrections are applied to 
the MC to account for differences in the individual iron target masses, 
the observed background rate, and the number of dead channels.
We also apply a correction for event pileup, which depends on the beam intensity and results in a loss of efficiency~\cite{Abe:2015awa}.
Table~\ref{tab:event_reduc} summarizes 
the number of events predicted 
by the MC simulation and observed in the data.

\begin{figure}[!h] 
   \centering
   \includegraphics[width=0.35\textwidth]{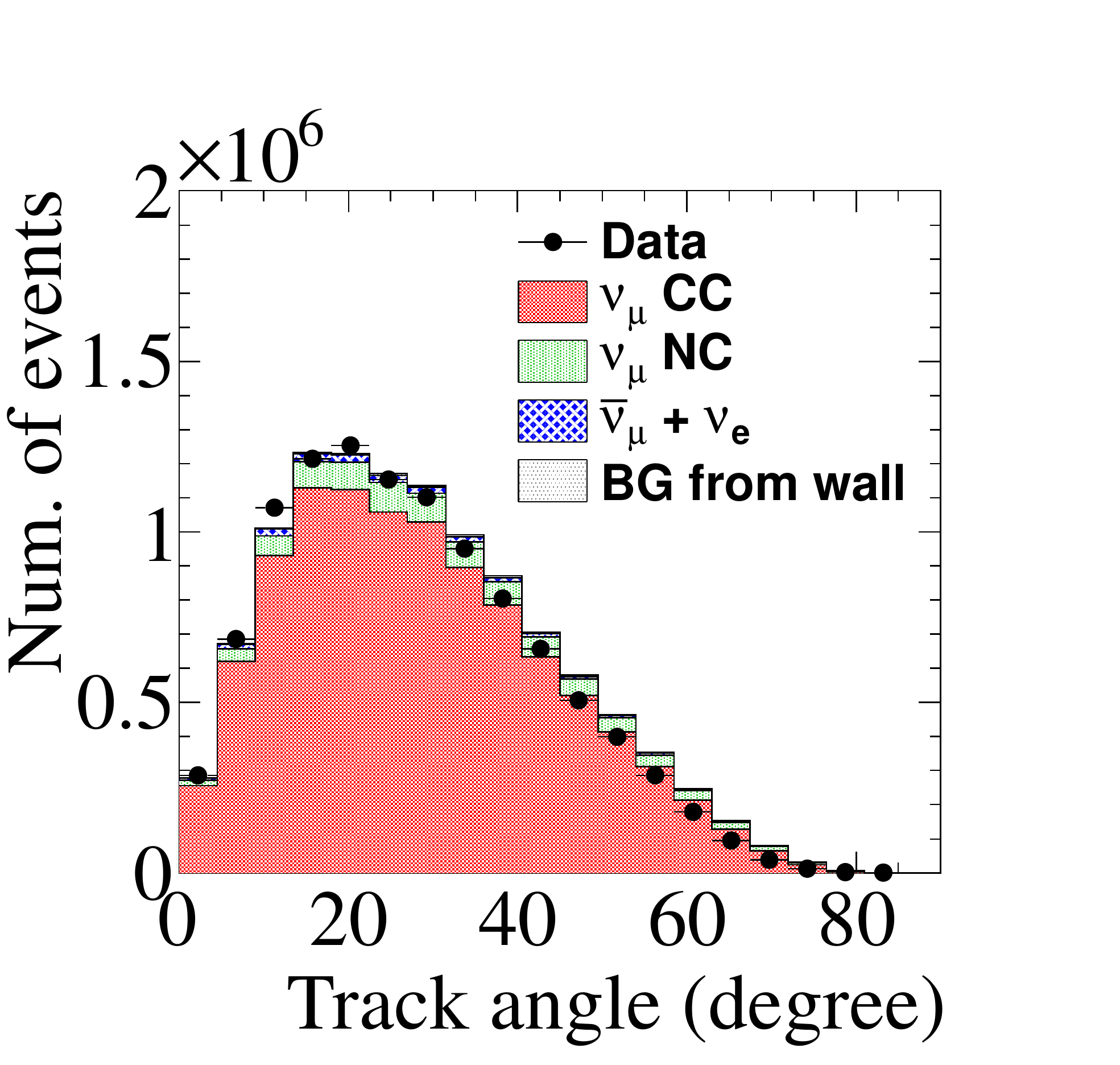} 
   \caption{
   Distribution of the reconstructed track angle 
   with respect to the beam direction
   after the event selection.
   The number of events shown in the figure is the total integrated
   over all modules.
   }
   \label{fig:after_select_b}
\end{figure}

The selected events contain the muon-neutrino signal 
as well as background events, such as 
$\bar\nu_{\mu}$ and $\nu_e$
interactions.
The other backgrounds come from
muons, neutrons and photons generated by neutrino interactions 
out of the FV or 
in the pit wall upstream of INGRID 
(hereafter called beam-related BG).
Since the contamination of $\bar{\nu}_e$ is negligible,
it is not counted in the MC.
The angular distribution of the lepton track for the selected events
is shown in Fig.~\ref{fig:after_select_b}.
Since the vertex is defined 
as the most upstream hit of the track,
the angular acceptance is limited to between 0$^{\circ}$ and 90$^{\circ}$.

\begin{figure}[!h] 
   \centering
    \includegraphics[width=0.35\textwidth]{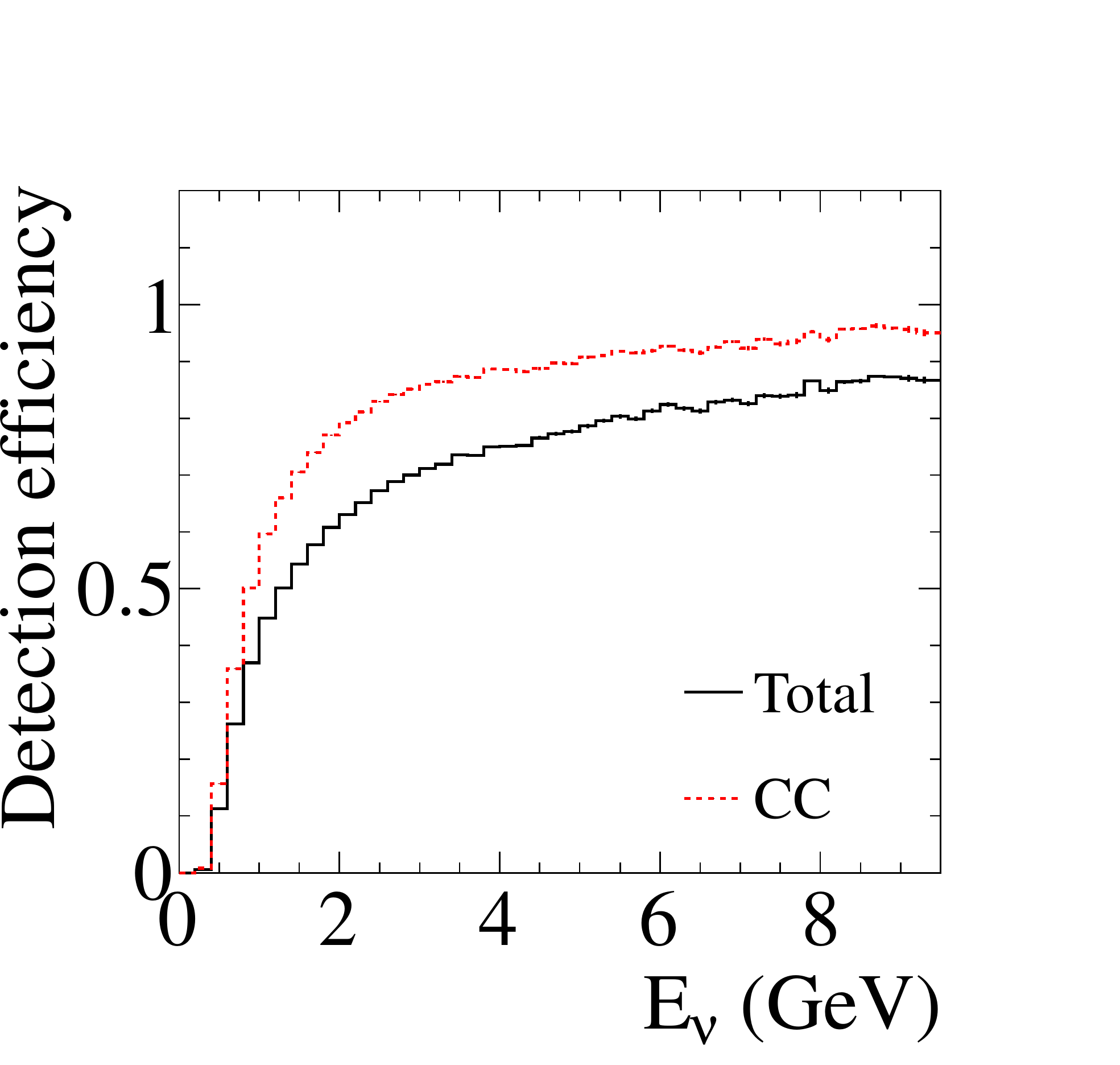} 
   \caption{
	Detection efficiency for CC+NC (solid line) and 
	CC (dashed line) events. 
	These efficiency curves are estimated
	 from the number of events 
	integrated over all modules.
	}
   \label{fig:after_select2_a}
\end{figure}

\begin{figure}[!h] 
   \includegraphics[width=0.35\textwidth]{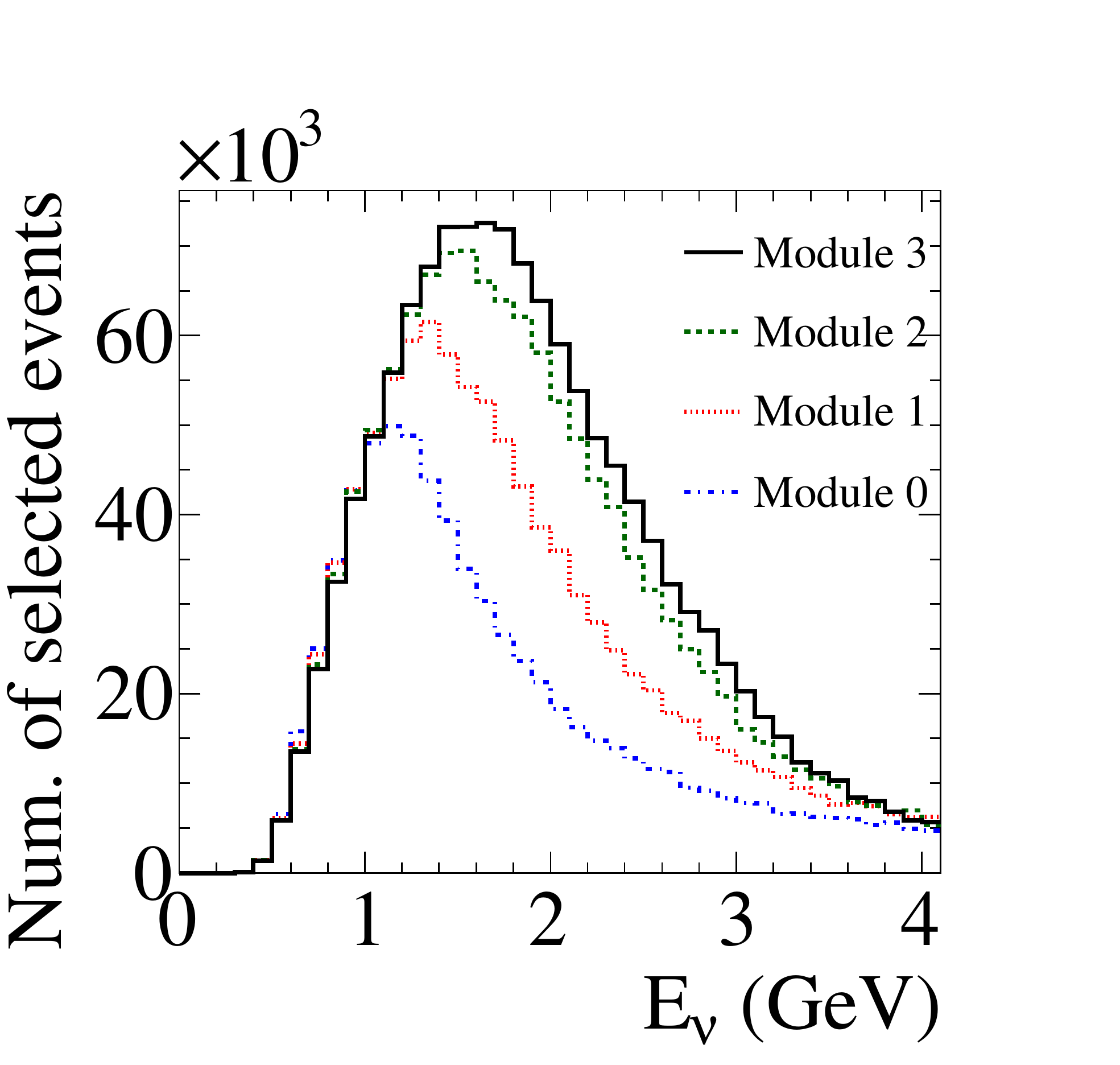} 
   \caption{
     Predicted energy spectrum of the reconstructed events at the different INGRID modules.}
   \label{fig:after_select2_b}
\end{figure}

The final selected event sample has $>$70\% efficiency for
CC interactions from neutrinos with energies $>$1~GeV,
as shown in 
Fig.~\ref{fig:after_select2_a}.
Around 5\% of the selection inefficiency 
in the higher energy region
is due to events 
where the muon is produced at a large angle to the Z axis of the detector.
This results in it escaping the module
before passing through two active planes.
The predicted energy spectrum of the reconstructed $\nu_\mu$ 
events at the INGRID modules
is shown in 
Fig.~\ref{fig:after_select2_b}.

%% file: categorization.tex
\subsection{Event topology}

\begin{figure}[hbt] 
   \centering
   \includegraphics[width=0.5\textwidth]{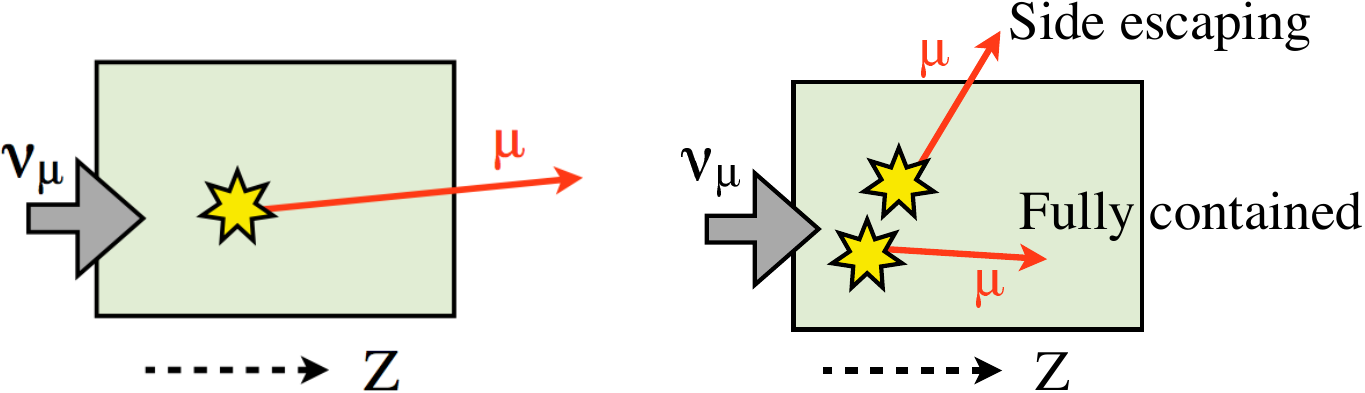} 
   \caption{Different event topologies. 
   If a track penetrates the most downstream plane, the event is categorized as DS-escaping (left). The other events are categorized as NonDS- escaping (right).}
   \label{fig:topology}
\end{figure}

To improve the sensitivity of this analysis to the energy of the neutrino,
the selected events are categorized into the following two topologies:
\begin{enumerate}
\item{Downstream (DS-) escaping}
\item{Non-downstream (NonDS-) escaping}
\end{enumerate}
If one of the tracks from the neutrino interaction
penetrates the most downstream plane,
as shown on the left in Fig.~\ref{fig:topology},
that event is categorized 
as DS-escaping.
All other
events,
i.e. both side escaping and fully contained events 
(see the right plot in Fig.~\ref{fig:topology})
are categorized 
as
NonDS-escaping.

Events are then further categorized according to the reconstructed vertex-Z position.
Vertex-Z is defined as the most upstream active plane number and 
ranges from 1 (most upstream) to 8 (most downstream).
Events whose vertex-Z is in the most downstream plane
are greatly affected by uncertainties in the GEANT4 hadron production model,
so only events with vertex-Z in the range 1-7 are used in this analysis.

\begin{figure}[hbt] 
   \centering
   \includegraphics[width=0.55\textwidth]{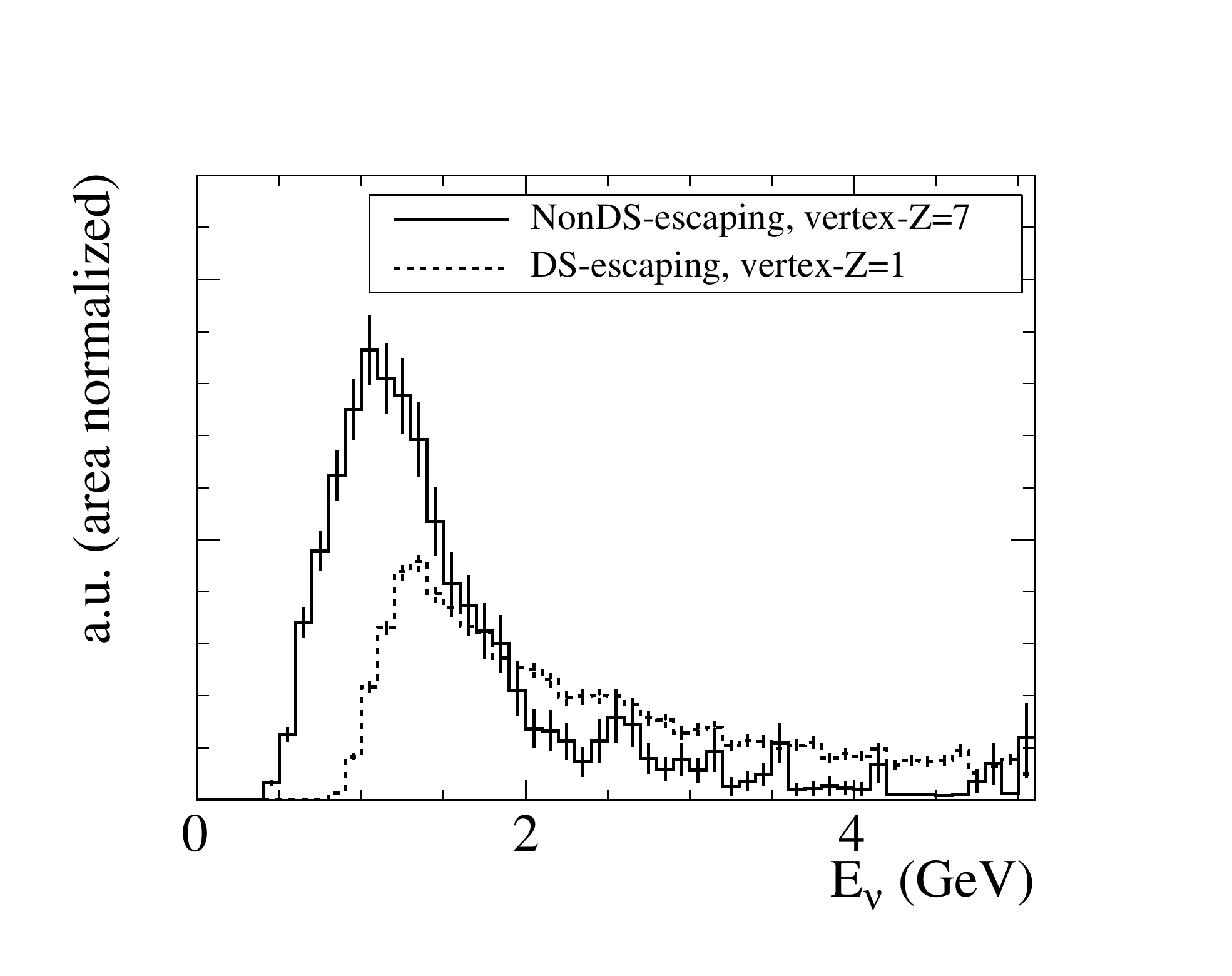} 
   \caption{
 Neutrino energy spectra for ``NonDS-escaping \& vertex-Z=7'' (solid line) events and ``DS-escaping \& vertex-Z=1'' (dashed line) events. 
   The energy spectra for module~0 are shown, normalized by area.
}
   \label{fig:enu_top}
\end{figure}

In total, there are 14 event topologies:
\begin{itemize}
\item DS-escaping: vertex-Z=1-7
\item NonDS-escaping: vertex-Z=1-7
\end{itemize}
Figure~\ref{fig:enu_top} shows the 
energy spectra 
of ``DS-escaping \& vertex-Z=1'' events and 
``NonDS-escaping \& vertex-Z=7'' events
for module~0.
The former has a more energetic $\mu$ track
and are generally produced by higher energy neutrinos.
The latter, on the other hand,
tend to have
muons produced at a larger angle to the neutrino beam
or with a lower energy, 
and so the majority come from lower energy neutrinos.

%% file: grouping.tex
\subsection{Module grouping}

A shift of the neutrino beam direction 
changes the peak of the neutrino energy spectra at the INGRID modules.
In order to reduce 
this effect, for the horizontal and vertical directions separately,
the two modules at beam-axis symmetric positions are grouped together.
This results in 7 module groups in total, defined in Table~\ref{tbl:group_def}.
\begin{table}[htb]
\begin{center}
\caption{Definition of the module groups}
\label{tbl:group_def}
\begin{tabular}{ccccc}
\hline\hline
Module & Module & Distance from  & Off-axis & Horizontal or\\
group & ID & the beam axis (cm) & angle & Vertical \\
\hline
1 & 0, 6  & 450 & 1.1$^{\circ}$ & Horizontal \\
2 & 7, 13 & 450 & 1.1$^{\circ}$  & Vertical \\
3 & 1, 5  & 300 & 0.7$^{\circ}$  & Horizontal \\
4 & 8, 12 & 300 & 0.7$^{\circ}$  &Vertical \\
5 & 2, 4  & 150 & 0.4$^{\circ}$  & Horizontal \\
6 & 9, 11 & 150 & 0.4$^{\circ}$  & Vertical \\
7 & 3, 10 & 0 & 0$^{\circ}$  & (Center) \\
\hline\hline
\end{tabular}
\end{center}
\end{table}

The number of selected events for each module group
and each topology is then defined as:
\begin{equation}
\label{eq:group_def}
N_{jg} = \frac{N_{jm} + N_{jm'}}{2} \;,
\end{equation}
where the indices $j$ and $g$ 
denote the $j^{th}$ topology and the $g^{th}$ module group 
($g=1,2,..,7$), respectively.
The $m$ and $m'$ indices
stand for the module numbers corresponding to each module group.

%% file: ana_detector_unc.tex
\subsection{Detector response uncertainties\label{subsec:det_unc}}

This section introduces two different 
kind of detector response uncertainty: 
those producing correlation among the event topologies and those that do not.
They are summarized as follow:
%
\begin{itemize}
\item Uncorrelated error sources
\begin{itemize}
\item Mass of iron plate
\item Pileup correction
\end{itemize}
\item Correlated error sources
\begin{itemize}
\item Event selection and reconstruction
\item MPPC noise rate
\end{itemize}
\end{itemize}

The treatment of these systematic uncertainties
in this analysis is described in Sec.~\ref{sec:systematics}

\subsubsection{Uncorrelated errors}

\begin{description}
\item[Iron mass]\mbox{}\\
The error on the measurement of the mass of each iron plate
and the machining tolerance for the plate area
are taken into account in the systematic error on the iron mass.
Since these errors are independent for each iron plate,
an uncorrelated error of 0.09\% assigned to the number of selected events.

\item[Pileup correction]\mbox{}\\
The number of selected events
is corrected to account for event pileup.
The correction factor is estimated 
using data sets at different beam intensities.
The uncertainty on the correction factor 
comes from the statistical error 
on the number of events,
so the uncertainty is uncorrelated between the event topologies.
An error of 0.5-2.0\%,
depending on the event topology,
is assigned 
in this analysis.
%

\end{description}


\subsubsection{Correlated errors}

\begin{description}
\item[MPPC noise]\mbox{}\\
MPPC noise 
hits sometimes result in mis-reconstruction 
of the event vertex or a miscounting of the number of active planes,
which produces a variation in
the neutrino event selection efficiency.
The systematic error 
caused
by the variation in the measured noise rate over time is evaluated by altering the noise rate in the MC.
As a result, a 0.1-1\% error is assigned to
the number of selected event in each topology.

\item[Event selection]\mbox{}\\
In this analysis, uncertainties in the following event selection steps are taken into account:
\begin{itemize}
\item 3D track matching
\item Vertexing
\item Veto cut
\item FV cut
\end{itemize}
The systematic error on the number of selected events is evaluated by varying the selection threshold for each step,
picking the loosest or tightest threshold.
The change from the nominal threshold 
to the loosest (tightest)  
is defined as the
$+1\sigma(-1\sigma)$ change.
The resultant fractional variation in the number of selected events ($\equiv \Delta N/N$) 
due to the $\pm1\sigma$ change is computed for both the data and MC.
Any difference in $\Delta N/N$ between the data and MC
is then taken as a systematic error.

\end{description}

Uncertainties on the hit efficiency of the tracking planes, the
contamination due to beam-related BGs, the
hit inefficiency of the upstream veto plane,
and the tracking efficiency 
were found to be negligible for this 
analysis and are not included
in the final result.
%
%

%% file: xsec_extraction.tex
\subsection{Cross-section extraction\label{subsec:xsec_extraction}}

This analysis uses the least $\chi^2$ method
to fit to the observed number of events
at each module group ($g^{th}$ bin: 1-7) and 
for each event topology ($j^{th}$ bin: 1-14):
\begin{align}
\label{eq:chisq}
\chi^2 = & \sum_j \sum_g \frac{\left\{ N^{obs}_{jg} - (N^{cc}_{jg} + N^{nc}_{jg} + N^{bg}_{jg}) \right\}^2 }
{(\sigma_{N_{jg}})^2} \nonumber \\
& + \sum_k \Delta (\vec{f}^k)^t (V_k)^{-1} \Delta \vec{f}^k \;,
\end{align}
where $N^{obs}$ 
is the observed number of events,
$N^{cc}$, $N^{nc}$, and $N^{bg}$ are the expected numbers of CC events,
NC, and BG events, and
$\Delta \vec{f}_k$ and $V_k$ are the systematic parameter
and the covariance for the $k^{th}$ error source, respectively.
For the covariance term uncertainties in the neutrino flux, neutrino interaction model
and the detector response are taken into account and are described in Sec.~\ref{sec:systematics}.
The denominator in the $\chi^2$ statistical term is composed of
the statistical error on the observed number of events ($N^{obs}_{jg}$),
the MC statistical error ($\sigma_{N^{mc}_{jg}}$), 
and the error on the detector response, which is uncorrelated among event topologies:
($\sigma_{N^{det}_{jg}}$):
\begin{equation}
\label{eq:chisq2_denom}
\sigma_{N_{jg}} = \sqrt{ N^{obs}_{jg} 
+ \left( \sigma_{N^{mc}_{jg}}\right)^2
+ \left(\sigma_{N^{det}_{jg}}\right)^2} \;.
\end{equation}


The expected number of CC events 
in the $g^{th}$ module group and for the $j^{th}$ event topology
is expressed as:
\begin{align}
& N^{cc}_{jg} 
\simeq  \sum_i \; \biggl[ (1+\Delta f^{d}_j + \Delta f^{cc}_j +\Delta f^b_{ig} + \Delta f_i) \nonumber \\
& \qquad \qquad  \quad \times \phi_{ig} \cdot \sigma^{cc}_i \cdot \epsilon^{cc}_{ij} \cdot T \biggr] \;, \label{eq:Npred2} \\ 
& \bullet \; \Delta f^{d} \;  \text{: systematic parameter for the detector response,} \nonumber   \\
&\bullet \; \Delta f^{cc}  \text{: systematic parameter for the CC interaction model,} \nonumber   \\
&\bullet \;  \Delta f^{b}  \; \text{: systematic parameter for the $\nu_\mu$ flux,} \nonumber \\
&\bullet \; \phi \quad \text{: $\nu_{\mu}$ flux,} \nonumber \\
&\bullet \; \sigma^{cc}  \; \text{: $\nu_{\mu}$ CC cross section,} \nonumber \\
&\bullet \; \epsilon^{cc} \;  \text{: detection efficiency for the CC interaction,} \nonumber \\
&\bullet \; T \quad \text{: the number of nucleons} \nonumber \\ 
& \qquad  \quad \text{ in the fiducial volume of the INGRID module.} \nonumber 
\end{align}
Here $i$ goes over 
the neutrino energy bins, described in Sec.~\ref{subsec:binning}, 
and $\Delta f_i$
 is the parameter being fit, which is used to represent fractional deviations of the CC inclusive cross section.
 The systematic parameters, 
 $\Delta f^d_j$, $\Delta f^{cc}_j$ and  $\Delta f^b_{ig}$, 
 are also fit to include the effect of these systematics into the cross-section result.
The INGRID modules are formed from both iron and plastic scintillator (CH).
The effect on this result coming from the different target nucleus for CH interactions is found to be small,
so the event rate per unit weight
on CH is assumed to be equal to that on iron.
$\Delta f^d_j$ and $\Delta f^{cc}_j$ are 
systematic parameters representing 
the uncertainty on the detector response and the CC 
cross-section model respectively
for the $j^{th}$ topology bin.
These uncertainties change the detection efficiency as a function of neutrino energy,
resulting in a variation in the number of selected events.
The 
difference in these uncertainties between the module groups 
is very small, therefore the same parameters are applied to all module groups.
$\Delta f^b_{ig}$ parameterizes the flux uncertainty,
changing the normalization 
of the neutrino flux,
in the $i^{th}$ energy 
bin of the $g^{th}$ module.
The $\Delta f^d_j$, $\Delta f^{cc}_j$ and  $\Delta f^b_{ig}$ 
parameters describe fractional deviations from the nominal MC and change the number 
of events in each
event topology, module group, and energy bin.

Since the fraction of NC events in the selected sample is very small,
the NC events are summed over the entire energy region
 and an averaged flux systematic parameter is applied to them:
\begin{equation}
\label{eq:Npred_nc_flux}
\Delta \bar{f}^b_{g} = \sum_i \Delta f^b_{ig} \cdot
\frac{\phi_{ig}}{\sum_{i'} \phi_{i'g}} \;.
\end{equation}
We express the number of the NC events as follows:
\begin{equation}
\label{eq:Npred_nc2}
N^{nc}_{jg} \simeq  \sum_i (1+\Delta f^d_j  + \Delta \bar{f}^b_{g} + \Delta f^{nc}_j) \;
\cdot \phi_{ig} \cdot \sigma^{nc}_i \cdot \epsilon^{nc}_{ij} \cdot T \;.
\end{equation}

The other BG events (the beam related BG and $\bar{\nu}_{\mu}$ and $\nu_e$ beam flux components)
 are summed for each module and for each topology ($N^{bg}_{jg}$).
The number of BG events in the sample,
and their associated errors,
are both small, 
so the errors on these BGs are neglected in this analysis.


%% file: binning.tex
\subsection{Energy binning\label{subsec:binning}}

The cross section is required to be continuous at the bin boundaries
and is linearly interpolated between boundaries,
so Eq.~(\ref{eq:Npred2}) is modified as follows:
\begin{align}
\label{eq:Npred6}
N^{cc}_{jg}  = &  (1+\Delta f^d_j) \cdot \sum_i (1+\Delta f^b_{ig}) \nonumber \\
& \times \sum_{l=0}^{L_i} \biggl[ \left (1+\Delta f_i+\frac{\Delta f_{i+1}-\Delta f_{i}}{L_i}\cdot l \right) \nonumber \\
& \;\; \qquad \times \phi_{ilg} \cdot \sigma^{cc}_{il} \cdot \epsilon^{cc}_{ilj} \cdot T \biggr ] \;.
\end{align}
To interpolate,
each $i^{th}$ energy bin is divided into fine bins ($l=0,1,..,L_i$).
The $i^{th}$ energy bin is defined as the ``global bin''
and the $l^{th}$ energy bin as the ``local bin'', respectively.
The energy range of the global bins and 
the binning for the local bins are 
summarized in Table~\ref{tbl:ebin_def}.
The $\Delta f_i$s are set at 
0.5, 0.8, 1.4, 2.6, and 4.0~GeV.

\begin{table}[hbt]
\begin{center}
\caption{Summary of the energy range of each global bin,
the size of each local bin, and the number of the local bins per global bin.}
\label{tbl:ebin_def}
\begin{tabular}{ccc}
\hline\hline
Energy range of & Size of each  & Number of  \\
global bin (GeV) & local bin (MeV) & local bins ($L_i$)\\
\hline
0-0.5  & 500 & 1 \\
0.5-0.8 & 100 & 3 \\
0.8-1.4 & 100 & 6 \\
1.4-2.6 & 100 & 12\\
2.6-4.0 & 100 & 14 \\
4.0-30.0 & 26000 & 1 \\
\hline\hline
\end{tabular}
\end{center}
\end{table}

Finally, the energy dependent CC inclusive
cross sections are extracted as follows.
After deriving the fit parameters using the least $\chi^2$ method,
the cross sections are obtained by 
multiplying those in the original model by 
$1+\frac{\Delta f_1 + \Delta f_2}{2}$,
$1+\frac{\Delta f_2 + \Delta f_3}{2}$, and 
$1+\frac{\Delta f_3 + \Delta f_4}{2}$.
Taking the average of neighboring parameters 
produces a measurement at the central energy between the bin boundaries,
since, as in Eq.~(\ref{eq:Npred6}),
a linear interpolation is applied between the neighboring $\Delta f_i$ parameters.
As a result of this averaging, the 
cross section
is measured at 
1.1, 2.0, and 3.3~GeV.
The final error on the cross section is 
smaller than those on the $\Delta f_i$ parameters
due to the negative correlations between the $\Delta f_i$'s.
This feature is a result of the cross section continuity 
requirement.

As seen in Fig.~\ref{fig:after_select2_a}, the INGRID detection efficiency for CC interactions falls rapidly for neutrinos with energies less than 0.5~GeV. 
Since the event samples are not sensitive to the cross section in this region
$\Delta f_0$ is not used in the final result.
For $E_\nu>4.0$~GeV
there is only a small difference in the neutrino energy spectra between the INGRID modules.
Therefore, sensitivity to the cross section for $E_\nu>4.0$~GeV
is expected to be worse compared to the lower energy regions.
For these reasons it was decided, before fitting the data, to use $\Delta f_1$-$\Delta f_4$ to measure the cross section at 1.1, 2.0, and 3.3~GeV.

\begin{figure*}[p]
\centering
\includegraphics[width=0.7\textwidth]{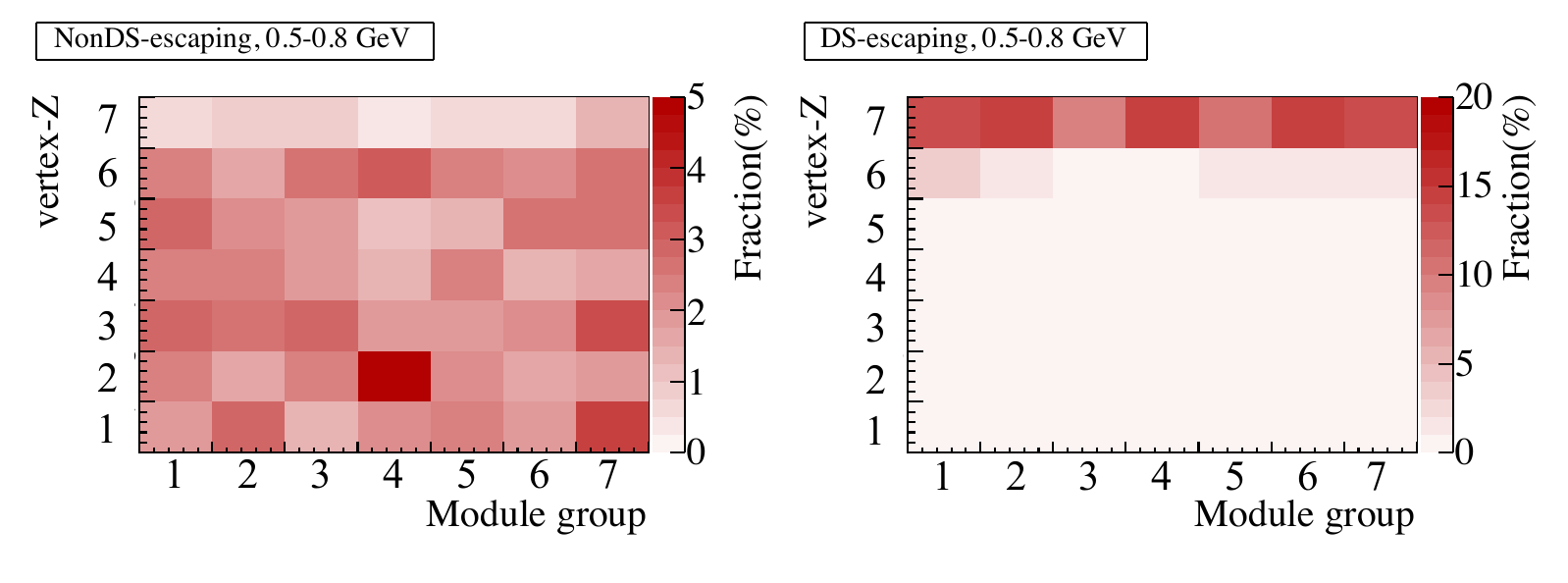}
\includegraphics[width=0.7\textwidth]{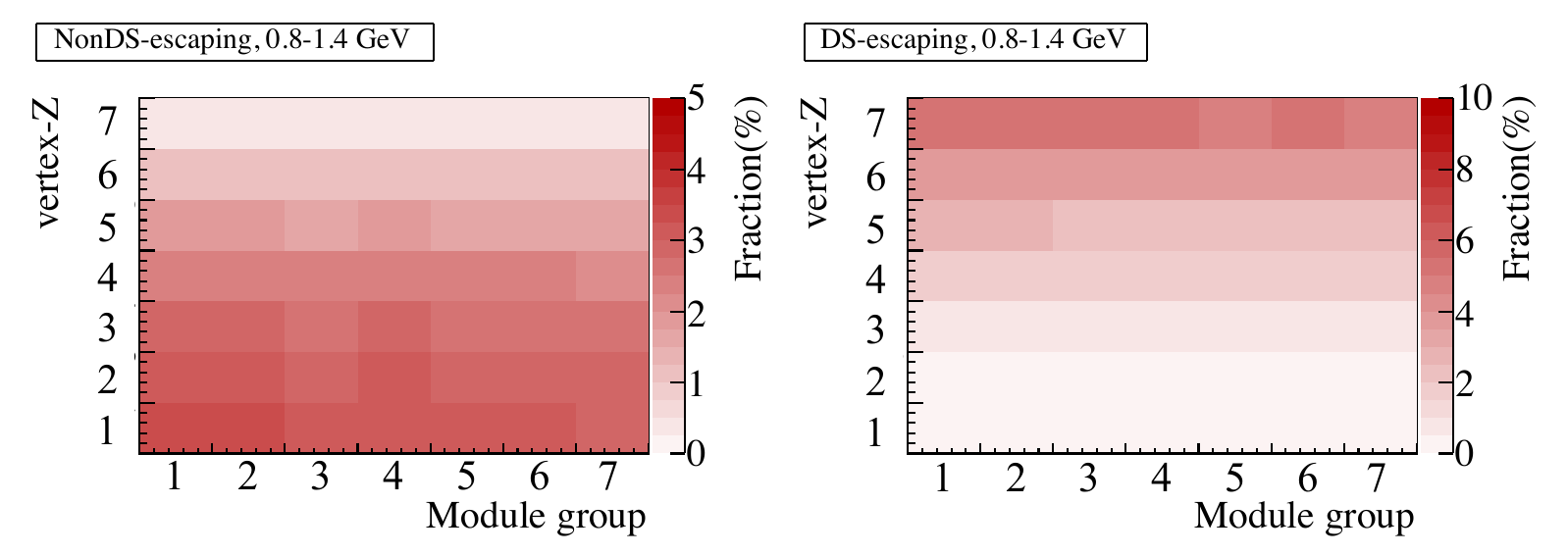}
\includegraphics[width=0.7\textwidth]{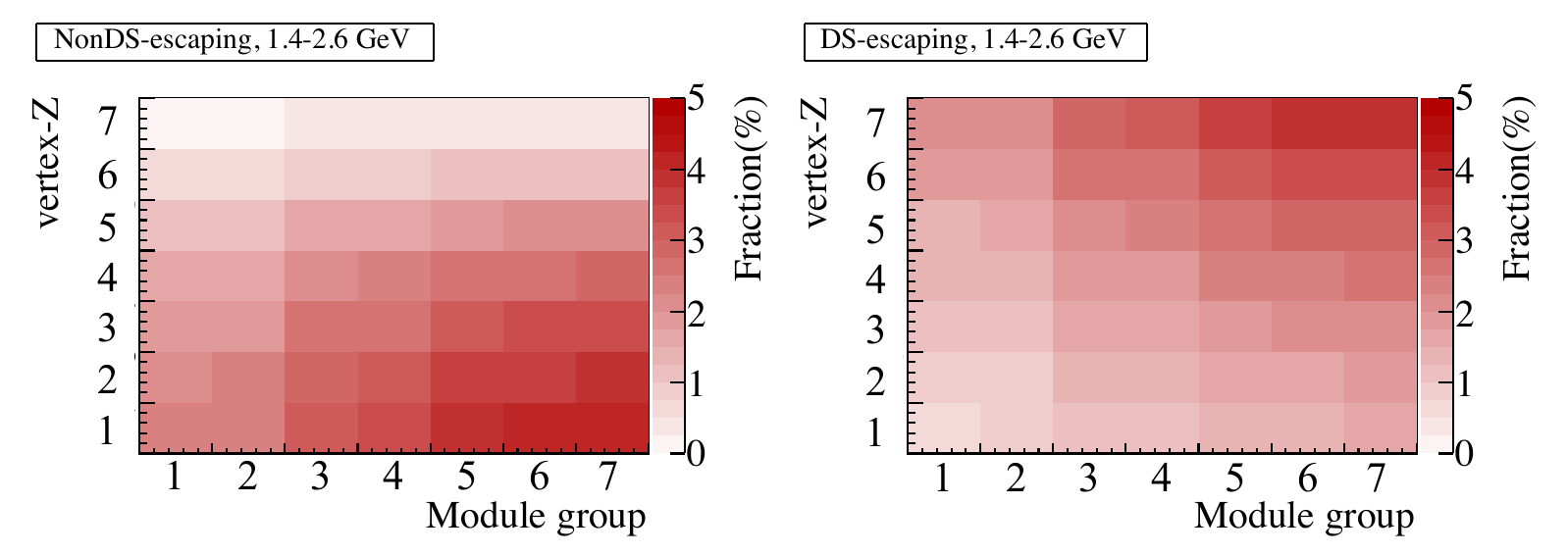}
\includegraphics[width=0.7\textwidth]{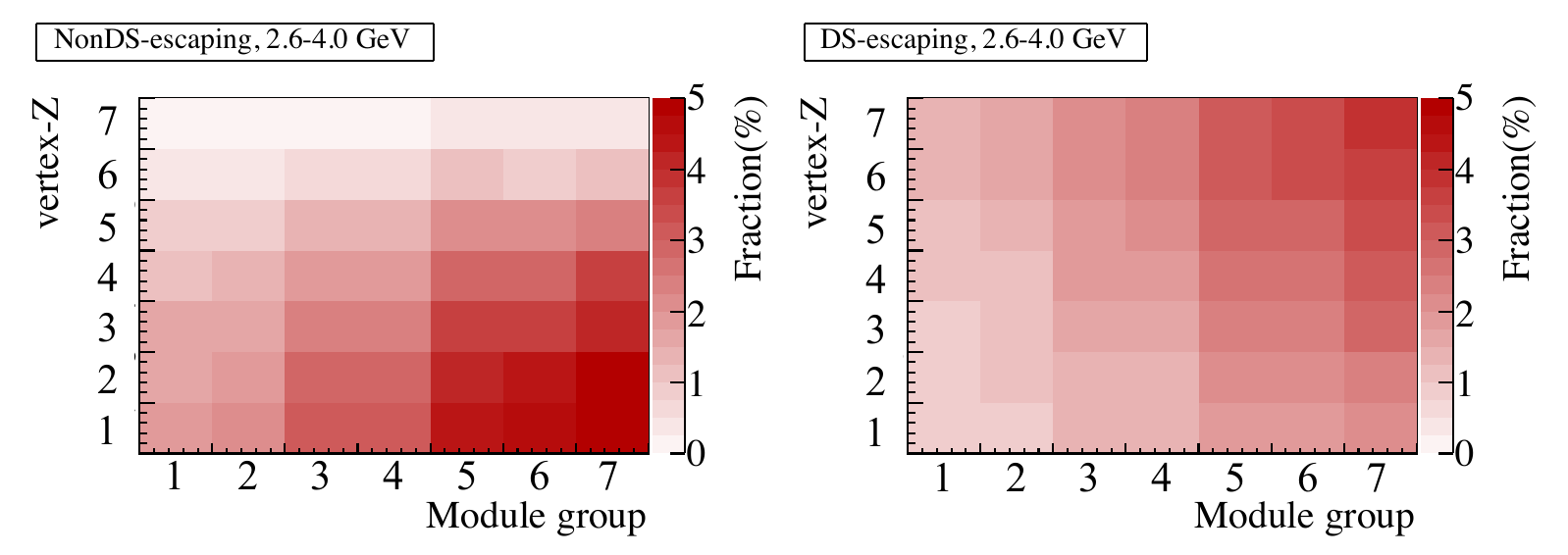}
\includegraphics[width=0.7\textwidth]{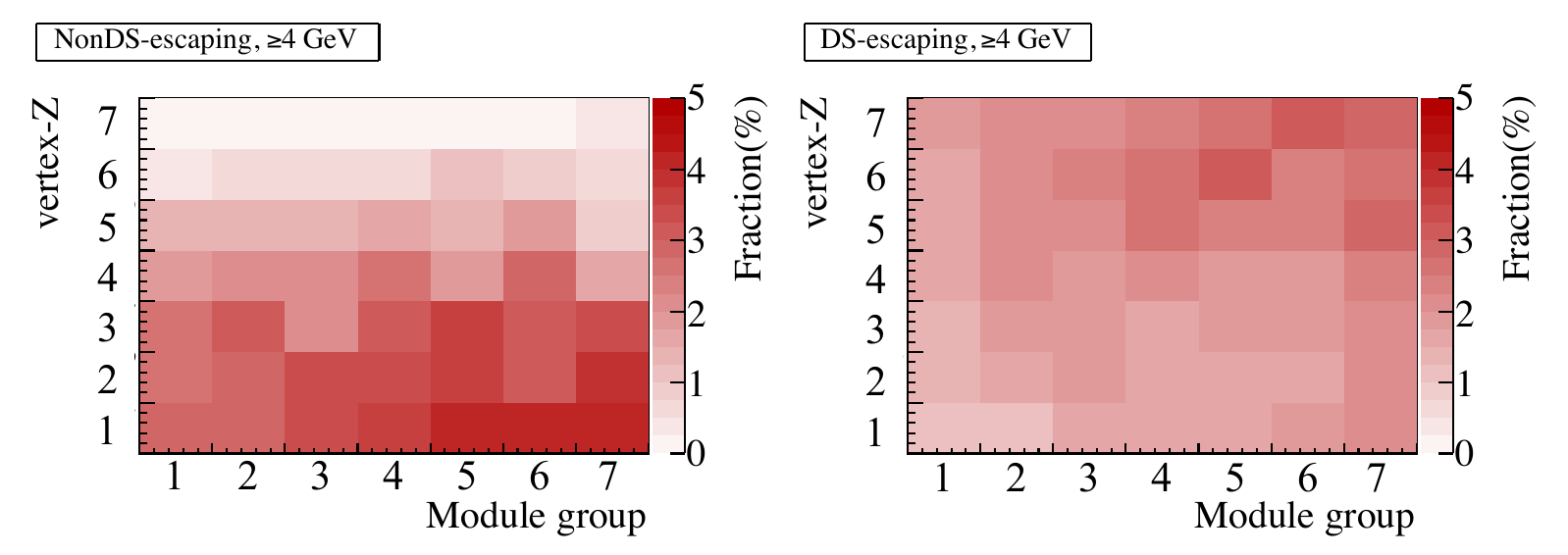}
\caption{Probability density function (PDF) for each energy region defined in Sec.~\ref{subsec:binning}:
$E_{\nu}$=0.5-0.8~GeV, 0.8-1.4~GeV, 1.4-2.6~GeV, 2.6-4.0~GeV,  $>$4.0~GeV.}
\label{fig:pdf}
\end{figure*}

The PDF of CC events
in the global energy binning
is shown in Fig.~\ref{fig:pdf}.
Here, the ``Fraction'' described by the z-axis of the figure is obtained for each energy region
by dividing the number of CC events in each bin
by the total number of CC events in that energy regions.
At lower neutrino energies
most of the CC events
are selected in the downstream vertex-Z bin for the DS-escaping topology 
whereas at higher energies
the DS-escaping events are distributed uniformly in vertex-Z.
NonDS-escaping CC events are selected in all vertex-Z bins for low energy neutrinos
but higher energy neutrinos tend to be located in upstream vertex-Z bins.
In addition, 
more high-energy neutrino events are selected 
in modules closer to the beam-axis.

%% file: systematics.tex
\section{Propagation of systematic uncertainties\label{sec:systematics}}

As described in Sec.~\ref{subsec:xsec_extraction},
the $\chi^2$
has terms with covariance matrices for 
systematic parameters.
In this section, 
we describe 
how the covariance matrices for
the neutrino flux,
the neutrino 
interaction model and the detector response,
which were introduced in Secs.~\ref{subsec:ingrid_nuflux},
~\ref{subsec:neut}
and \ref{subsec:det_unc}, respectively,
are constructed.

\input{flux_unc}
\input{nuint_unc}

\input{detector_unc}

\input{other_unc}

%% file: flux_unc.tex
\subsection{Neutrino flux uncertainties\label{subsec:flux_unc}}

\begin{figure}[htb] 
   \centering
   \includegraphics[width=0.45\textwidth]{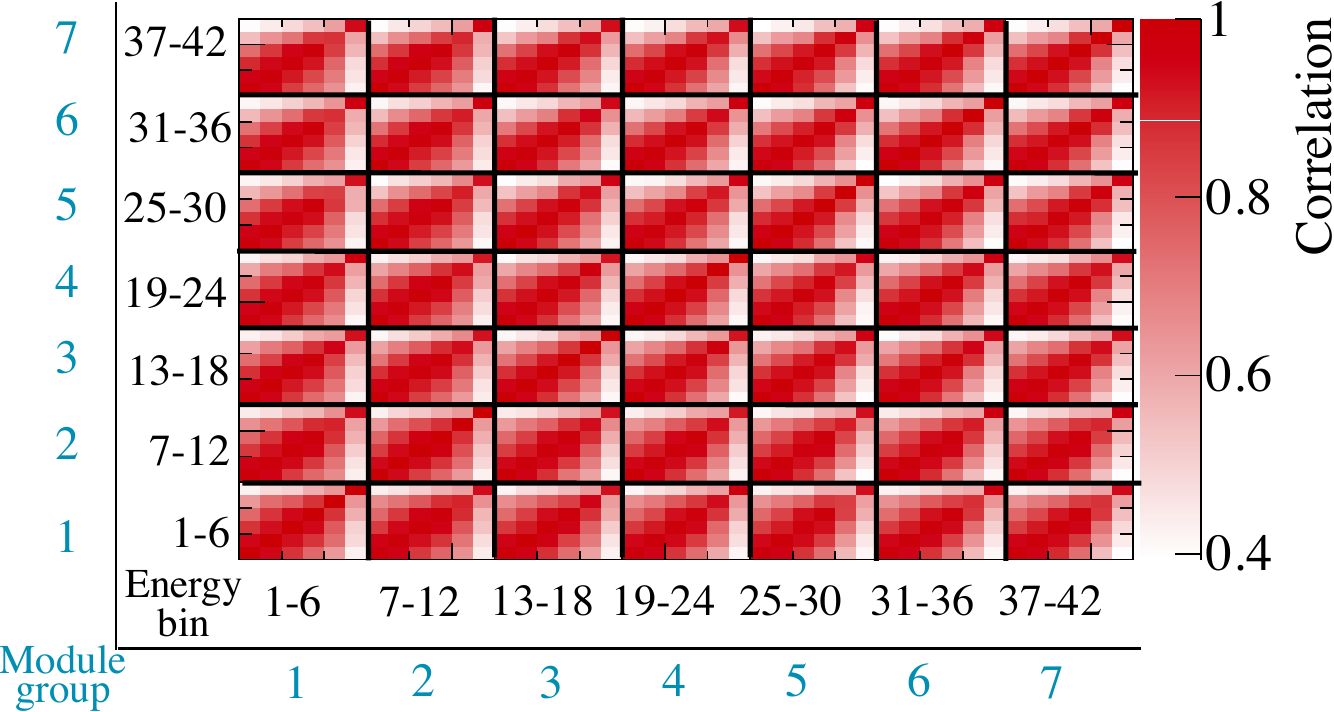} 
   \caption{ Correlation matrix between module groups for the flux error. 
   The energy binning used in this matrix is same as the ``lobal bin'' defined in Table~\ref{tbl:ebin_def}.}
   \label{fig:flux_cor}
\end{figure}

The covariance matrix for each source 
of error on the neutrino flux perdiction, such as the horn current uncertainty,
is calculated by taking the variation of the flux due to that error.
The total covariance matrix is obtained by summing all these matrices and
Figure~\ref{fig:flux_cor} shows the correlation matrix derived from it.
The energy binning in the covariance matrix 
is the same as that used to define the ``global bin''
in Table~\ref{tbl:ebin_def}.
One can see that it is largely positively correlated.
This correlation comes mainly from 
the uncertainty associated with hadron production at the target,
which varies the neutrino flux in the same way at all INGRID modules.

%% file: nuint_unc.tex
\subsection{Neutrino interaction model uncertainties~\label{subsec:neut_unc}}

Any systematic error in the CC interaction model
would, by definition, alter the CC inclusive cross section itself.  
This is not the case for NC interaction model uncertainties, and
so the
systematic errors associated with these two processes are
evaluated separately.

\subsubsection{\label{sec:NCint_syst} Systematics uncertainty on NC interactions}

\begin{figure}[!h]
\centering
\includegraphics[width=0.4\textwidth]{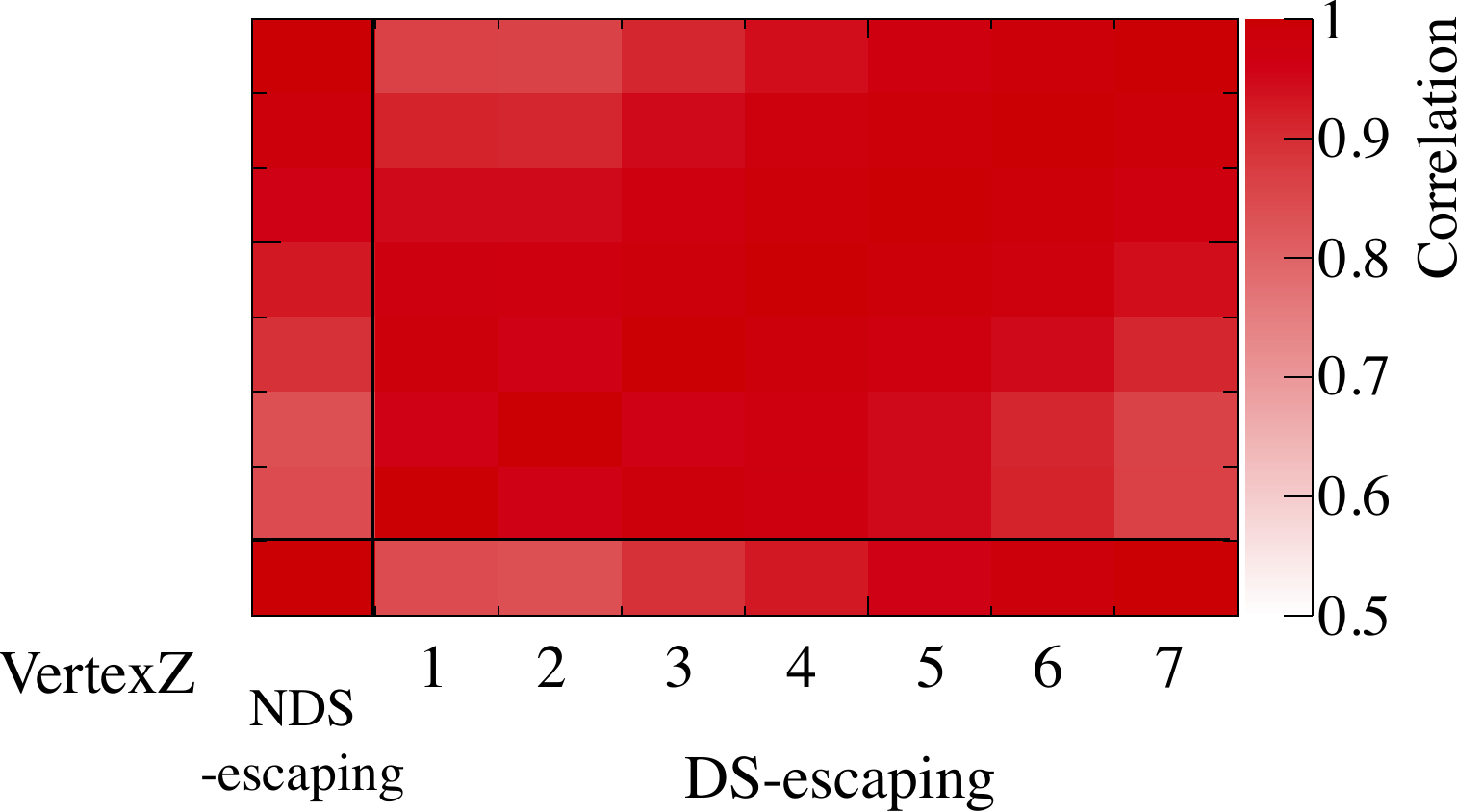}
\caption{
 Correlation matrix for the errors coming from NC interaction uncertainties.
The binning on the y-axis is identical to that on the x-axis.
}
\label{fig:ncint_err_a}
\end{figure}

The uncertainty in the number of NC interactions in each bin 
is expressed by the normalization parameter:
\begin{equation}
f^{nc}_{jg} \equiv \frac{N'^{nc}_{jg}}{N^{nc}_{jg}} \;,
\end{equation}
where $N^{nc}_{jg}$ is the predicted number of NC events
for the $j^{th}$ topology and the $g^{th}$ module group.
$N'^{nc}_{jg}$ is the predicted number of events in the same bin
but for the case where one of the NC systematic parameters has been changed by 1$\sigma$.
The number of events is altered not only 
by the change in the cross section
but also by changes in the event detection efficiency.
The same normalization parameter, $f^{nc}_{jg}$,
is used for all module groups, so the number of predicted events,$N^{nc}_{jg}$, becomes:
\begin{equation}
N^{nc}_{jg} \rightarrow f^{nc}_{j} \cdot N^{nc}_{jg} \qquad (f^{nc}_j = f^{nc}_{jg})\;.
\end{equation}
$f^{nc}_j$ is estimated by combining 
NC events from all module groups.
The fractional covariance for the topology bins
is then calculated by varying each NC systematic parameter 
by $\pm1\sigma$.
\begin{align}
V^{nc}_{ij}=& \frac{1}{2}
\Biggl[
\left(\frac{N^{nc}_i - N^{nc,+1\sigma}_i}{N^{nc}_i} \cdot
\frac{N^{nc}_j - N^{nc,+1\sigma}_j}{N^{nc}_j} \right) \nonumber \\
& + \left(\frac{N^{nc}_i - N^{nc,-1\sigma}_i}{N^{nc}_i} \cdot
\frac{N^{nc}_j - N^{nc,-1\sigma}_j}{N^{nc}_j} \right)
\Biggr] \;,
\end{align}
where $N^{nc}_{i(j)}$ is obtained by summing over all module groups:
\begin{equation}
N^{nc}_{i} = \sum_g N^{nc}_{ig} \;.
\end{equation}
We found that the total NC error is fully correlated
between all of the ``NonDS-escaping'' bins.
Therefore,
the 7 ``NonDS-escaping'' bins are merged into a single bin.
Figure~\ref{fig:ncint_err_a}
shows the correlation matrix for the NC interactions uncertainty,
demonstrating that the event topology bins are almost fully correlated with each other.
This correlation comes mainly from the NC normalization error (see Table~\ref{tbl:nuint_syst}).

\begin{figure}[!h]
\centering
\includegraphics[width=0.4\textwidth]{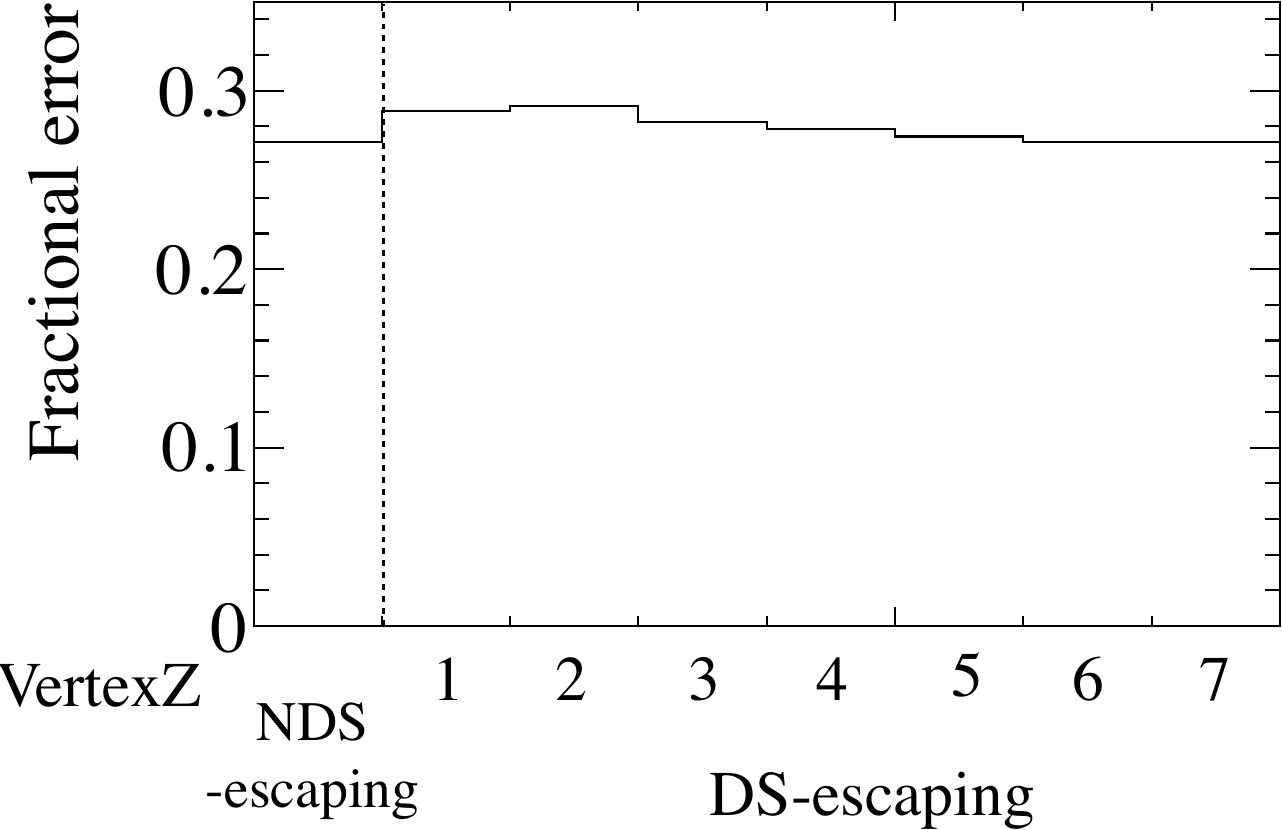}
\caption{
 Fractional error from NC interaction uncertainties
 on the number of NC events.}
\label{fig:ncint_err_b}
\end{figure}

  The total normalization error on the number 
of NC events is 27-30\%,
which is dominated by the NC normalization error,
which is shown in
Fig.~\ref{fig:ncint_err_b}.
This gives a maximum error size of 5\% on the total (CC+NC) number of events.

\subsubsection{\label{sec:CCint_syst}Systematic uncertainty on CC interactions}

\begin{figure}[!h]
\centering
\includegraphics[width=0.4\textwidth]{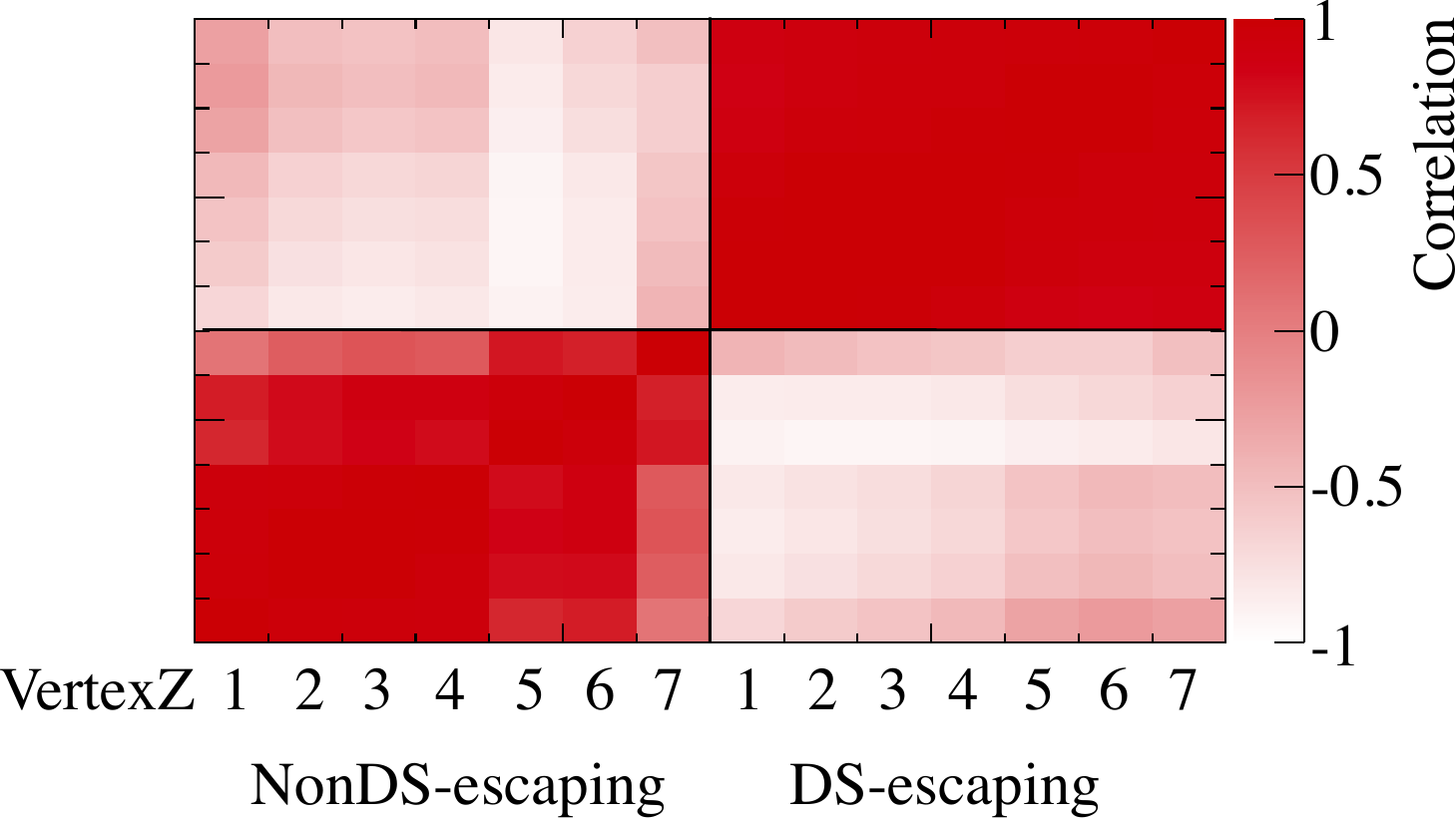}
\caption{
 Correlation matrix for the errors coming from CC interaction uncertainties.
The binning on the y-axis is identical to that on the x-axis.}
\label{fig:ccint_err_a}
\end{figure}

\begin{figure}[!h]
\centering
\includegraphics[width=0.4\textwidth]{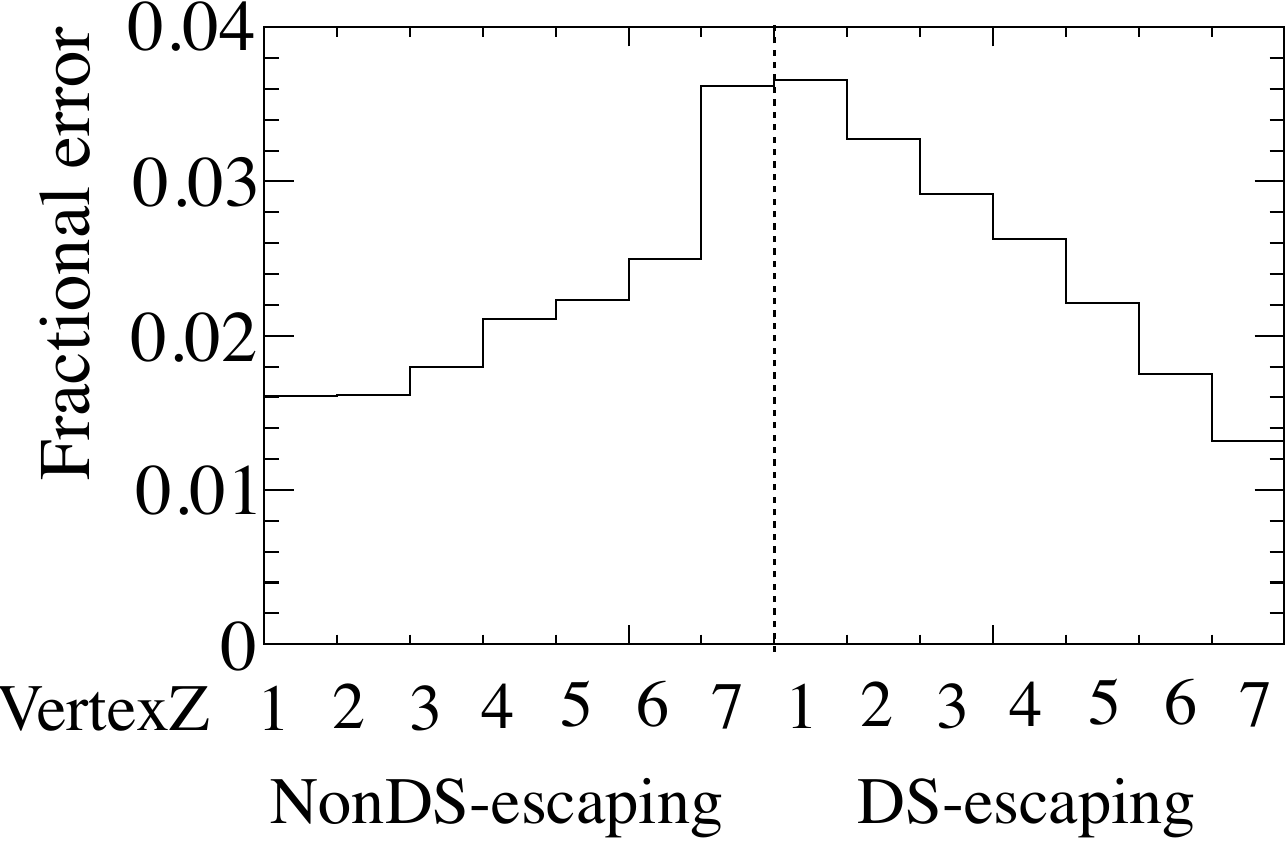}
\caption{
 Fractional error from CC interaction uncertainties.}
\label{fig:ccint_err_b}
\end{figure}

Varying the CC interaction parameters
results in a change in both the neutrino cross section 
and 
the detection efficiency.
In this analysis, for the systematic uncertainty on CC interactions,
only the latter change is taken into account, 
since the result of the analysis will be the cross section itself.
A $1\sigma$ variation is applied to a CC interaction parameter and
the new selection efficiency, $\epsilon'^{cc}_{ij}$ is calculated.
The change in the detection efficiency 
is then given by the
ratio of the new efficiency
to the nominal one:
\begin{equation}
w_{ij} = \frac{\epsilon'^{cc}_{ij}}{\epsilon^{cc}_{ij}}\;,
\end{equation}
where 
the indices $i$ and $j$ denote 
the $i^{th}$ energy bin
and the $j^{th}$ topology bin, respectively.
The predicted number of CC events is then modified
using $w_{ij}$:
\begin{equation}
\label{eq:corrected_Npred_cc}
N'^{cc}_{jg} =\sum_i \phi_{ig} \cdot \sigma^{cc}_i \cdot \left( w_{ij} \epsilon^{cc}_{ij} \right) \cdot T\;,
\end{equation}
A fractional covariance between topology bins
is then calculated
for each CC interaction parameter
using the modified ($N'^{cc}$) and nominal 
number of CC events.  This is performed in
the same way as for the NC interaction uncertainty, described earlier.
The total covariance matrix is computed
by summing up the individual matrix from each CC interaction parameter.
The obtained correlation matrix and fractional error
between the event topology bins
are shown in Figs.~\ref{fig:ccint_err_a}~and~\ref{fig:ccint_err_b}.

\subsubsection{\label{sec:FSI_syst}
Systematic uncertainty on FSI}

For the pion FSI parameters,
uncertainties on the absorption, charged exchange, quasi-elastic, and inelastic scattering cross sections
of the pion are taken into account.
These systematic errors are treated in a different to the previous interaction uncertainties
because there are correlations between them.
The INGRID data is fitted with $N'^{cc}$,
obtained by changing each FSI parameter by $1\sigma$,
and the difference between the fitted and the nominal result is taken as the systematic error.
The effect of each FSI parameter on the measured cross section 
was 
found to be negligible, 
except for the pion absorption uncertainty, 
which was then added in quadrature to the final result.



%% file: detector_unc.tex
\subsection{Detector response uncertainties\label{subsec:det_unc2}}

As described in Sec.\ref{subsec:det_unc},
each error source is categorized according to
whether it produces correlation among topology bins or not.
For the uncorrelated error sources, the iron mass and pileup correction,
the individual systematic errors are summed quadratically,
 and the total inserted into the denominator of the $\chi^2$ statistical term,
($\sigma_{N^{det}_{jg}}$, in Eq.~(\ref{eq:chisq2_denom})).

\begin{figure}[!h]
\centering
\includegraphics[width=0.4\textwidth]{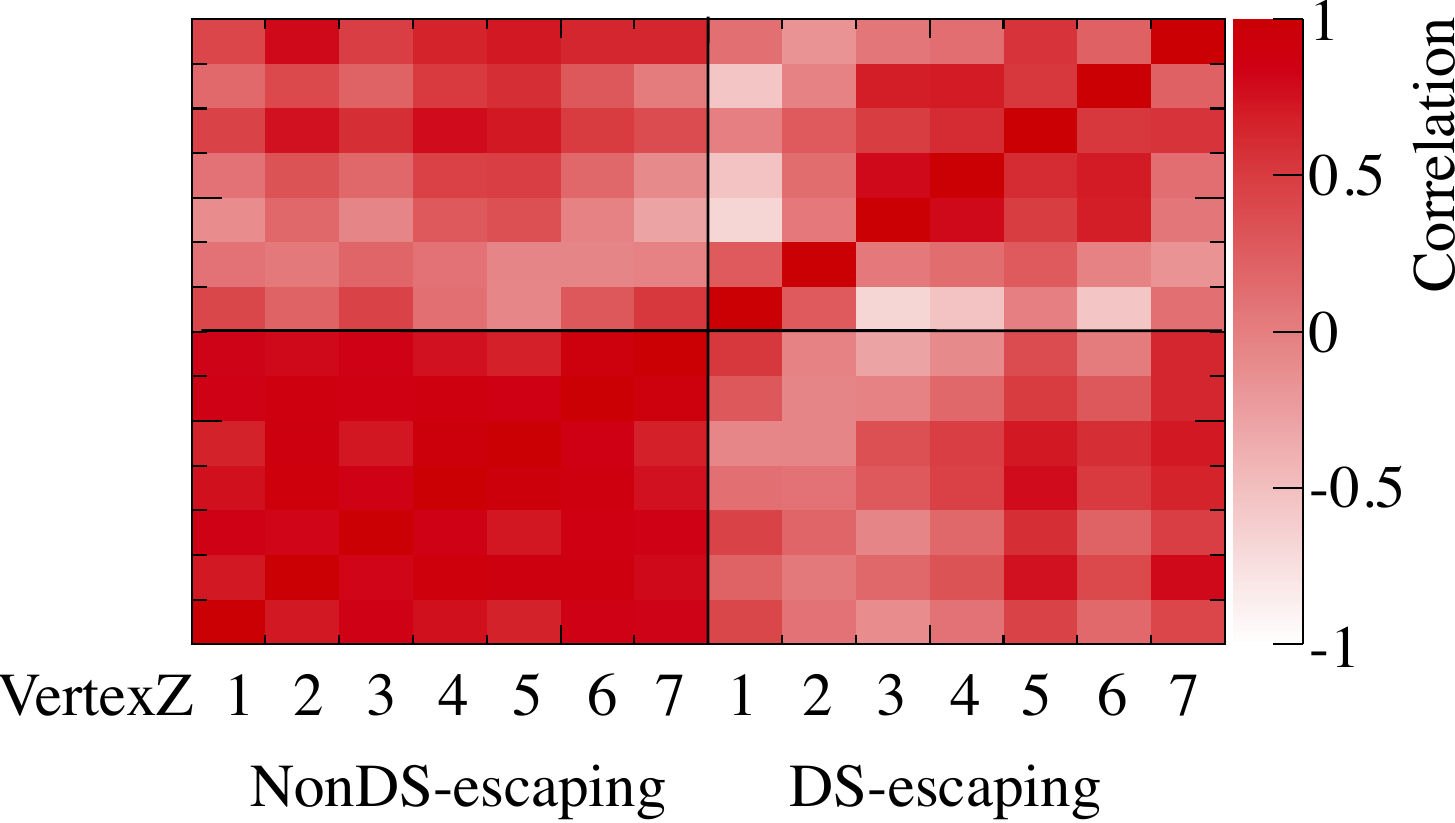}
\caption{
Correlation matrix from the uncertainties in the detector response.
The binning on the y-axis is identical to that on the x-axis.
}
\label{fig:det_syst_errmat_a}
\end{figure}

\begin{figure}[!h]
\centering
\includegraphics[width=0.4\textwidth]{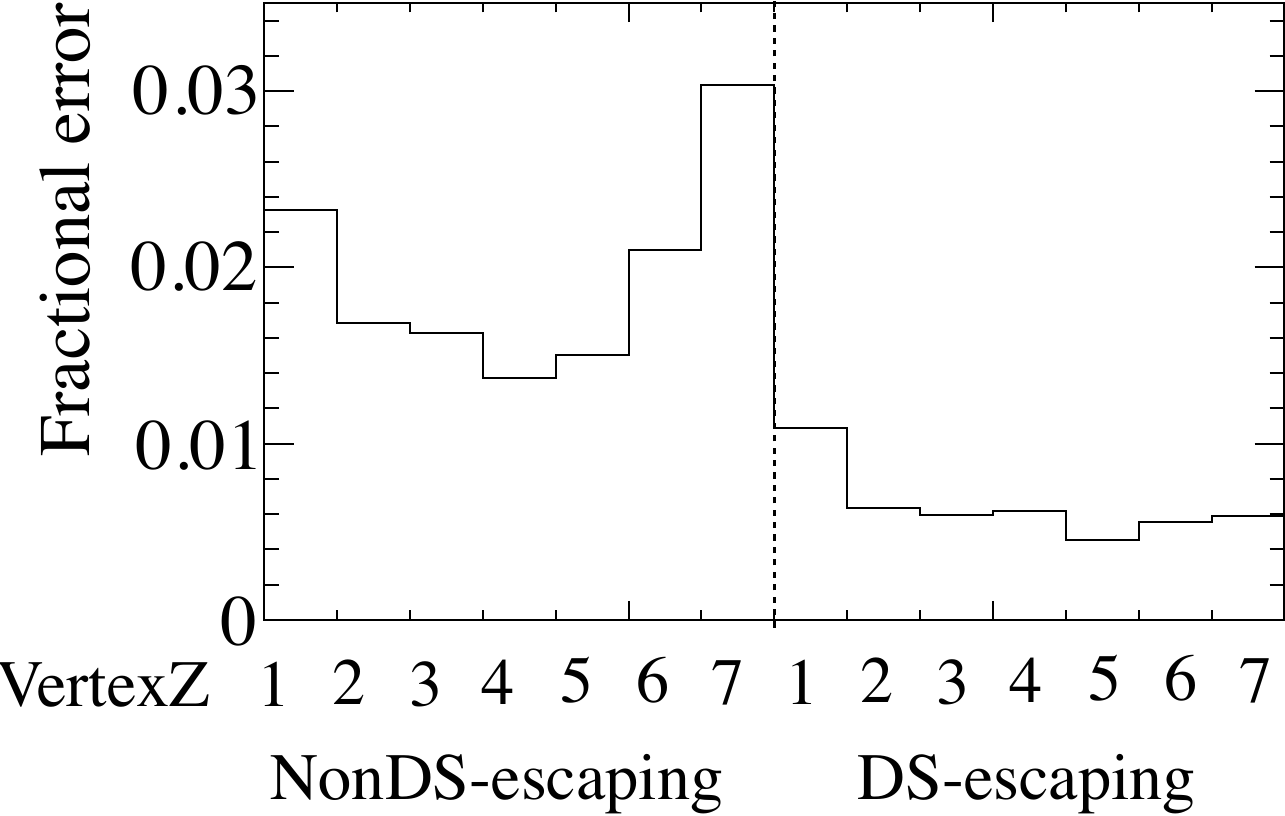}
\caption{
 Fractional error for the topology bins coming from uncertainties in the detector response.}
\label{fig:det_syst_errmat_b}
\end{figure}

For the correlated error sources, 
the size of the error does not to vary between each module, so
a covariance matrix is constructed from the topology bins using the average change over all modules.
Namely, the errors 
are assumed to be fully correlated 
between module groups.
For uncertainties from the event selection,
the systematic error is evaluated by varying each selection threshold 
by $1\sigma$,
and 
the resultant fractional variation in the number of selected events ($\equiv \Delta N/N$) 
computed for data and MC.
The difference in $\Delta N/N$
between data and MC is then taken as the systematic error,
calculated as:
\begin{equation}
\label{eq:diff_data_mc_det}
\Delta_{j} = 
\left( \frac{\Delta N_{obs}}{N_{obs}} \right)_{j}
 - \left( \frac{\Delta N_{exp}}{N_{exp}} \right)_{j} \;,
\end{equation}
where the index $j$ denotes the $j^{th}$ topology bin.
If both a $+1\sigma$ and $-1\sigma$ variation are applied to
the event selection then the
following covariance is calculated:
\begin{equation}
\label{eq:calc_det_cov1}
V_{ij}=\frac{1}{2}\left\{ 
\left(\Delta_i \cdot \Delta_j\right)_{+1\sigma} + 
\left(\Delta_i \cdot \Delta_j\right)_{-1\sigma}\right\} \;.
\end{equation}
If only a $+1\sigma$ change can be applied the 
the covariance becomes:
\begin{equation}
\label{eq:calc_det_cov2}
V_{ij}=\Delta_i \cdot \Delta_j \;.
\end{equation}
The statistical error of $\Delta_i$ is also calculated 
and added to Eq.~(\ref{eq:calc_det_cov1}) (or Eq.~(\ref{eq:calc_det_cov2})). 
Finally, the total covariance is calculated by summing 
each individual covariance.
Figures~\ref{fig:det_syst_errmat_a}~and~\ref{fig:det_syst_errmat_b} show the  
correlation matrix obtained and the size of the fractional error for each event topology bin.


\begin{table}[!h]
\centering
\caption{
Systematic error size on topology bins due to
 uncertainties in the detector response.}
\label{tbl:det_syst_summary}
\begin{tabular}{c c}
\hline\hline
Error type & Error size (at maximum)\\
\hline
Correlated error & 3\% \\
Uncorrelated error & 2\% \\
\hline\hline
\end{tabular}
\end{table}

Table~\ref{tbl:det_syst_summary} summarizes
the size of the detector systematic error for 
each error type.

%% file: other_unc.tex
\subsection{Uncertainty in pion multiplicities and secondary interactions}

\begin{table*}[t]
\caption{
Summary of pion-nucleus scattering data
used to evaluate the pion SI uncertainty.
The reactive cross section is defined as the sum of all the inelastic cross sections.
}
\begin{center}
\begin{tabular}{l l l l l}
\hline\hline
& Hadrons & Targets & $p_{lab}$ (MeV/c) & Interaction type \\
\hline
K. Nakai~\textit{et al.}~\cite{Ashery:1981tq} & $\pi^+$/$\pi^-$ 
& Al, Ti, Cu, Sn, Au & 83-395 & ABS \\
D. Ashery~\textit{et al.}~\cite{Nakai:1980cy} &  $\pi^+$/$\pi^-$ 
& Li, C, O, Al, Fe, Nb, Bi & 175-432 & Reactive, Elastic, QEL, ABS, SCX\\
M.K. Jones~\textit{et al.}~\cite{Jones:1993ps} & $\pi^+$
& C, Ni, Zr, Sn, Pb & 363-624 & QEL, ABS, SCX\\
G.J. Gelderloos~\textit{et al.}~\cite{Gelderloos:2000ds} & $\pi^-$
& Li, C, Al, S, Ca, Cu, Zr, Sn, Pb & 479-616 & Reactive \\
B.W. Allardyce~\textit{et al.}~\cite{Allardyce:1973ce} & $\pi^+$/$\pi^-$
& C, Al, Ca, Ni, Sn, Ho, Pb & 710-2000 & Reactive \\
\hline\hline
\end{tabular}
\end{center}
\label{tab:pixsec_data}
\end{table*}%

Uncertainties associated with pion multiplicities
and pion secondary interactions (SI)
are treated in a different way to the systematics described above.
These uncertainties are evaluated 
by comparing the underlying model with external data.
Any observed difference is used to correct
the nominal MC sample, and then the $\chi^2$ fit is performed using
this ``corrected'' MC.
The differences in the fitted values between the nominal and corrected MC are then taken as the systematic error on the final result.
\begin{description}
\item[Pion multiplicity]\mbox{}\\
In this analysis,
 the number of events is determined from the number of reconstructed vertices, which are sometimes missed due to the pile-up of tracks from multiple neutrino interactions. 
Events with large numbers of tracks usually contain pions, therefore
the uncertainty associated with the pion multiplicity in these events needs to be considered.
This uncertainty
is estimated by following the method described in Ref.~\cite{Koba:1972ng}.
The probability of an event having a pion multiplicity of $n$ is 
expressed as:
\begin{align}
 P(z; A,B,c) &= \frac{1}{\langle n \rangle} 
 \frac{2 e^{-c} c^{cz+1}}{\Gamma(cz+1)}\;,  \label{eq:KNOstandard} \\
 \langle n \rangle &= A + B\log W^2 \;, \label{eq:mean_multiplicity}
\end{align}
where $z$=$n$/$\langle n \rangle$;
$\langle n \rangle$ is the mean pion multiplicity and can be 
expressed by an approximate formula as in Eq.~(\ref{eq:mean_multiplicity}).
The $W$ is the hadronic invariant mass,
and is expressed as:
\begin{align}
W^2 &= (E_\nu+E_N-E_\mu)^2-(\vec{p}_\nu+\vec{p}_N-\vec{p}_\mu)^2 
\end{align}
with $E_{\nu_\mu}(\vec{p}_{\nu_\mu})$ 
and $E_\mu(\vec{p}_\mu)$ denoting the
energy (3-momentum) of the $\nu_\mu$ and $\mu$ respectively;
$\vec{p}_N$ denoting the Fermi momentum of the nucleon and $E_N$ 
the nucleon energy.
$A$, $B$, and $c$ 
are derived by
fitting two external data sets~\cite{Zieminska:1983bs,Allen:1981vh} with Eq.~(\ref{eq:KNOstandard}).
These fitted parameters are 
compared with those used in NEUT,
and the differences assigned as 
the systematic uncertainty.
The parameters are used to produce a corrected MC sample
which is input to the $\chi^2$ fit.  The differences in
the fitted values coming from the corrected and the nominal MC
are taken as the systematic error on the final result due to
pion multiplicity uncertainties.
\item[Pion SI]\mbox{}\\
Hadrons produced in neutrino interactions can also interact whilst traveling through the detector, 
a process known as ``secondary interaction (SI)''.
In the INGRID simulation the pion SI processes are controlled by GEANT4.
In order to evaluate the uncertainty 
in the pion SI model, the
following interaction modes are considered:
\begin{description}
\item[Quasi-elastic scattering (QEL)]\mbox{}\\
The final state pion is the same type as the incoming pion.
\item[Absorption (ABS)]\mbox{}\\
The incident pion is absorbed by the nucleus, 
resulting in there being no pions in the final state.
\item[Single charge exchange (SCX)]\mbox{}\\
A charged pion interacts such that there is only one $\pi^0$ and no other pions in the final state.
\end{description}

The existing experimental data
used to evaluate this uncertainty
is summarized in Table~\ref{tab:pixsec_data}. 
In the table, the reactive cross section 
is defined as $\sigma_\text{{total}} - \sigma_\text{{elastic}}$,
where $\sigma_\text{{total}}$ is the total cross section
and $\sigma_\text{{elastic}}$ is the elastic cross section.
As seen in the table, 
D. Ashery~\textit{et al.} provides
various cross sections across a range of pion momenta
and target nuclei,
including iron.
The other data do not include measurements
on iron.
For these data, 
an $A$-dependent scaling
is applied
in order to extract the cross section
on iron.
To evaluate the systematic uncertainty coming from pion SI,
we first tune the pion cross section in the momentum region covered by the data in Table~\ref{tab:pixsec_data}.
Second, for the lower energy region not covered by data,
the ABS $\pi^+$($\pi^-$) cross section $<$20(30)~MeV is kept constant, motivated by the microscopic calculation from~\cite{Salcedo:1987md}.
The QEL cross section is extrapolated to 0 at 0~MeV.
For the higher energy region, 
each of the cross sections above is tuned
on the basis
that 
the size of the reactive cross section is conserved,
since the cross sections predicted by GEANT4
are in agreement with the experimental data in this energy region.
This study 
gives 4 corrected MC samples in total, each of which
is then used 
to fit the T2K data.
Finally, 
the size of the systematic error on the $f_k$ parameters
due to the uncertainty on pion FSI
is calculated as follows:
\begin{equation}
\Delta f_k = \sqrt{ \frac{1}{4}\sum_{i=1}^{4} ( f^{nom}_k - f^i_k )^2 } \;,
\label{eq:pionSI_error}
\end{equation}
where 
the index $i$ denotes the $i^{th}$ 
corrected MC sample
and 
$f_k$ is the fitted normalization parameter for the $k^{th}$ energy bin.


 \end{description}

%% file: result.tex
\section{Result\label{sec:result}}

In this section, 
we present the result of this
$\nu_\mu$ inclusive CC cross-section measurement. 
Section~\ref{subsec:data_set}
shows the data set used in this analysis, whilst
Secs.~\ref{subsec:xsec_fit} and \ref{subsec:summary}
describe the output from the $\chi^2$ fit
and give a summary of this result.

\input{data_set}


\input{xsec_fit}


\input{summary}

%% file: data_set.tex
\subsection{Data set\label{subsec:data_set}}

\begin{figure*}[!p] 
   \centering
   \includegraphics[width=0.3\textwidth]{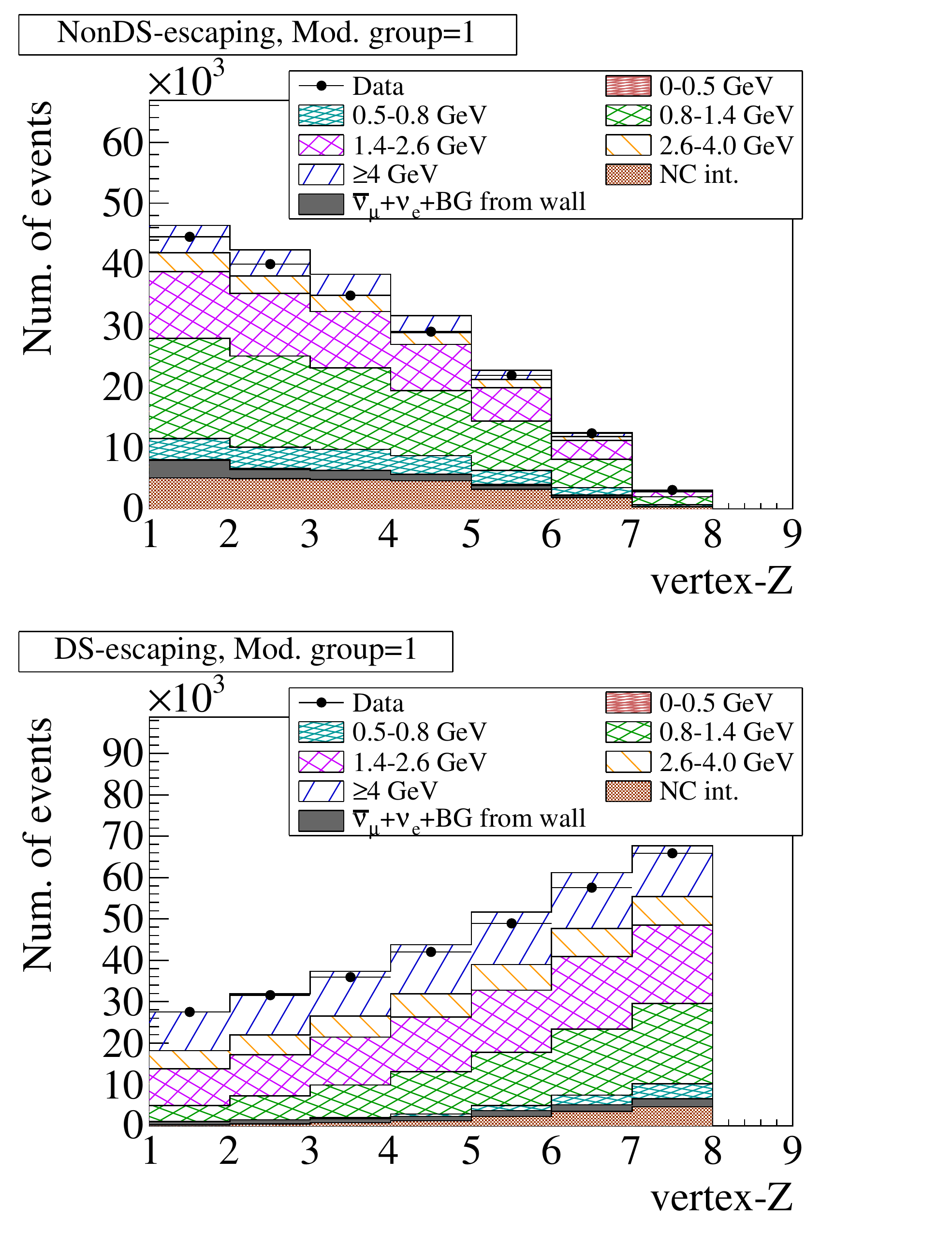} 
   \includegraphics[width=0.3\textwidth]{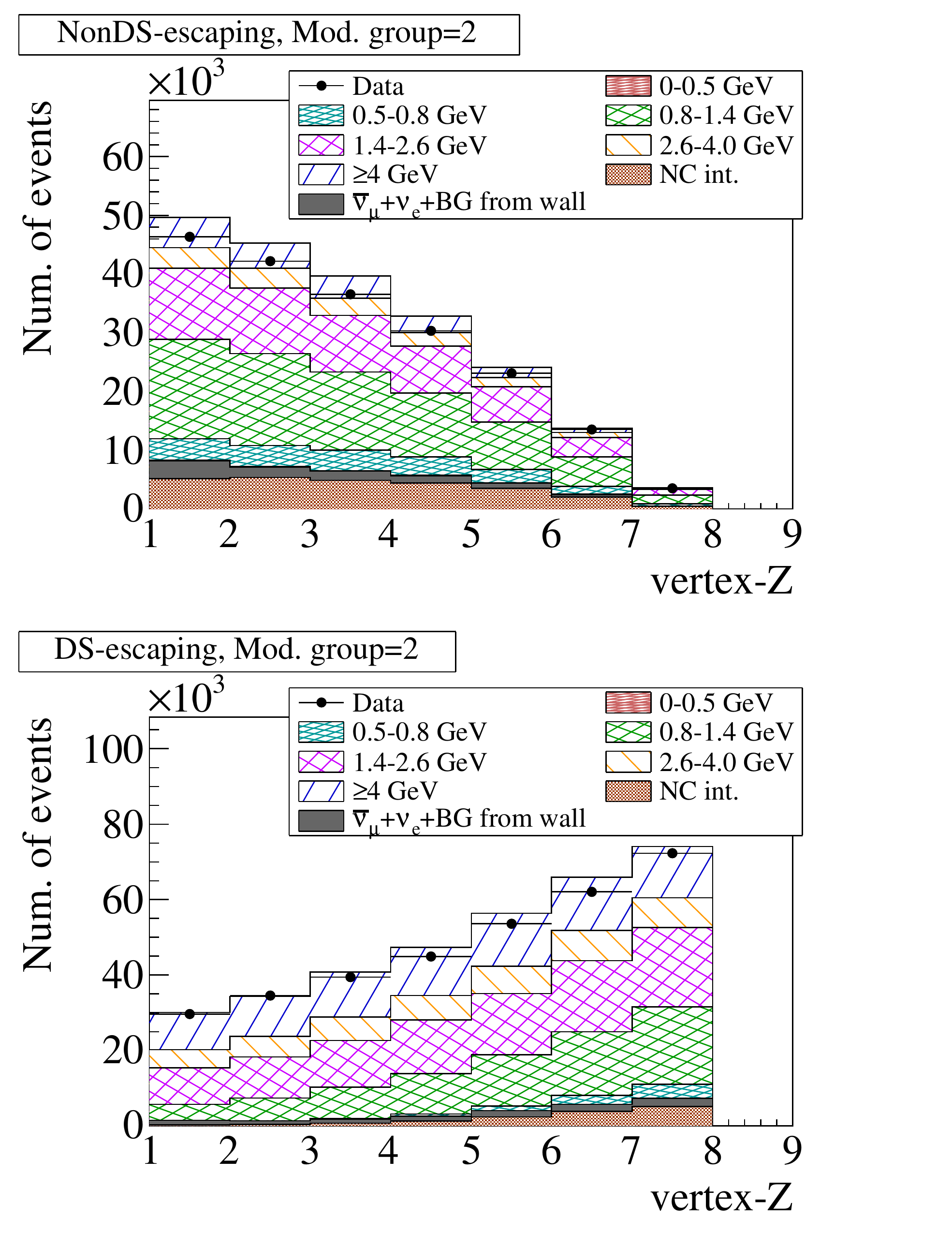} 
   \includegraphics[width=0.3\textwidth]{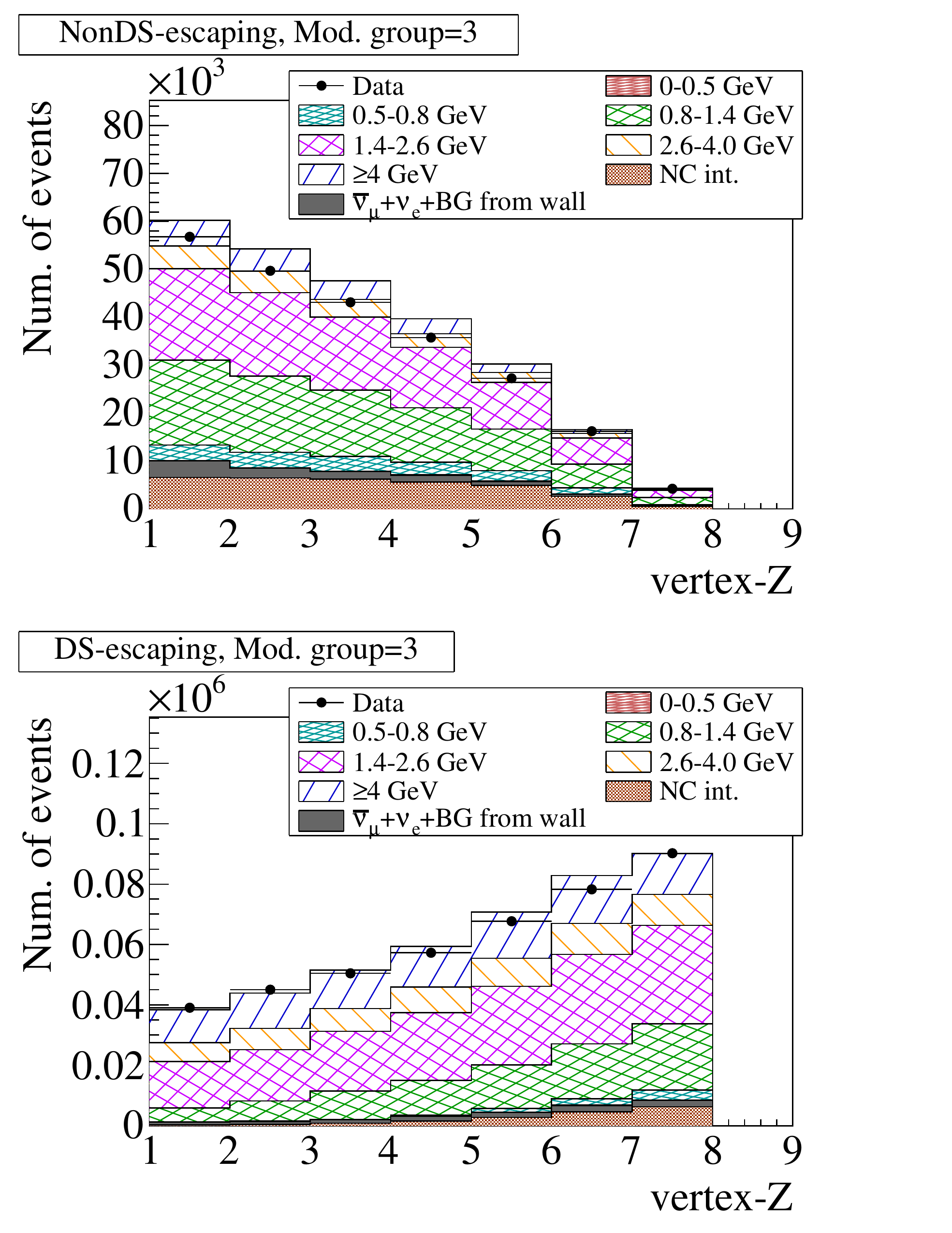} 
   \includegraphics[width=0.3\textwidth]{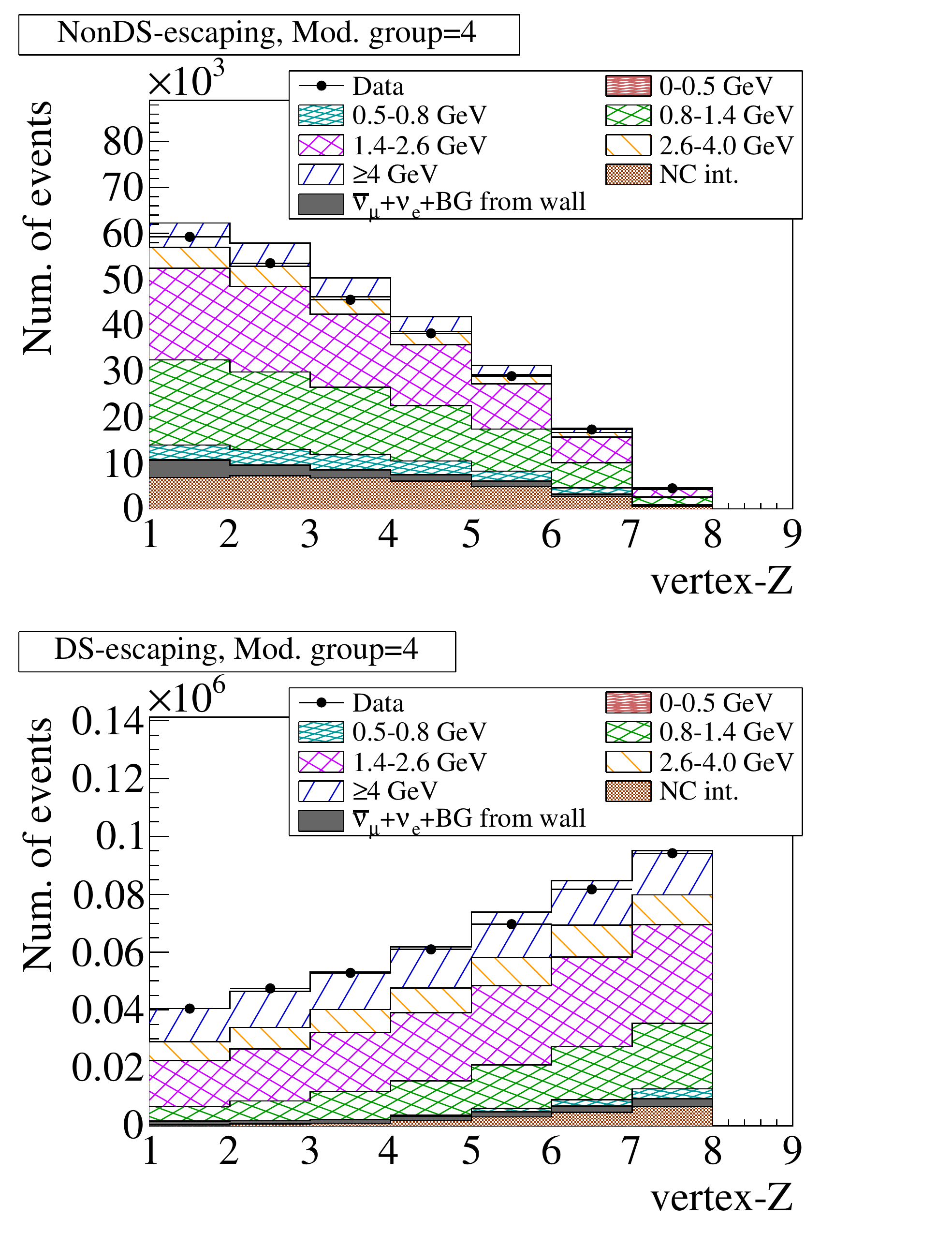} 
   \includegraphics[width=0.3\textwidth]{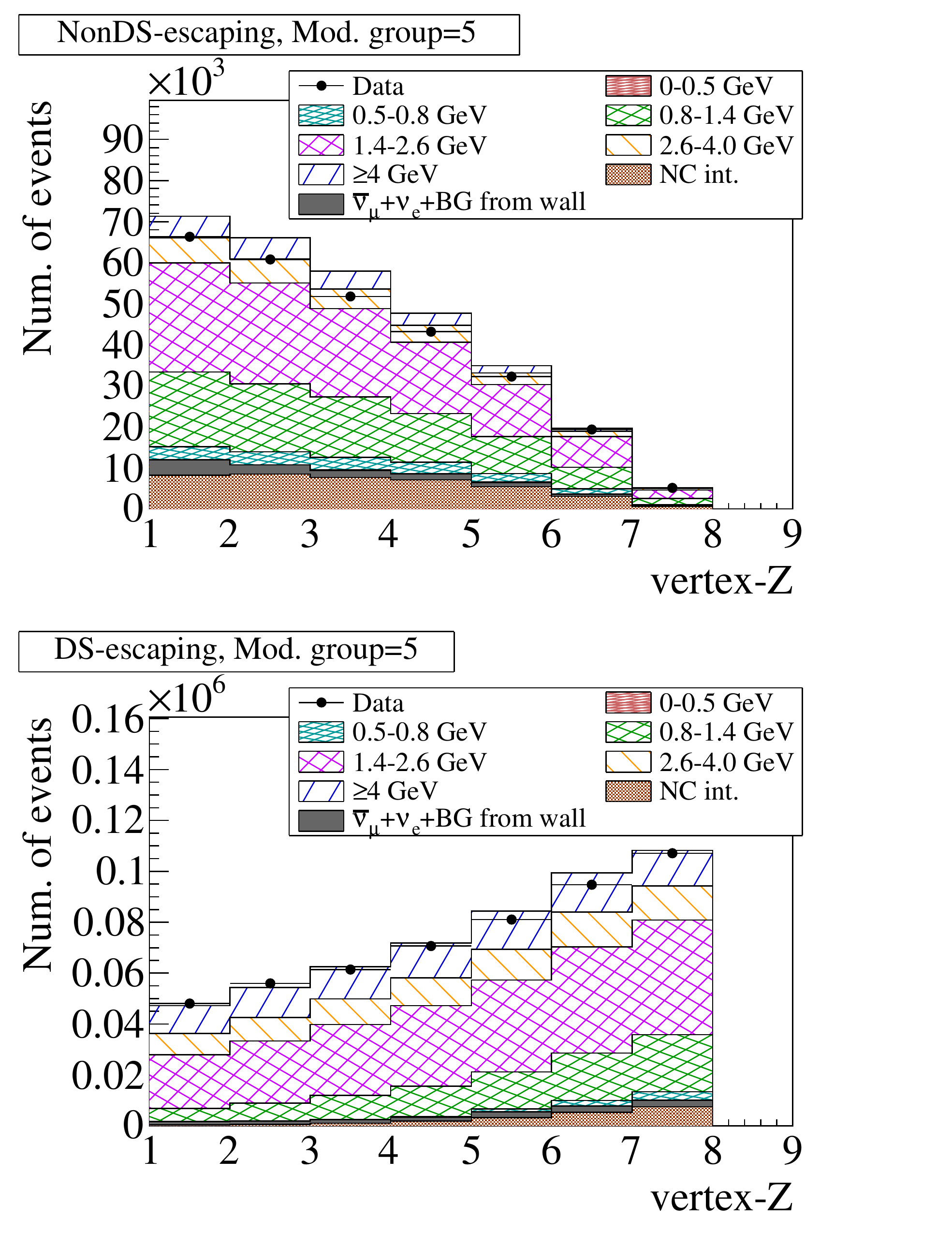} 
   \includegraphics[width=0.3\textwidth]{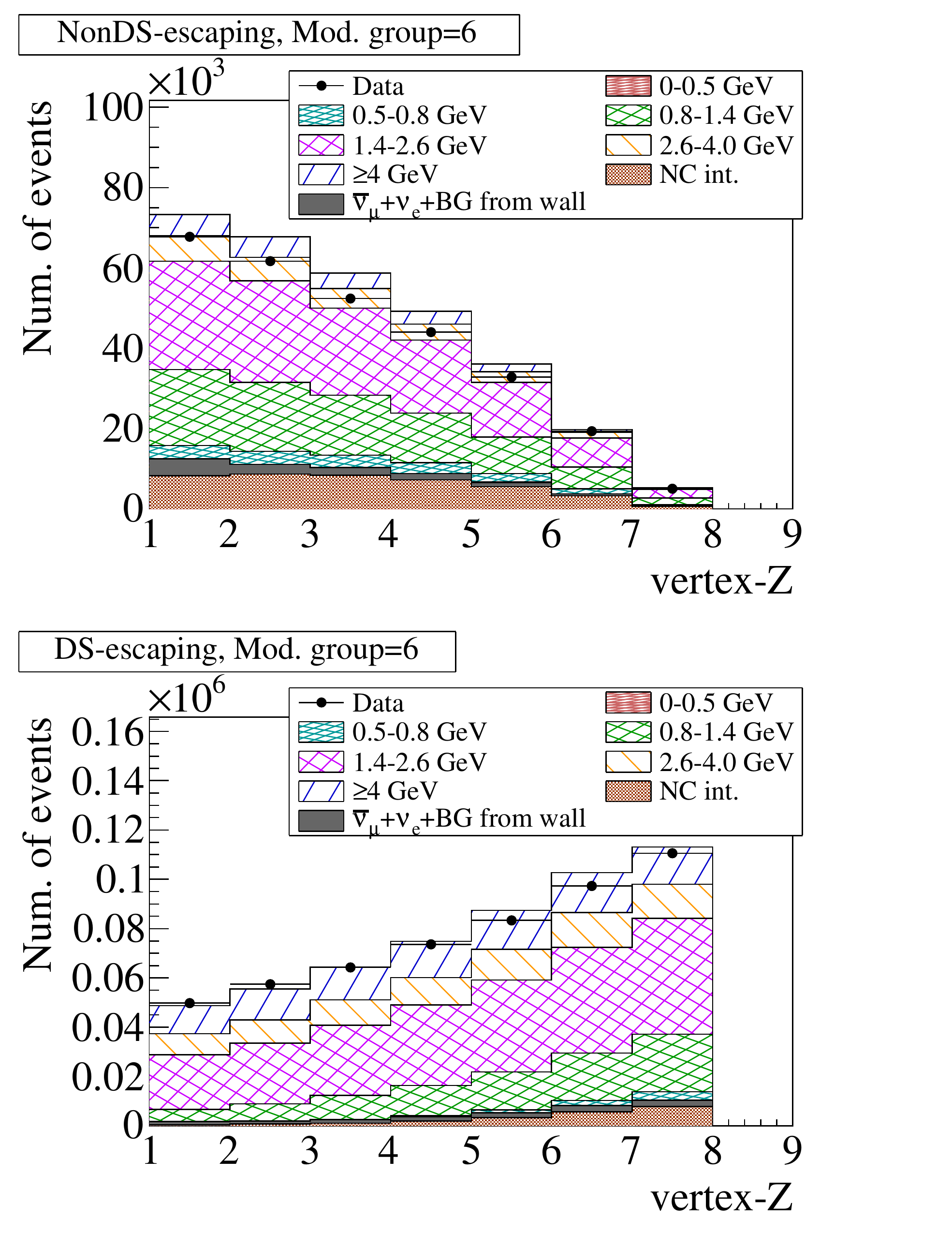} 
   \includegraphics[width=0.3\textwidth]{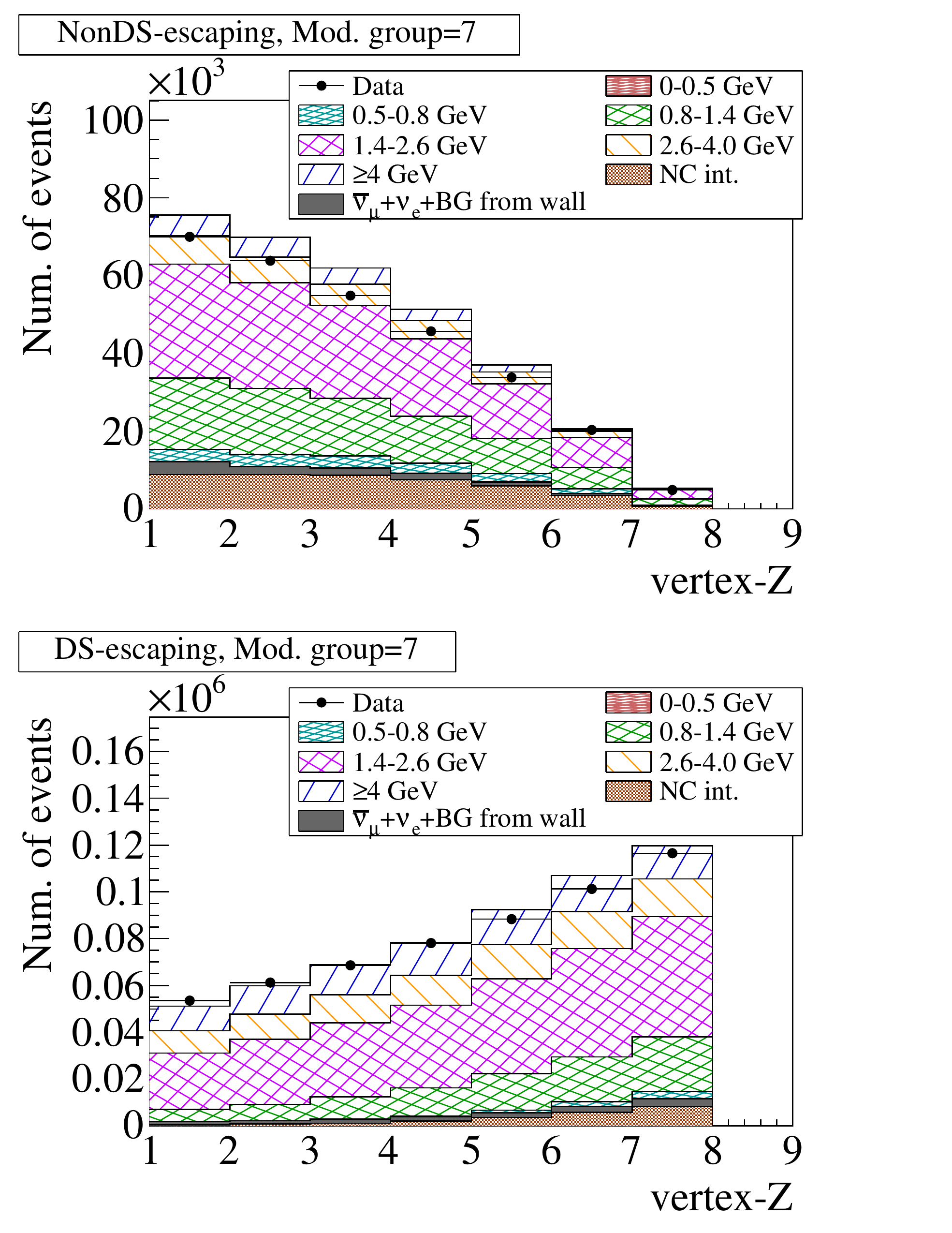} 
   \caption{Topology distribution for NonDS-escaping (top) and DS-escaping (bottom)
   for all module groups.
The predicted events, before the fit, are categorized as CC events, NC
events, and either intrinsic beam $\bar{\nu}_\mu + \nu_e$ backgrounds or backgrounds from the wall, and shown as a stacked histogram. 
The CC events are subdivided into 6 true neutrino energy regions.}
   \label{fig:before_fit_tot}
\end{figure*}

Figure~\ref{fig:before_fit_tot} shows
the observed and predicted topology distributions
in all module groups
for the data set used in this analysis, corresponding to $6.27\times10^{20}$~p.o.t.
The number of observed events for the NonDS-escaping topology is 3-10\% smaller than expected.

%% file: xsec_fit.tex
\subsection{Cross section fit\label{subsec:xsec_fit}}
\begin{figure*}[p]
\centering
\includegraphics[width=0.3\textwidth]{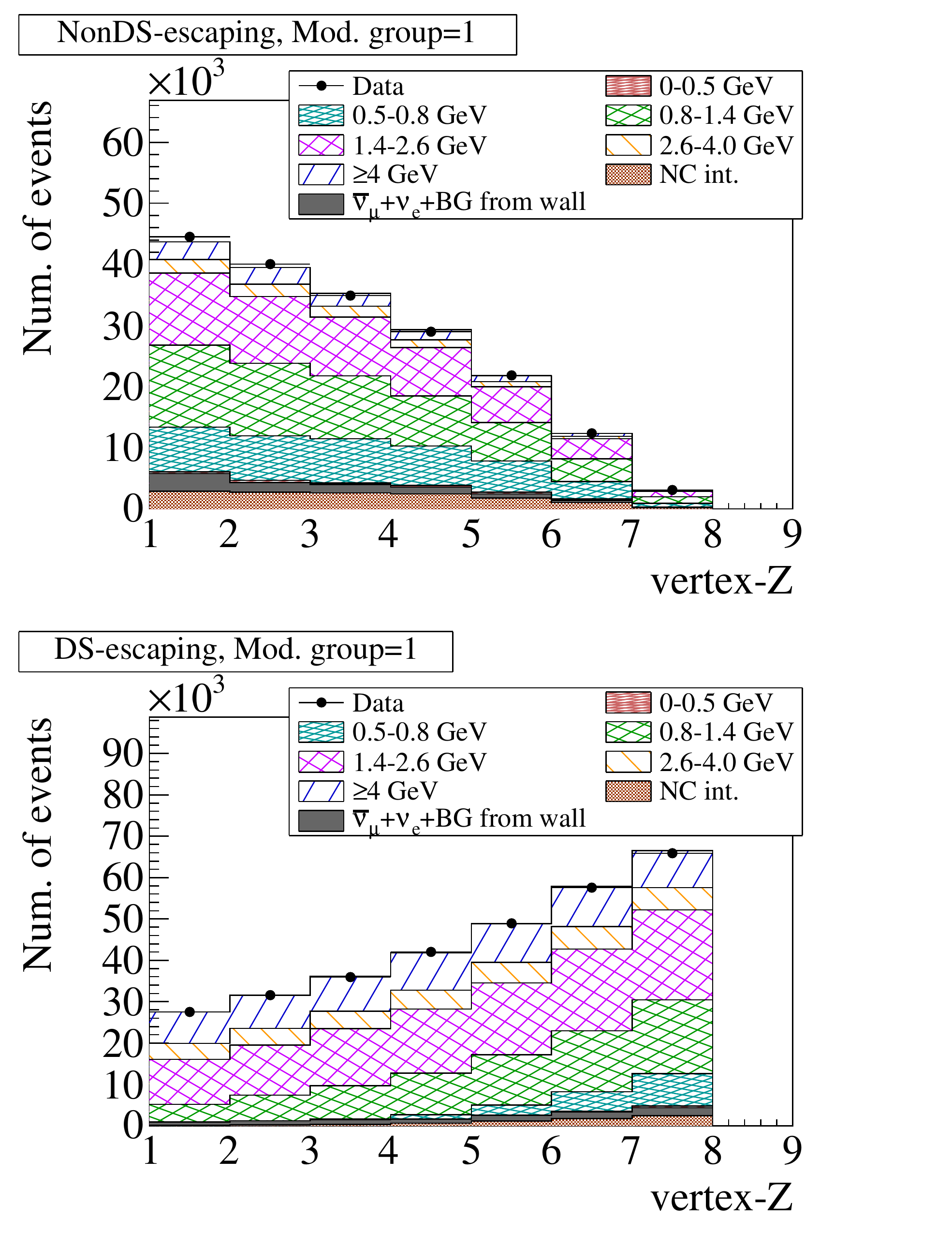}
\includegraphics[width=0.3\textwidth]{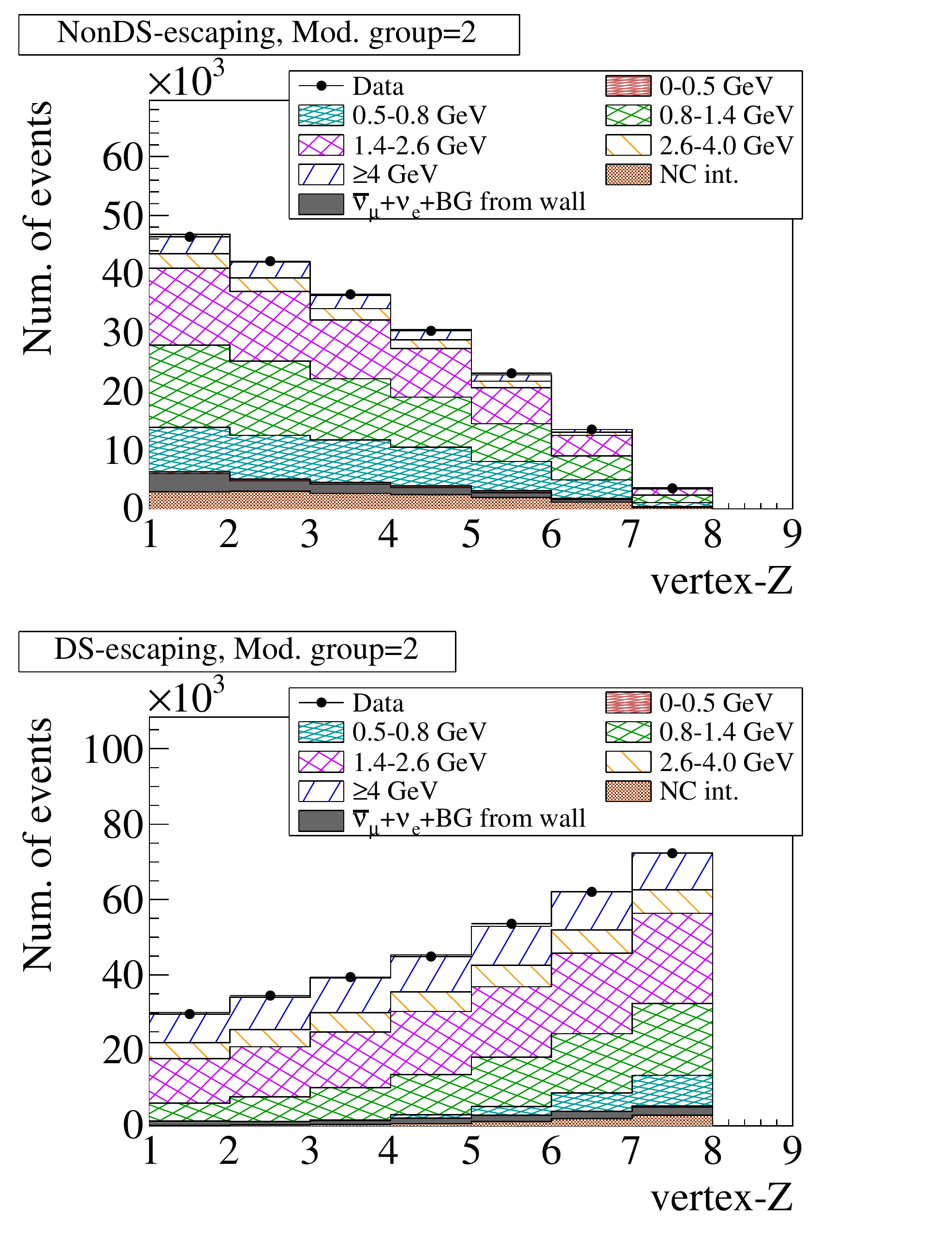}
\includegraphics[width=0.3\textwidth]{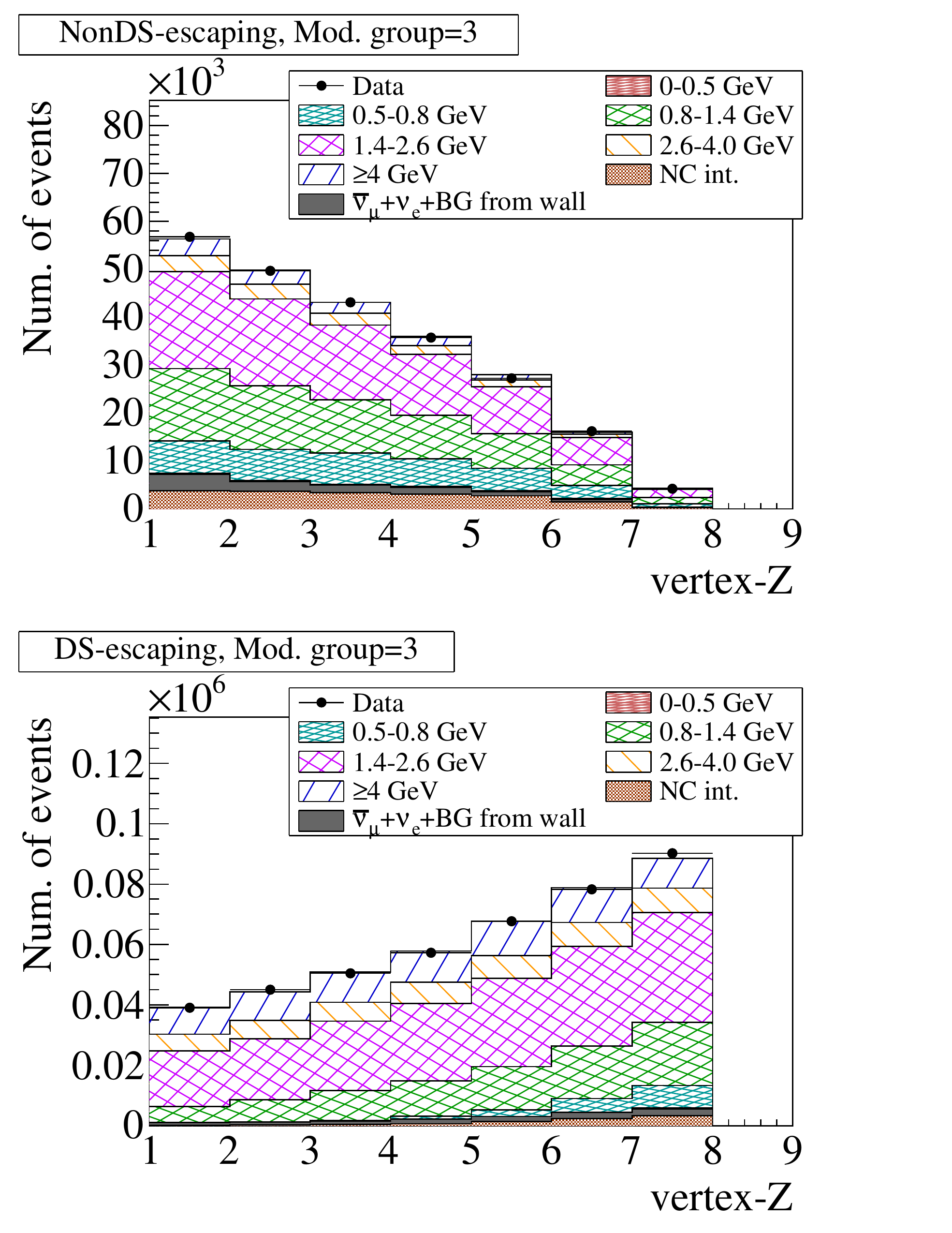}
\includegraphics[width=0.3\textwidth]{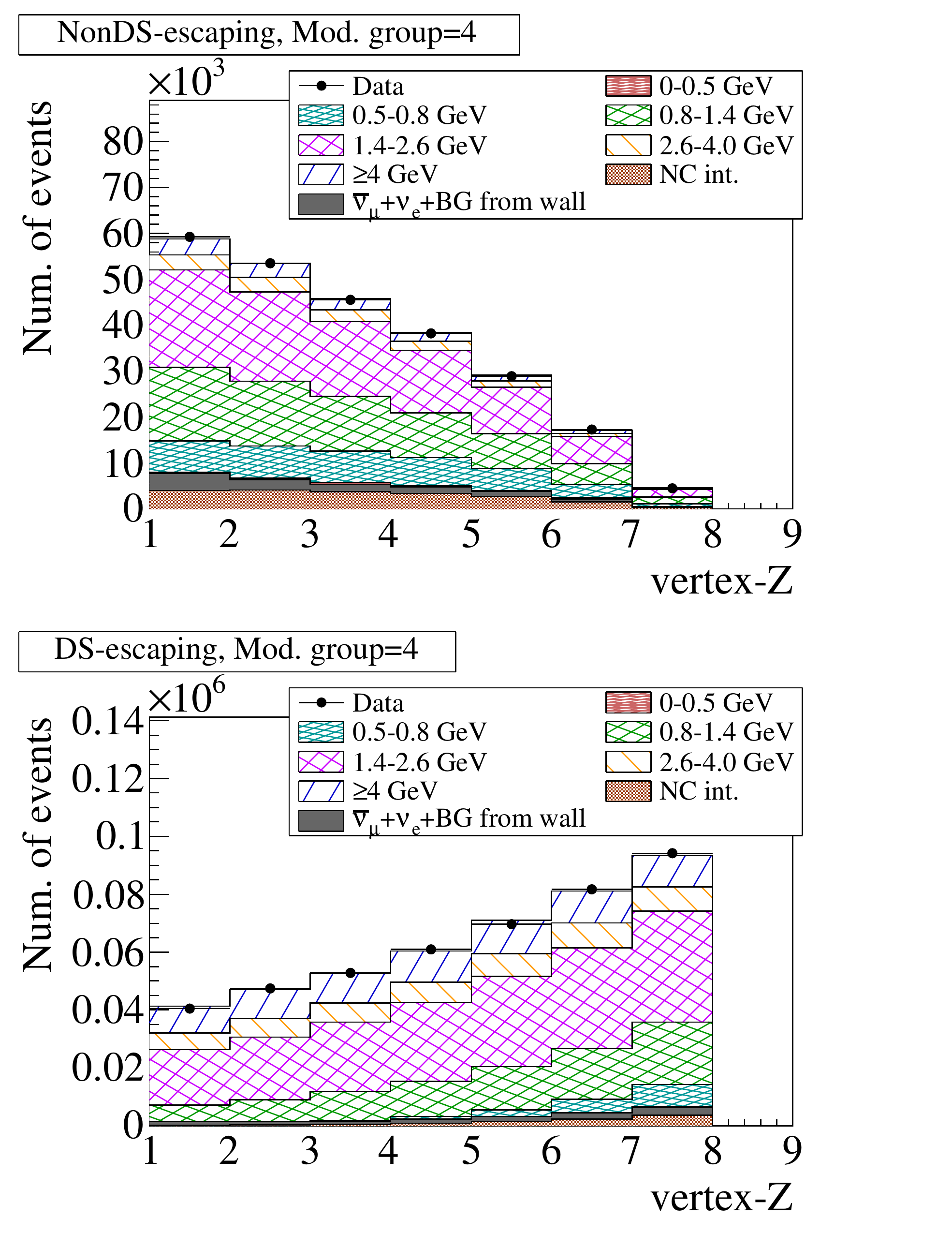}
\includegraphics[width=0.3\textwidth]{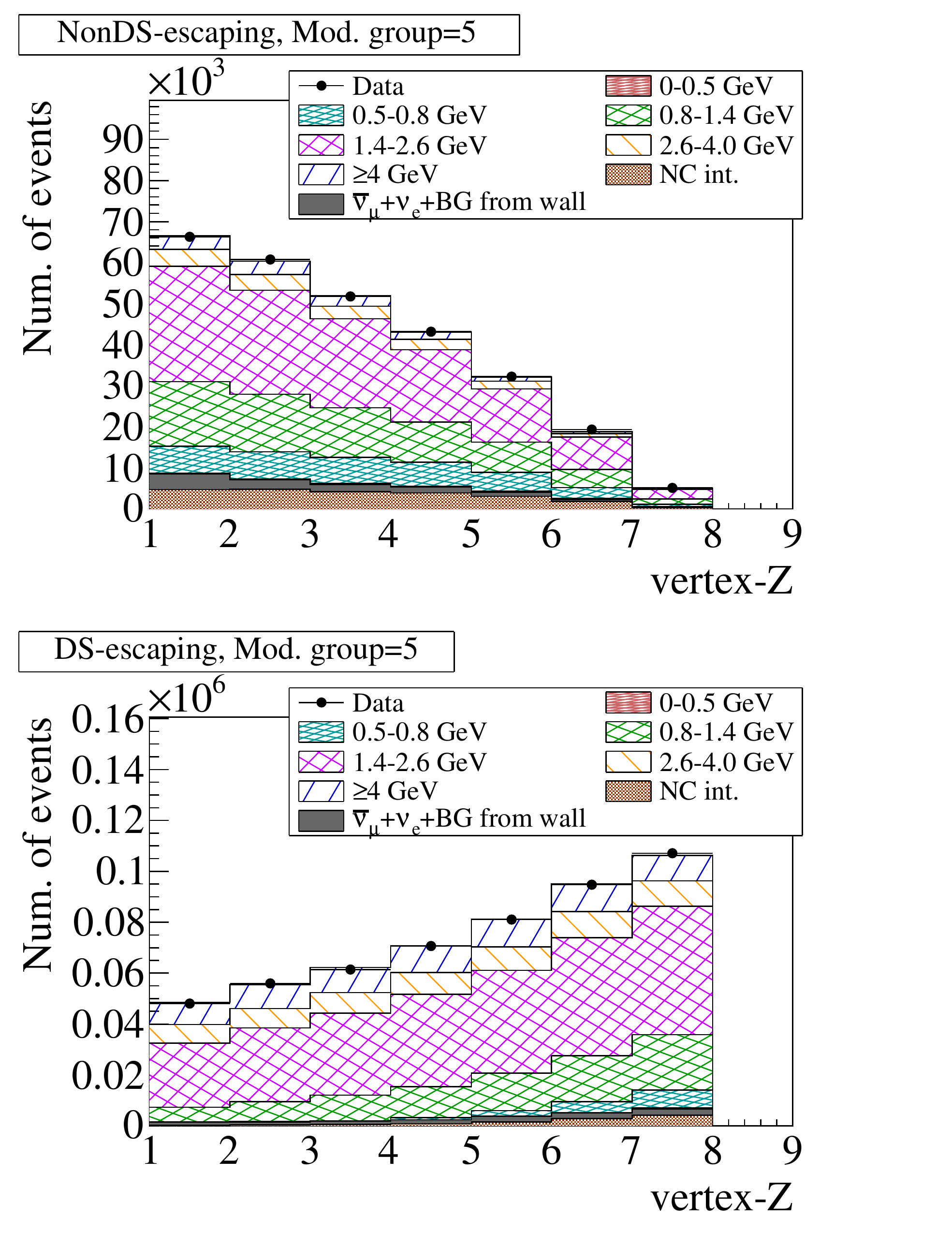}
\includegraphics[width=0.3\textwidth]{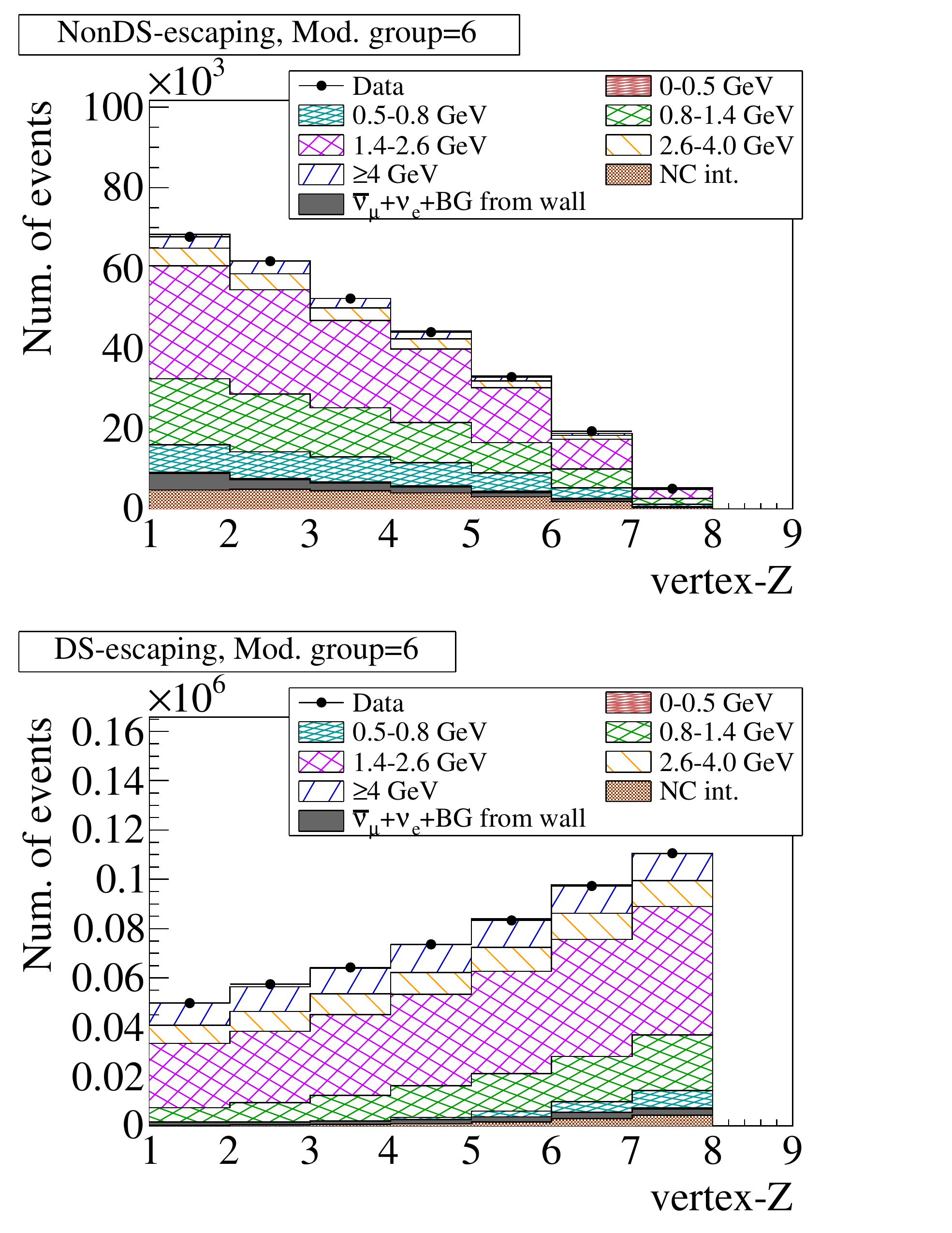}
\includegraphics[width=0.3\textwidth]{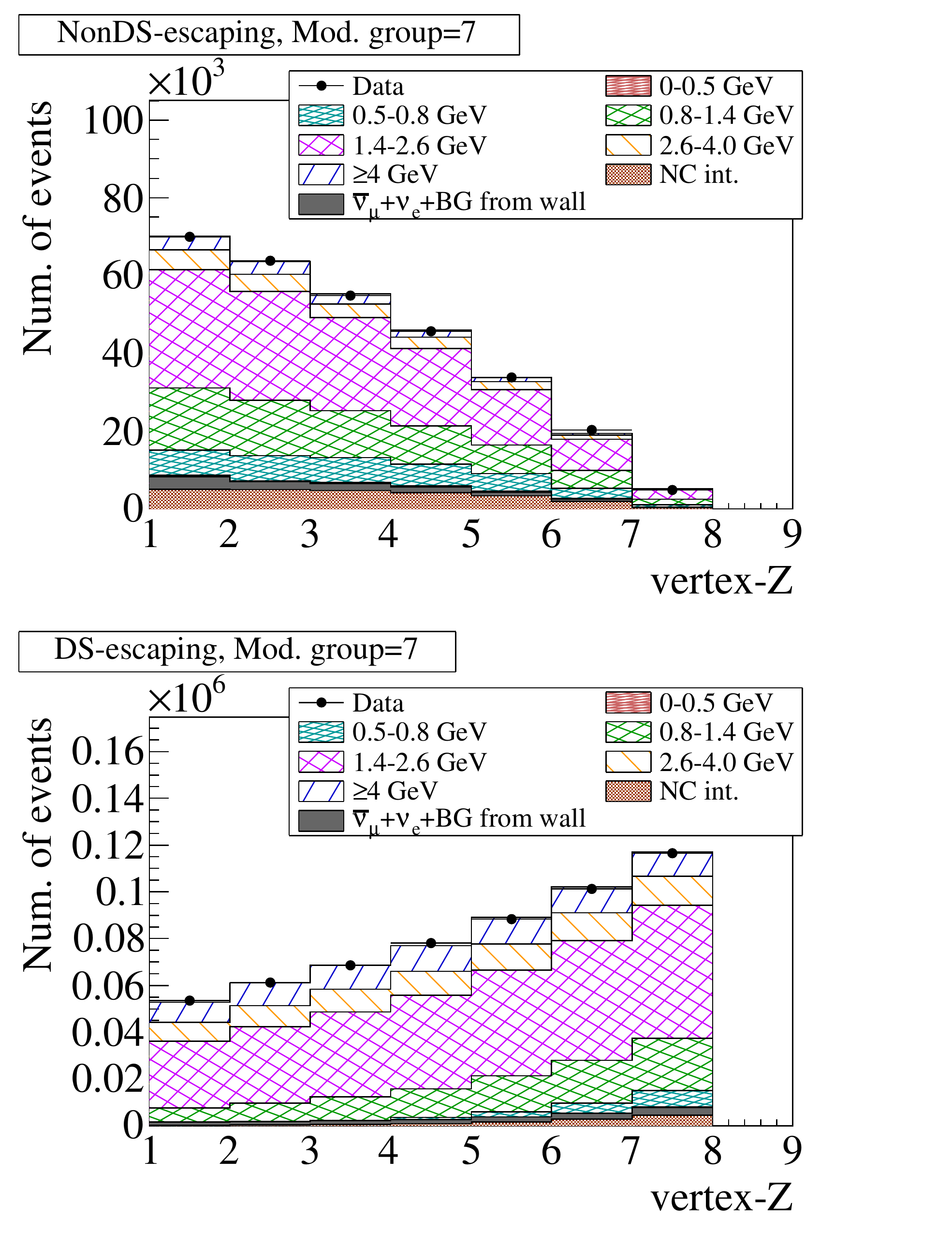}
\caption{Topology distribution for NonDS-escaping (top) and DS-escaping (bottom) events
for all module groups
after
the data fit.
The predicted events are categorized as CC events, NC
events, and either intrinsic beam $\bar{\nu}_\mu + \nu_e$ backgrounds or backgrounds from the wall, and shown as a stacked histogram. 
The CC events are subdivided into 6 true neutrino energy regions.}
\label{fig:after_fit_tot}
\end{figure*}

The topology distributions after the data fit
are shown in 
Fig.~\ref{fig:after_fit_tot}.
As seen in the figure,
the predicted topology distributions
with the fitted cross-section normalization parameters applied
agree well with the observed distributions.

Table~\ref{tbl:fit_result}
shows the 
cross-section normalization parameters,
$f_i = 1 + \Delta f_i$ (i=0-4),
obtained by fitting
the INGRID data,
where the $\Delta f_i$'s are those from
Eq.~(\ref{eq:Npred2}).

\begin{table}[htbp]
\begin{center}
\caption{Fitted values of the cross-section normalization parameters,
$f_i = 1+\Delta f_i$.
}
\label{tbl:fit_result}
\begin{tabular}
{c|c}
\hline\hline
Fit parameter & Fit result \\
\hline
$f_0$ (0.5 GeV) & 3.560 $\pm$ 0.508 \\
$f_1$ (0.8 GeV) & 0.637 $\pm$ 0.180 \\ 
$f_2$ (1.4 GeV) & 1.324 $\pm$ 0.181 \\
$f_3$ (2.6 GeV) & 0.800 $\pm$ 0.211 \\
$f_4$ (4.0 GeV) & 0.712 $\pm$ 0.120 \\
\hline
$\chi^2$ & 155.4\\
\hline\hline
\end{tabular}
\end{center}
\end{table}

\begin{figure}[!h]
\centering
\includegraphics[width=0.45\textwidth]{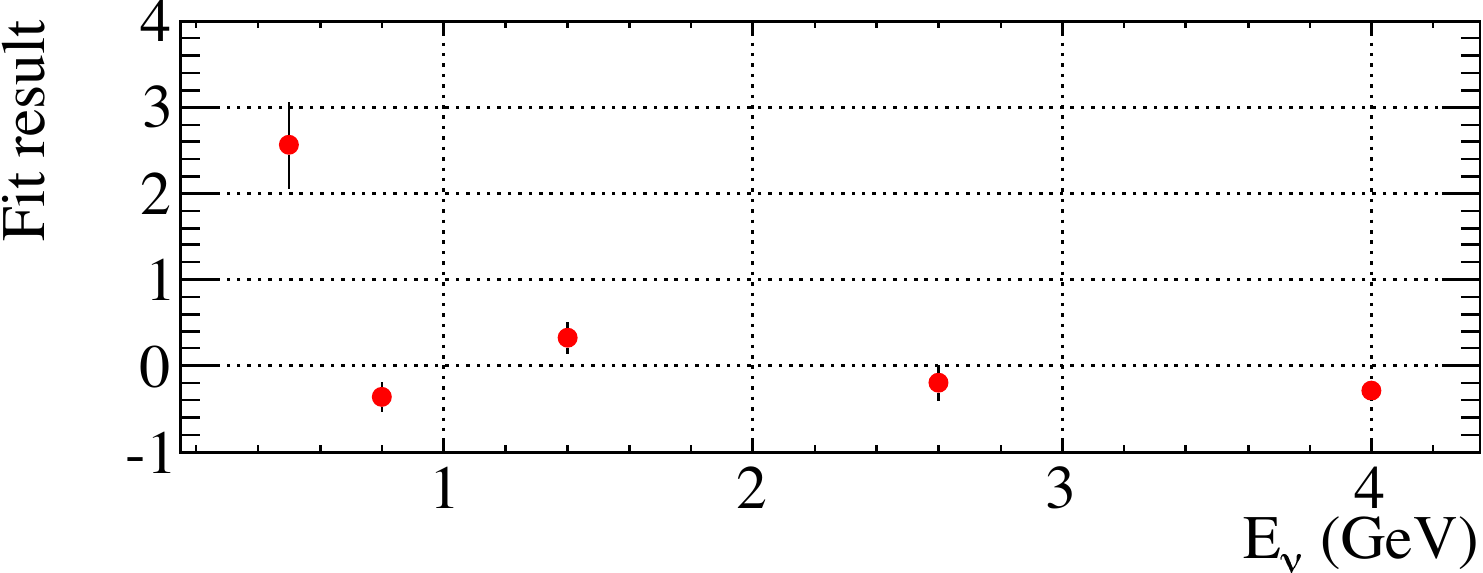}
\includegraphics[width=0.45\textwidth]{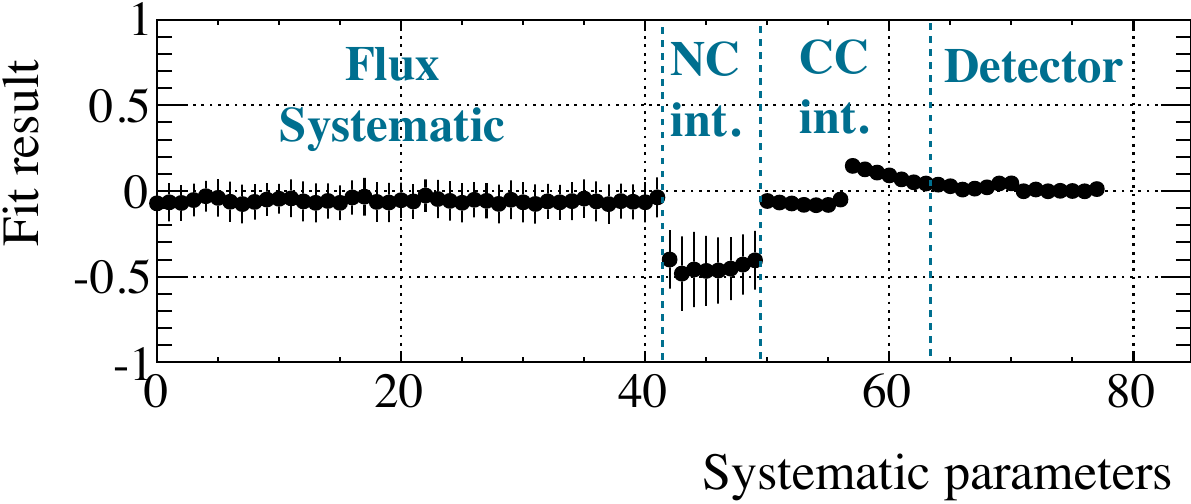}
\caption{
Fitted values of the cross-section normalization (top)
and systematic (bottom) parameters.
}
\label{fig:fit_result_all_param}
\end{figure}

The fitted values for the
flux, detector, CC interaction, and NC interaction uncertainty parameters
are shown in Fig.~\ref{fig:fit_result_all_param}.
A large deviation from 0 is seen 
for all the NC interaction uncertainty parameters.
As described in Sec.~\ref{subsec:data_set},
the prediction overestimates 
the number of NonDS-escaping events
by 3-10\%.

The fitter preferentially reduces
the number of NC events
to match the predicted topology distribution
to the observed one.
Since the NC interaction uncertainty parameters
are almost fully correlated among topologies, 
as shown in Fig.~\ref{fig:ncint_err_a},
all the parameters move toward negative values.
There are jumps seen
in the CC interaction and detector uncertainty parameters.
Both of the jumps appear at the boundary between
parameters for NonDS-escaping and DS-escaping events.

In order to derive the normalization factor for the cross section,
we take the average of the neighboring fitted $f_i$ parameters.
The obtained cross-section normalizations are:
\begin{align*}
f(1.1\text{~GeV}) &= \frac{f_1 + f_2}{2} \\ 
&= 0.980 \pm 0.115 \;, \\
f(2.0\text{~GeV}) &= \frac{f_2 + f_3}{2} \\
&= 1.062 \pm 0.123 \;, \\
f(3.3\text{~GeV}) &= \frac{f_3 + f_4}{2} \\
&= 0.756 \pm 0.136 \;, \\
\end{align*}

Table~\ref{tab:error_summary_fit} 
summarizes the uncertainty on the fitted cross-section normalization 
parameters, broken down by error source.
The errors on the combined normalization 
parameters are summarized in Table~\ref{tab:error_summary_xsec}.
The largest systematic error source
is the flux uncertainty, which gives a 8-9\% error on the cross-section normalization.
The cross-section normalization at 2.0~GeV
is less affected by most of the systematic uncertainties than the other normalizations, as shown in Table~\ref{tab:error_summary_xsec}.
The reason for this is as follows.
In Fig.~\ref{fig:pdf},
one can see that 
the PDF is well differentiated 
for both module group and event topology
at $E_\nu$=1.4-2.6~GeV ($f_2$-$f_3$) and $E_\nu$=2.6-4.0~GeV ($f_3$-$f_4$),
while poor differentiation is seen for $E_\nu\geq$4.0~GeV ($f_4$),
which results in weak correlation with the other cross-section normalizations.
Thus, 
sensitivity to the cross-section normalization
is good for $E_\nu$=2.0~GeV ($=\frac{f_2 + f_3}{2}$)
but limited for $E_\nu$=3.3~GeV ($=\frac{f_3 + f_4}{2}$).

\begin{table}[h]
\caption{Contribution to the uncertainty on the fitted parameters ($f_0$-$f_4$) from each error source.}
\centering
\scalebox{0.8}{
\begin{tabular}{l c c c c c}
\hline\hline
Error source & $f_0$      & $f_1$      & $f_2$      & $f_3$       & $f_4$  \\
             & (0.5~GeV)  & (0.8~GeV)  & (1.4~GeV)  & (2.6~GeV)   & (4.0~GeV) \\
 \hline
Statistical error & 18.7\% & 6.0\% & 4.5\% & 4.8\% &1.4\% \\
Flux + Stat. & 26.0\% & 7.9\% & 12.8\% & 14.5\% & 9.3\%\\
Detector + Stat. & 33.8\% & 10.0\% & 7.2\% & 7.0\% & 3.0\%\\
Interaction (cc) + Stat. & 30.6\% & 9.3\% &6.8\% & 7.2\% & 3.8\%\\
Interaction (nc) + Stat. & 22.6\% & 6.6\% & 6.4\% & 5.8\% & 2.0\%\\
\hline
pion FSI & $^{+0.4\%}_{-0.2\%}$ & $^{+0.3\%}_{-0.8\%}$ 
& $^{+2.0\%}_{-3.3\%}$ &$^{+4.0\%}_{-3.6\%}$ & $^{+3.5\%}_{-5.2\%}$\\
pion multiplicity & 2.6\% & 8.8\% & 0.7\% & 12.4\% & 9.4\% \\
pion SI & 7.3\% & 9.5\% & 9.4\% & 11.4\% & 2.9\% \\
\hline\hline
\end{tabular}
}
\label{tab:error_summary_fit}
\end{table}%

\begin{table}[h]
\caption{Contribution to the uncertainty on the cross-section normalization at 1.1, 2.0, and 3.0~GeV from
each error source.}
\begin{center}
\begin{tabular}{l c c c }
\hline\hline
 Error source &  1.1~GeV  & 2.0~GeV  & 3.3~GeV \\
 \hline
Statistical error & 2.0\% & 0.6\% & 2.3\% \\
Flux + Stat. & 7.6\% & 9.0\% & 8.4\% \\
Detector + Stat. & 4.3\% & 0.9\% & 3.9\% \\
Interaction (cc) + Stat. & 3.7\% & 0.8\% & 4.8\%\\
Interaction (nc) + Stat. & 2.4\% & 0.9\% & 3.2\%\\
\hline
pion FSI & $^{+1.0\%}_{-1.9\%}$ &  0.5\%  & $^{+3.7\%}_{-2.9\%}$\\
pion multiplicity & 3.3\% & 5.1\% & 2.1\% \\
pion SI & 5.6\% & 2.0\% & 6.9\% \\
\hline\hline
\end{tabular}
\end{center}
\label{tab:error_summary_xsec}
\end{table}%

\begin{figure}[!h] 
   \centering
   \includegraphics[width=0.23\textwidth]{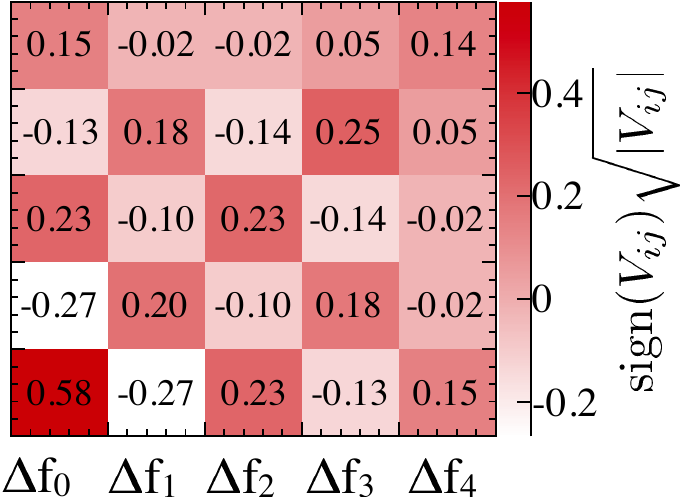}
       \includegraphics[width=0.22\textwidth]{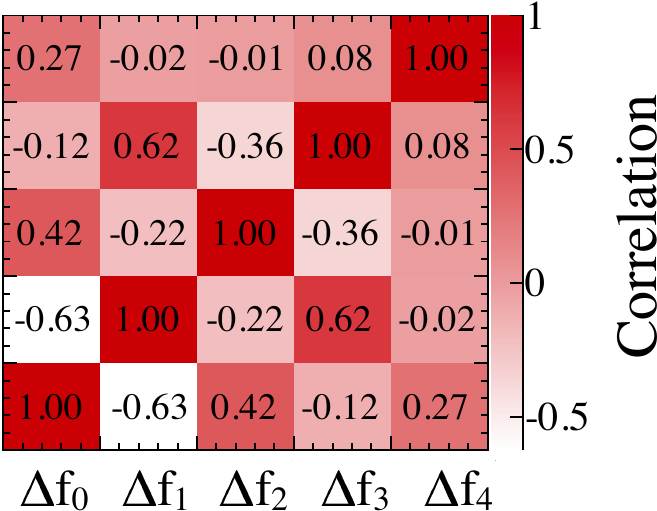}
   \caption{Error (left) and correlation (right) matrices for the 5 fitted parameters.
   In both of the matrices, the binning on the y-axis is identical to that on the x-axis.}
   \label{fig:cov_5p}
\end{figure}

\begin{figure}[!h] 
   \centering
   \includegraphics[width=0.23\textwidth]{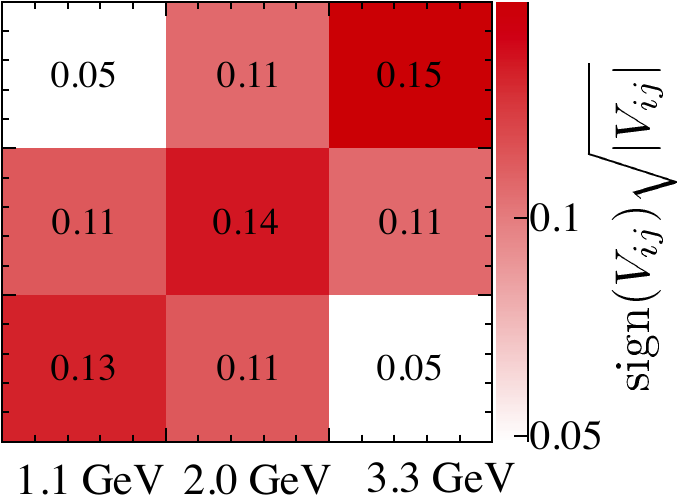}
   \includegraphics[width=0.22\textwidth]{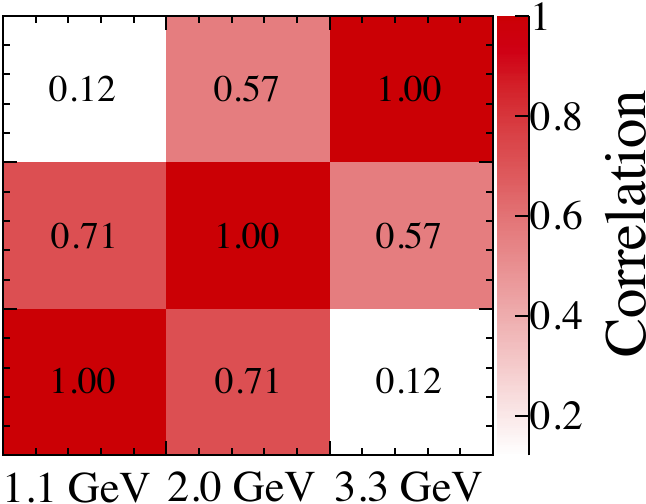}
   \caption{Error (left) and correlation (right) matrices for the cross-section normalization
   at 1.1, 2.0, and 3.3~GeV.
   In both of the matrices, the binning on the y-axis is identical to that on the x-axis.
   }
   \label{fig:cov_3p}
\end{figure}

Figures~\ref{fig:cov_5p} and \ref{fig:cov_3p}
show the error and correlation matrices
for the 5 fitted parameters ($\Delta f_0$-$\Delta f_4$)
and the cross-section 
normalizations
at 1.1, 2.0, and 3.3~GeV, respectively.

%% file: summary.tex
\begin{figure*}[t] 
   \centering
   \includegraphics[width=0.8\textwidth]{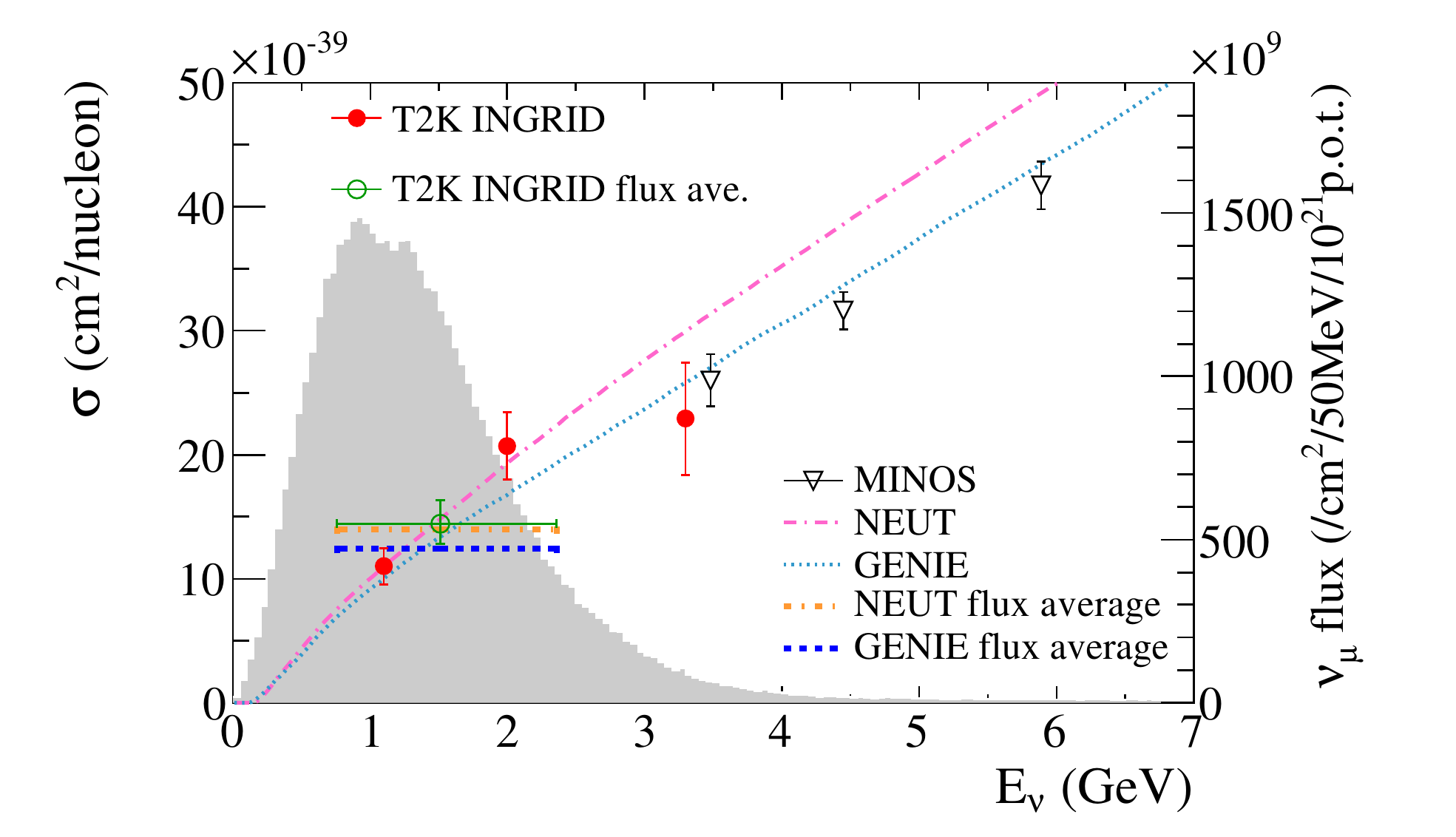}
   \caption{Results of the $\nu_\mu$ CC inclusive cross section on Fe.
The energy dependent cross section measured by the MINOS near detector~\cite{Adamson:2009ju}
and the flux-averaged cross section from INGRID~\cite{Abe:2014nox}
 are shown with the NEUT (v.5.1.4.2) and GENIE (v.2.8.0) predictions.
The T2K on-axis $\nu_\mu$ flux is shown in gray.
The T2K INGRID flux-averaged cross-section measurement 
and this result are consistent with one another.}
 \label{fig:xsec_result}
\end{figure*}

\subsection{Summary\label{subsec:summary}}

In the previous section,
five individual fitting parameters were extracted
with the least $\chi^2$ method,
and used to calculate
the following 
measured cross sections at energies of 1.1, 2.0, and 3.3~GeV:
\begin{align*}
\sigma^{cc}(1.1\text{~GeV}) 
&= 1.10 \pm 0.15 \; (10^{-38}\text{cm}^2/\text{nucleon}) \;, \nonumber \\
\sigma^{cc}(2.0\text{~GeV}) 
&= 2.07 \pm 0.27 \; (10^{-38}\text{cm}^2/\text{nucleon}) \;, \nonumber \\
\sigma^{cc}(3.3\text{~GeV}) 
&= 2.29 \pm 0.45 \; (10^{-38}\text{cm}^2/\text{nucleon}) \;. \nonumber \\
\end{align*}
Figure~\ref{fig:xsec_result} shows 
these results
compared to other measurements \cite{Adamson:2009ju,Abe:2014nox}
 and the neutrino event generators, NEUT (ver.5.1.4.2) and GENIE (ver.2.8.0).
 These measurements are consistent with 
 the energy dependent cross section measured by the MINOS near detector 
 and the previous, flux-averaged, cross section measured by INGRID,
which used a subset 
of the data included in this analysis.

This analysis utilizes the different off-axis angle technique together with the final state kinematics of the out-going lepton
to enhance sensitivity to the neutrino energy.  Using final state lepton kinematics in this analysis makes the result sensitive
to uncertainties in the neutrino interaction model, which increases the uncertainty on the final measurement.  As a result, the
errors achieved are not small enough to distinguish between the neutrino models used in the different event generators.
Nevertheless, the result seems to prefer NEUT for $E_\nu\leq$2~GeV and agrees with the MINOS data point and GENIE at
$E_\nu$=3.3~GeV.  Further reduction of the systematic error size would help in differentiating the neutrino models at the
higher energy transition. This reduction could be made by utilizing neutrino beams covering a wider range of off-axis angles,
which would provide a ``model independent'' way to infer the neutrino energy.  This analysis demonstrates the feasibility of
using different off-axis samples from the same neutrino beam to measure the energy dependence of neutrino interactions, which
will provide useful information for future neutrino oscillation analyses.
%
%

%% file: conclusion.tex
\section{Conclusion\label{sec:conclusion}}

In this paper, we have reported the measurement of the energy dependent inclusive $\nu_\mu$ charged current cross section on iron using the T2K INGRID detector
and the T2K neutrino beam.
The unique variation in the neutrino flux across the INGRID modules, along with event kinematic information, was used to produce event samples sensitive to neutrinos with energies from 1-3~GeV. 
These were used to extract the inclusive CC muon neutrino interaction cross section on iron at 1.1, 2.0 and 3.3 GeV, 
using data corresponding to $6.27\times10^{20}$~p.o.t.
This result is consistent with the predictions of the NEUT and GENIE neutrino interaction generators.
This is the first measurement to use the off-axis effect to measure neutrino cross sections 
as a function of energy and demonstrates the potential of this technique to provide useful information for future neutrino experiments.